\RequirePackage{lineno}

\documentclass[onecolumn,floatfix,superscriptaddress,a4paper,
               showpacs,showkeys,nofootinbib,preprint]{revtex4}

\usepackage{epsfig}
\usepackage{latexsym}
\usepackage{xspace}
\usepackage{hyperref}
\usepackage[latin2]{inputenc}
\usepackage{indentfirst}
\usepackage{enumerate}

\usepackage{amsmath}
\usepackage{amssymb}
\usepackage[english]{babel}
\usepackage{url}

\begin{document}
\today
\vspace*{1cm}

\title{ Search for critical behavior of strongly interacting
        matter \\ at the CERN Super Proton Synchrotron }
\author{M. Gazdzicki}
 \affiliation{Goethe--University, Frankfurt, Germany}
 \affiliation{Jan Kochanowski University, Kielce, Poland}
\author{P. Seyboth}
\affiliation{Max-Planck-Institut fuer Physik, Munich, Germany}
 \affiliation{Jan Kochanowski University, Kielce, Poland}

\begin{abstract}
History, status and plans of the search for critical behavior of
strongly interacting matter created in nucleus-nucleus collisions
at the CERN Super Proton Synchrotron is reviewed.
In particular, it is expected that the search should
answer the question whether the critical point of strongly interacting matter
exists and, if it does, where it is located.

First, the search strategies are presented and a short introduction is given
to expected fluctuation signals and to the quantities used by experiments to detect them.
The most important background effects are also discussed.

Second, relevant experimental results are summarized and discussed.
It is intriguing that both the fluctuations of 
quantities integrated over the full experimental acceptance
(event multiplicity and transverse momentum) as well as
the bin size dependence of the second factorial moment of pion and proton
multiplicities
in medium-sized Si+Si collisions
at 158$A$~GeV/c suggest critical behaviour of the created matter.

These results provide strong motivation for the ongoing systematic
scan of the phase diagram by the NA61/SHINE experiment at the SPS
and the continuing search at the Brookhaven Relativistic Hadron Collider.
\end{abstract}



\maketitle

\newpage

\tableofcontents
\newpage

\vspace{3cm}

\section{Introduction}

The structure of the phase diagram of strongly interacting matter is
one of the most important topics in nuclear and particle physics.
We know that at low densities strongly interacting particles are
hadrons and thus the matter is in form of a hadron gas or liquid.
Since the discovery of sub-hadronic particles, quarks and gluons,
it was speculated that
at high temperature and/or pressure densely packed hadrons will
''dissolve'' into a new phase of
quasi-free quarks and gluons, the quark-gluon-plasma (QGP)~\cite{qgp}. 

Many years of intense experimental and theoretical studies of
high energy nucleus-nucleus (A+A) collisions led to the conclusion
that the quark-gluon plasma exists in nature.
This conclusion is based on a wealth of systematic data on A+A collisions
at very high energies from the CERN Large Hadron Collider (LHC)~\cite{QM2014:procs}
and the BNL Relativistic Heavy Ion Collider (RHIC) 
(see e.g. Ref.~\cite{Adams:2005dq}), 
and, very importantly
the observation of the transition
between hadronic matter and quark-gluon plasma 
at the CERN Super Proton Synchrotron energies~\cite{NA49:ood} 
(for recent review see~Refs.~\cite{Gazdzicki:2010iv,Gazdzicki:2014sva}).
 
Thus the current key question in the study of the phase diagram
of strongly interacting matter is the structure of the phase
transition region between the hadron gas and the quark--gluon plasma. 

\begin{figure}[!htb]
\includegraphics[width=0.80\textwidth]{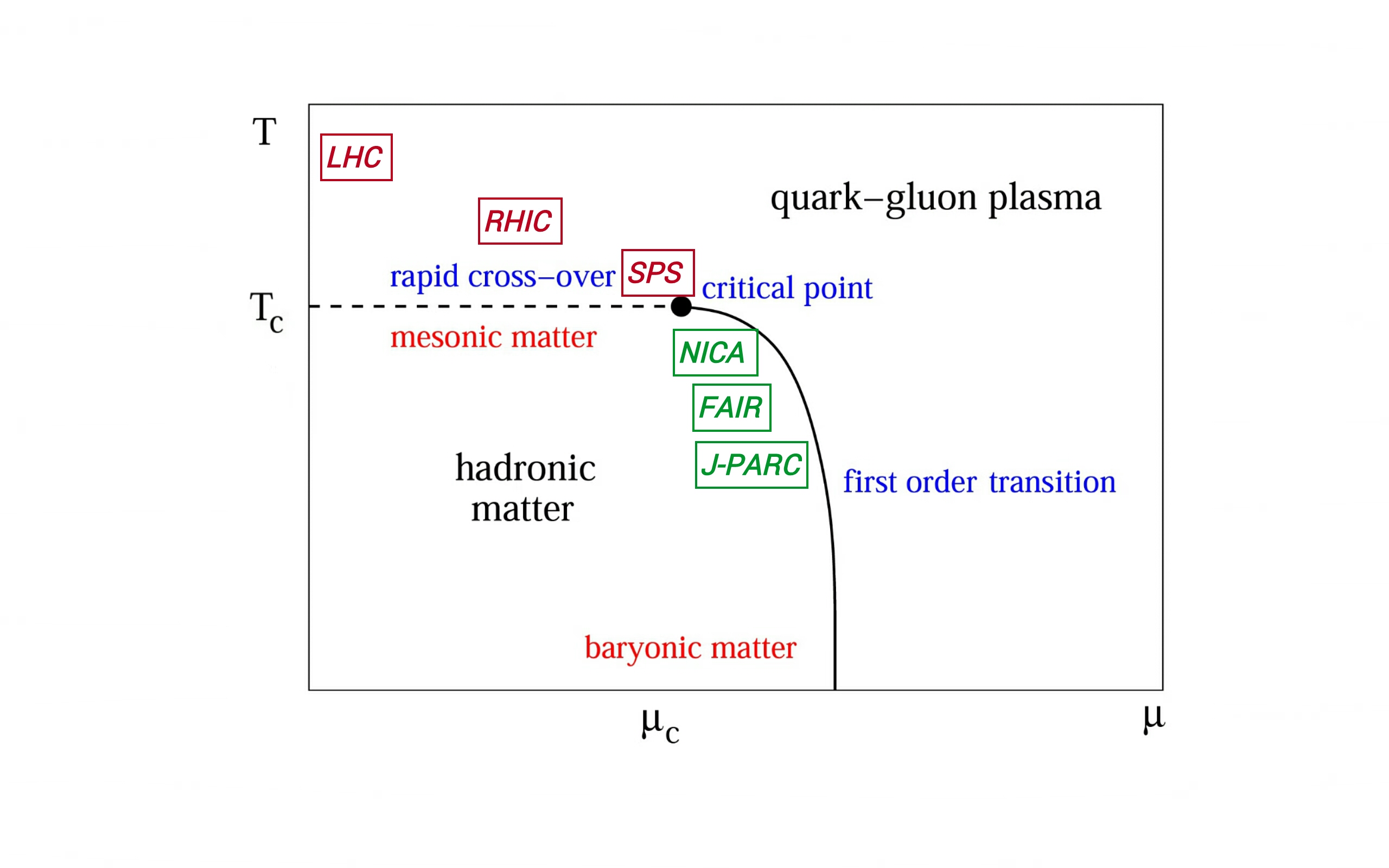}
\caption{
Sketch of the phase diagram of strongly interacting
matter. Italic labels added to the sketch  
show  regions probed
by the early stage of heavy ion collisions studied in current 
(in red) and future (in green) experimental programmes. 
\label{phase_diag}
}
\end{figure}

The most popular suggestion is shown schematically in
Fig.~\ref{phase_diag}. The transition at
low temperature $T$ and high baryochemical potential $\mu_B$ is believed to be 
of the first order and happen along a line which ends with decreasing
$\mu_B$ in a critical point (of the second order) and then turns
into a crossover region. 
The non-trivial structure of the phase transition region was first
suggested by Asakawa, Yazaki~\cite{Asakawa:1989bq} 
and Barducci, Casalbuoni, De Curtis, Gatto, Pettini~\cite{Barducci:1989wi}.
Experimental studies of the features of the phase diagram were strongly 
motivated by predictions of measurable effects. The pioneering work of
Wosiek~\cite{Wosiek:1988} pointed out that
intermittent~\cite{Bialas:1985jb} behaviour is naturally expected 
at a phase transition of the second order. 
Soon after the conjecture was further developed  by Satz~\cite{Satz:1989vj},
Antoniou et al.~\cite{Antoniou:1990vf} and Bialas, Hwa~\cite{Bialas:1990xd}.
This initiated  experimental studies of the structure
of the phase transition region via studies of particle multiplicity fluctuations
using scaled factorial moments. 
Later additional measures of fluctuations were   
also proposed as probes of critical behaviour~\cite{Stephanov:1998dy,Stephanov:1999zu}.
The results of the KLM~\cite{Holynski:1988mr} and 
NA49~\cite{Seyboth:ismd2013} experiments suggest
that effects related to the critical point may have been observed in
collisions of medium size nuclei at the top SPS energy.
This motivated the NA61/SHINE experiment to perform a systematic
scan in collision energy and system size.
The new measurements should answer the general question about the nature of
the transition region and, in particular, the question: 
{\bf does the critical point of strongly interacting matter
exist in nature and, if it does, where is it located?}
The most recent experimental and theoretical status of the exploration
of the phase diagram is reviewed at the
regular workshops on the Critical Point and Onset of Deconfinement~\cite{www.cpod}.

\begin{figure}[!htb]
\includegraphics[width=0.43\textwidth]{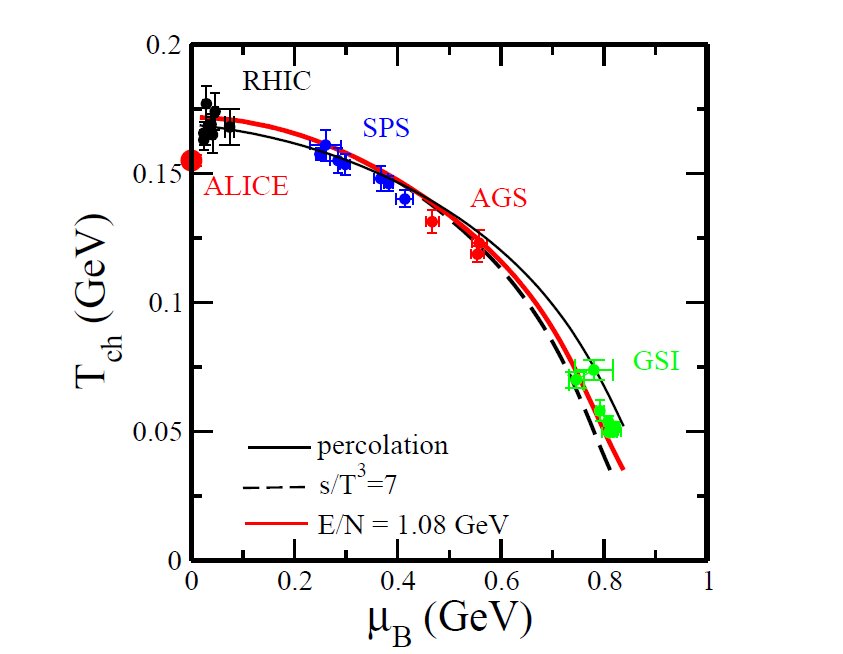}
\includegraphics[width=0.53\textwidth]{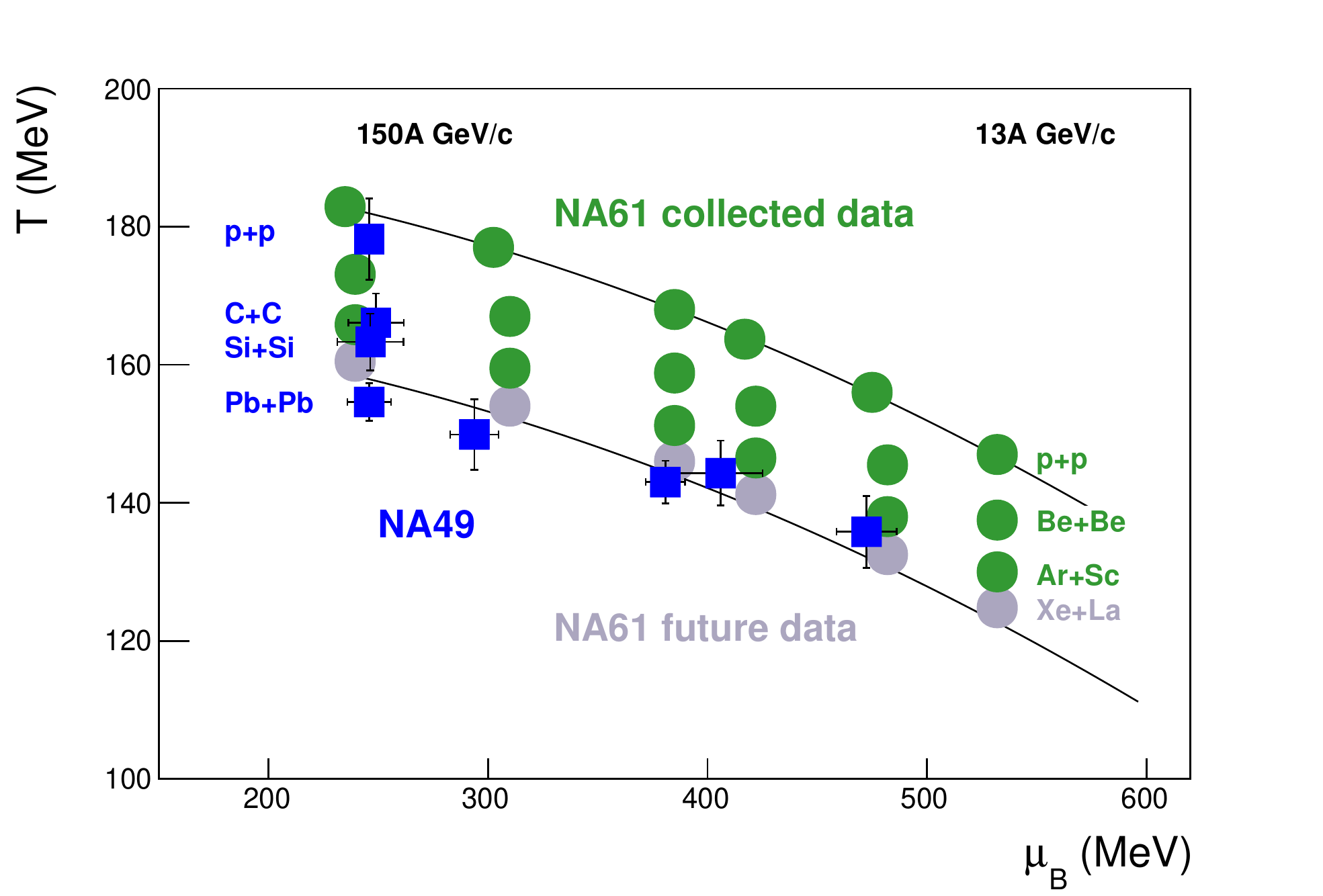}
\caption{
{\it Left:} Compilation of chemical freeze-out points 
of central Pb+Pb (Au+Au) collisions~\cite{Cleymans:2014xha}.
{\it Right:} Chemical freeze-out points of nucleus-nucleus collisions
studied with the NA49~\cite{Becattini:2005xt} 
and NA61/SHINE~\cite{Antoniou:2006mh} programmes 
at the CERN SPS.
\label{fig:phase}
}
\end{figure}

The structure of the transition region is explored 
experimentally by studying the final states produced in
nucleus-nucleus collisions. 
By changing collision energy and size of colliding nuclei one
changes temperature $T$ and chemical potentials $\mu_B$
of matter at the freeze-out stage~\cite{Becattini:2005xt}.
In particular, by increasing collision energy more and
more pions per colliding baryon are produced which is
the main cause for the decrease of the baryon chemical potential
with collision energy. 
With increasing size of the colliding nuclei the volume of created
matter increases and consequently
the role of hadron-hadron interactions at the late stage of
matter evolution becomes more important. This leads to a decrease of the freeze-out
temperature with increasing size of the colliding nuclei.
Thus scanning in collision energy and system size one hopes
to be able to move the freeze-out close to the transition region. 
This is illustrated in Fig.~\ref{fig:phase}. The left plot
shows the collision
energy dependence of the freeze-out parameters in
central Pb+Pb collisions~\cite{Cleymans:1999cb,BraunMunzinger:2003zd,Becattini:1997ii}, 
whereas the right one presents
their energy and system size dependence 
at the CERN SPS~\cite{Becattini:2005xt}.

The experimental search for the critical point by investigating
nuclear collisions is promising only at energies higher
than the energy of the onset of deconfinement,
which experimentally was located to be at 
the low SPS energies~\cite{Alt:2007aa, Gazdzicki:2014pga}.
This is because the energy density at the early stage of
the collision, which is required for the onset of deconfinement
is higher than the energy density at freeze-out, which
is relevant for the search for the critical point.

\begin{figure}[!htb]
\begin{center}
\begin{minipage}[b]{0.8\linewidth}
\includegraphics[width=1.0\linewidth]{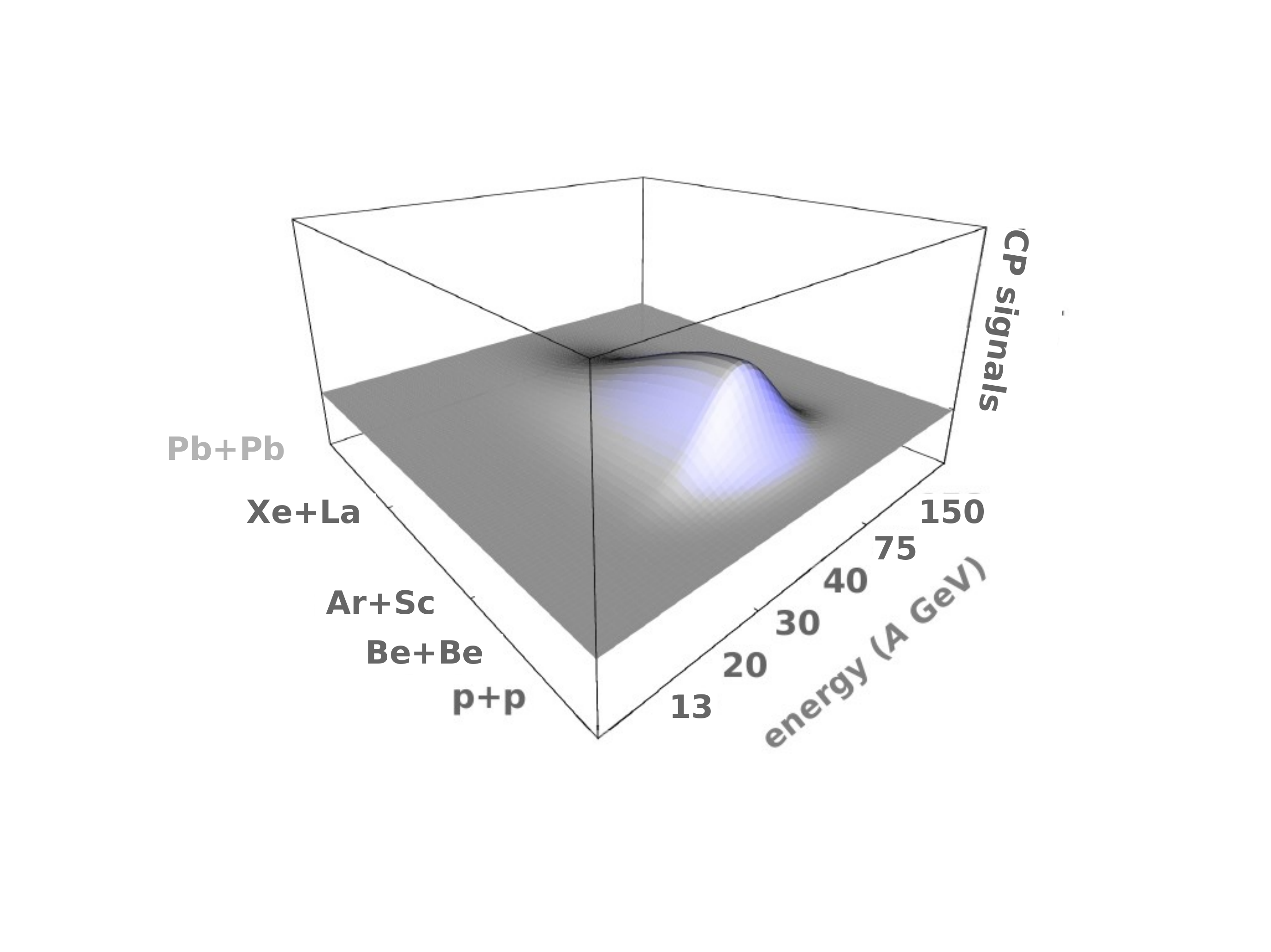}
\end{minipage}
\caption{
Sketch of
the ''critical hill'' expected in the search for the critical point
in the two dimensional plane (system size) - (collision energy).
At the hill the characteristic fluctuation signals of the critical point
are maximal, see Sec.~\ref{sec:hill} for details.
\label{fig:hill}
}
\end{center}
\end{figure}

A characteristic feature of a second order phase transition
(the critical point or line) is the
divergence of the correlation length.
The system becomes scale invariant. 
This leads to large fluctuations in particle multiplicity.
Moreover these fluctuations have specific 
characteristics~\mbox{\cite{Wosiek:1988,Bialas:1990xd}}.  
Also other properties of the system should be sensitive to the
vicinity of the critical point~\cite{Stephanov:1999zu}.
Thus when scanning the phase diagram a region of increased  
fluctuations may signal the critical point
or the critical line.
This is illustrated schematically in Fig.~\ref{fig:hill} which presents
an updated version of the plot shown first in Ref.~\cite{Gazdzicki:2006fy}.

The study of fluctuations and correlations is significantly more
difficult than the study of single particle spectra and mean multiplicities.
In general, results on fluctuations  are sensitive to
conservation laws, resonance decays and many of them also to the
unavoidable volume fluctuations of colliding nuclear matter.
Moreover, they cannot
be corrected for a limited experimental acceptance.

This review is organized as follows.
In Sec.~\ref{sec:strategies} experimental strategies, as well as
techniques and problems for the search for the
critical point are briefly presented.
Search results from experiments at the CERN SPS, in particular
NA49 and NA61/SHINE, 
are reviewed in Sec.~\ref{sec:results}.
Conclusions and an outlook in Sec.~\ref{sec:outlook} close the paper.

\section{Search strategies, techniques and problems}\label{sec:strategies}

This section reviews basic ideas and tools relevant for the
experimental search for the critical behaviour of strongly
interacting matter at the CERN SPS.
The most important background effects are listed and examples are discussed.

\subsection{Onset of deconfinement versus critical point}\label{sec:cpod}

\begin{figure}[!htb]
\begin{center}
\begin{minipage}[b]{1.0\linewidth}
\includegraphics[width=1.0\linewidth]{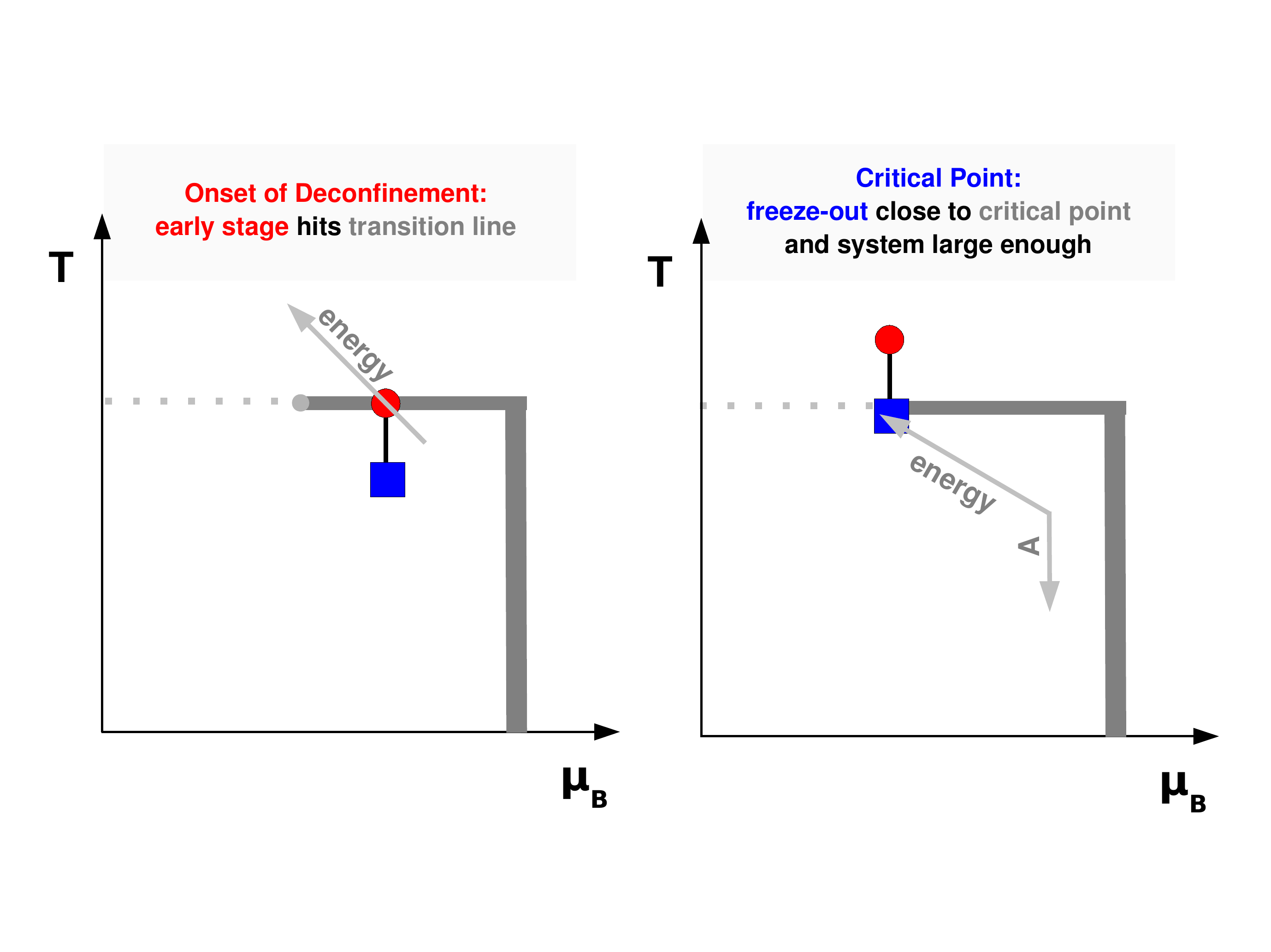}
\end{minipage}
\caption{\label{fig:cpod}
Sketch of the search strategy for the onset of deconfinement ({\it left}) and
the critical point ({\it right}). The location in the phase diagram of the
matter at the early stage is indicated by the red dot and at freeze-out
by the blue square. The solid (dotted) lines show the first order (rapid
crossover) boundary and the critical point is the grey dot at the end of
the first order transition line. 
}
\end{center}
\end{figure}

Here relations between the onset of deconfinement,
the critical point of strongly interacting matter and the
possibilities of their experimental study in relativistic nucleus-nucleus
collisions are discussed. The two sketches presented in Fig.~\ref{fig:cpod}
should help to understand the basic ideas.

The onset of deconfinement refers to the beginning of
the creation of a deconfined state of strongly interacting matter
(ultimately a quark-gluon plasma)  at the early stage of
nucleus-nucleus collisions when increasing the collision energy.
With increasing collision energy the energy density of matter created
at the early stage of A+A collisions 
increases\footnote{The correlation between collision energy and the early stage
energy density is expected to be strong for central collisions of large nuclei
and weak for collisions of low mass nuclei. In the latter case energy and/or
multiplicity of produced particles may help select collisions with a similar
energy density.}.
Thus, if there are two phases\footnote{
The discussed two phase diagram is the simplest one
which allows to introduce the concepts  of the
onset of deconfinement and the transition region.
There are numerous suggestions of phase diagrams with a
much richer structure (see e.g., Ref.~\cite{Alford:1997zt}).}
of matter separated by the transition
region (solid and dotted lines)  as indicated in Fig.~\ref{fig:cpod}~{\it left}
the early stage (the red point) has to cross the transition region.
Therefore, the existence of the onset of deconfinement is
the most straightforward consequence of the existence of
two phases of strongly interacting matter, i.e. confined matter
and QGP.
The experimental observation of the onset of deconfinement
required a one dimensional scan in collision energy with
heavy ions as performed by NA49~\cite{Alt:2007aa}.
Signals of the onset of deconfinement 
relate to the difference in properties of confined
matter and QGP. They are weakly sensitive to the structure of the
transition region.

Discovery of the onset of deconfinement implies the existence of
QGP and of a transition region between confined and QGP phases.
Recent experimental results~\cite{Gazdzicki:2014pga} indicate that the transition region
(the mixed phase or rapid cross-over) ranges from $\sqrt{s_{NN}} \approx$~8~GeV
to $\sqrt{s_{NN}} \approx$~12~GeV,
where $\sqrt{s_\mathrm{NN}}$ denotes collision energy per nucleon pair 
in the centre of mass system. 

Numerous possibilities concerning the structure of the transition
region are under discussion (see e.g., Ref.~\cite{Bowman:2008kc}). The
most popular one~\cite{Stephanov:2004wx,Stephanov_2006}, sketched in Fig.~\ref{fig:cpod}, 
claims that a 1$^{st}$ order phase transition (thick gray line) separates
both phases in the high baryonic chemical potential domain. In the
low baryonic chemical potential domain a rapid crossover is
expected (dotted line). The end point of the 1$^{st}$ order phase
transition line (grey dot in Fig.~\ref{fig:cpod}) is the critical point (of the second order).

The characteristic signatures of the critical point can be
observed if the freeze-out point (blue square in
Fig.~\ref{fig:cpod}~{\it right}) is located close to the critical
point. The analysis of the existing experimental data~\cite{Becattini:2005xt}
indicates that the location of the freeze-out point in the phase
diagram depends on the collision energy and the mass of the
colliding nuclei. This dependence is schematically indicated in
Fig.~\ref{fig:cpod}~{\it right} and quantified in Fig.~\ref{fig:phase}~{\it right}.
Thus the experimental search for the
critical point requires a two-dimensional scan in collision energy
and size of the colliding nuclei. The \mbox{NA61/SHINE}
experiment~\cite{Gazdzicki:2006fy,Antoniou:2006mh} at the CERN SPS
started this scan in 2009 and completion is expected within the coming few
years. Note, that a two dimensional scan is actually required for
any study of the structure of the transition region, independent
of the hypothesis tested.

The transition region can be studied experimentally in
nucleus-nucleus collisions only at $T$, $\mu_B$ values which
correspond to collision energies higher than the energy of the
onset of deconfinement. This important conclusion is easy to
understand when looking at Fig~\ref{fig:cpod}. Signals of the transition region
can be observed provided the freeze-out point is close to it
(see Fig.~\ref{fig:cpod}~{\it right}). 
Furthermore,
the energy density at the early stage of the collision is, of course,
higher than the energy density at freeze-out. 
Thus, the condition that the freeze-out point is near the
transition region implies that the early stage of the system
is above (or on) it. This in turn means that 
the optimal energy range for the search for the critical point
(or, in general, for the study of properties of the transition region)
lies above the energy of the onset of deconfinement 
(see Fig.~\ref{fig:cpod}~{\it left}).
This general condition limits the search for the critical point to
the collision energy range $E_{LAB} > 30A$~GeV ($\sqrt{s_{NN}} \approx$~8~GeV).

\subsection{Scaled factorial moments}\label{intermittency}

In the grand canonical ensemble the correlation length $\xi$  diverges
at the critical point (or second order phase transition line) and
the system becomes scale invariant~\cite{Wosiek:1988,Satz:1989vj}. 
This leads to large multiplicity fluctuations with special properties.
They can be conveniently exposed using scaled 
factorial moments $F_r(\delta)$~\cite{Bialas:1985jb} of rank (order) $r$: 

\begin{equation}
F_r(\delta)=\frac{ \langle \displaystyle{\frac{1}{M}\sum_{i=1}^{M}} 
N_i(N_i-1)...(N_i-r+1) \rangle }
{\langle \displaystyle{\frac{1}{M}\sum_{i=1}^{M}} N_i \rangle^r } ~,
\label{eq:facmom}
\end{equation}
where $\delta$ is the size of the subdivision intervals of
the momentum phase space region $\Delta$ and
$M = \Delta / \delta$ is the number of intervals. 
$N_i$ refers to particle multiplicity in the interval $i$ and
$\langle ... \rangle$ indicates averaging over the analyzed collisions.

For a non-interacting (ideal) gas of Boltzmann  particles 
in the grand canonical ensemble (IB-GCE) one gets $F_r(\delta) = 1$
for all values of $r$ and $\delta$ provided the
mean particle multiplicity is proportional to $\delta$.
The latter condition is trivially obeyed for a subdivision in configuration
space where the particle density is uniform throughout the
gas volume. For the case of subdivision in momentum
space the subdivision should be performed using so-called 
cumulative kinematic variables~\cite{Bialas:1990dk} in which 
the particle density is uniform.

At the second order phase transition the matter properties 
strongly deviate from the ideal gas.  The system is a simple
fractal and $F_r(\delta)$ possess
a power law dependence on $\delta$:
\begin{equation}
F_r(\delta) = F_r(\Delta) \cdot (\Delta / \delta)^{\phi_r} ~.
\label{eq:p1}
\end{equation}
Moreover the exponent (intermittency index) $\phi_r$ satisfies the relation:
\begin{equation}
\phi_r = ( r - 1 ) \cdot d_r ~,
\label{eq:p2}
\end{equation}
with the anomalous fractal dimension $d_r$ being independent 
of $r$~\cite{Bialas:1990xd}.
These results are valid when
cumulative variables~\cite{Bialas:1990dk} are used
to define the intervals $\delta$.
The properties represented by~Eqs.~\ref{eq:p1} and~\ref{eq:p2} are called in this paper
the critical behaviour of the scaled factorial moments.


An experimental search for the properties~\ref{eq:p1} and~\ref{eq:p2}
in high energy collisions requires significant additional input.
In particular, one has to decide on:
\begin{enumerate}[(i)]
\item
dimension, size and location of the momentum phase space region $\Delta$,
\item
selection of collisions used in the analysis,
\item
selection of particles used in the analysis.
\end{enumerate}

Concerning (i) it was shown by 
Bialas, Seixas~\cite{Bialas:1990gu}
(see also Ochs, Wosiek~\cite{Ochs:1988ky,Ochs:1990mg})
that unbiased results on the critical behaviour of scaled factorial
moments can be obtained only by performing the analysis in variables and dimensions
in which the singular behaviour appears.
Any projection procedure is likely to to remove, at least partly,
the critical fluctuation signal. 

Concerning (iii) QCD-inspired considerations~\cite{Antoniou:2000ms,Stephanov:2004wx}
suggest that
the order parameter of the phase transition is 
the chiral condensate $\langle \bar{q}q\rangle$ ($q$ is the quark field).
The quantum state carrying the quantum numbers as well as the critical 
properties of the chiral condensate is the isoscalar $\sigma$-field. Assuming that this state 
can be formed in high energy collisions there are two possibilities for 
the observation of the properties~\ref{eq:p1} and~\ref{eq:p2}:
\begin{enumerate}[(i)]
\item
Directly from its decay products~\cite{Antoniou:2005am}. The condensate will decay into 
$\pi^+ \pi^-$ pairs with invariant mass just above twice the pion mass. Detection 
of the expected fluctuations requires reconstruction of the pion pairs.
One expects here $d = \phi_2 = 2/3$~\cite{Antoniou:2005am}.
\item
Through measuring the fluctuations of the proton number.  
The net-baryon density mixes with the chiral condensate transferring the
critical fluctuations to the net-baryon 
density~\cite{Fukushima:2010bq,Hatta:2002sj,Stephanov:2004wx,
Antoniou:2008vv,Karsch:2010ck,Skokov:2010uh,Morita:2012kt},
which is an equivalent order parameter of the phase transition. The resulting fluctuations
are predicted to be present also in the net-proton number as well as in the proton and 
anti-proton numbers separately~\cite{Hatta:2003wn}. 
One expects here $d = \phi_2 = 5/6$~\cite{Antoniou:2006zb}.
\end{enumerate}

\begin{figure}[!htb]
\begin{center}
\begin{minipage}[b]{0.8\linewidth}
\includegraphics[width=0.49\linewidth]{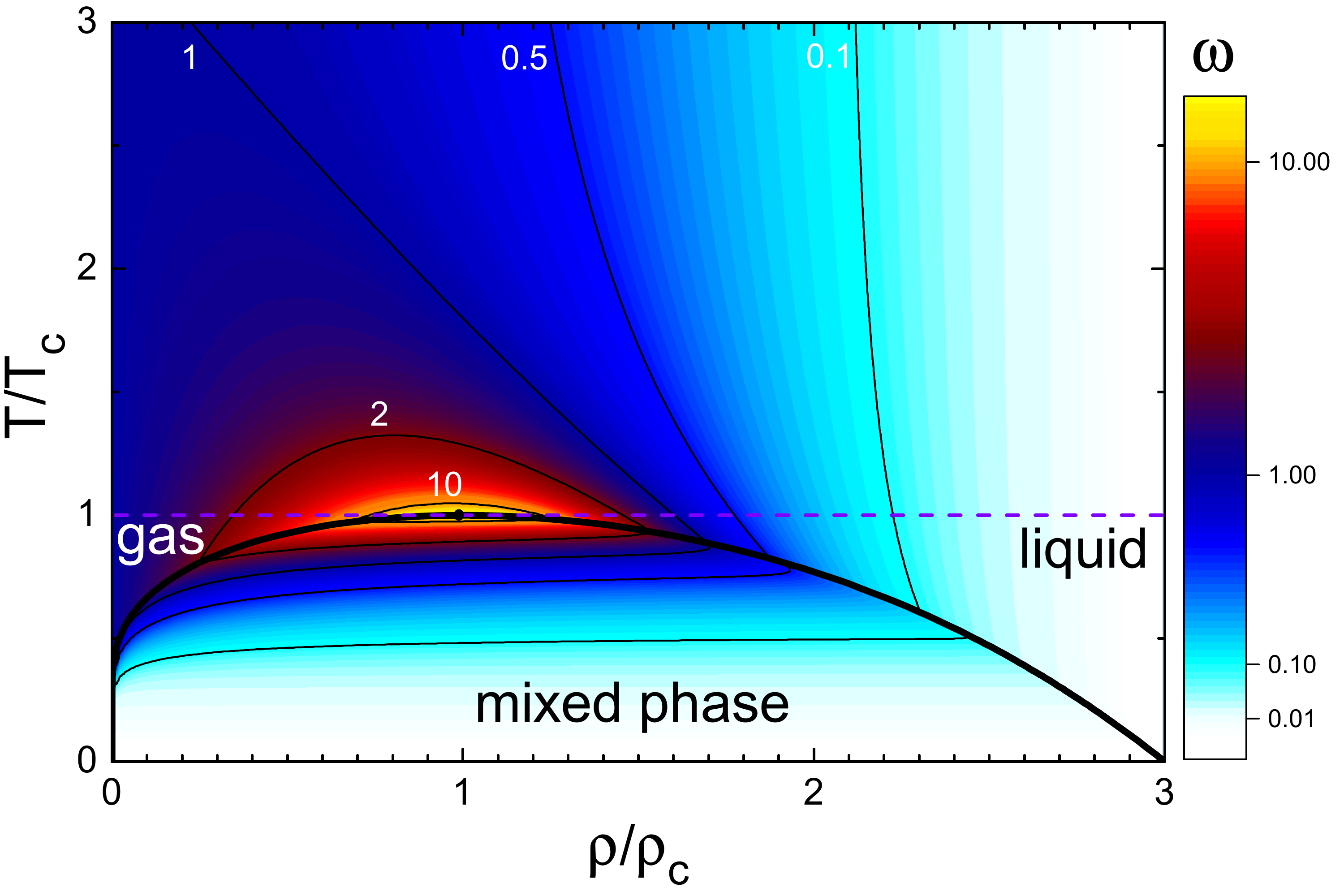}
\includegraphics[width=0.49\linewidth]{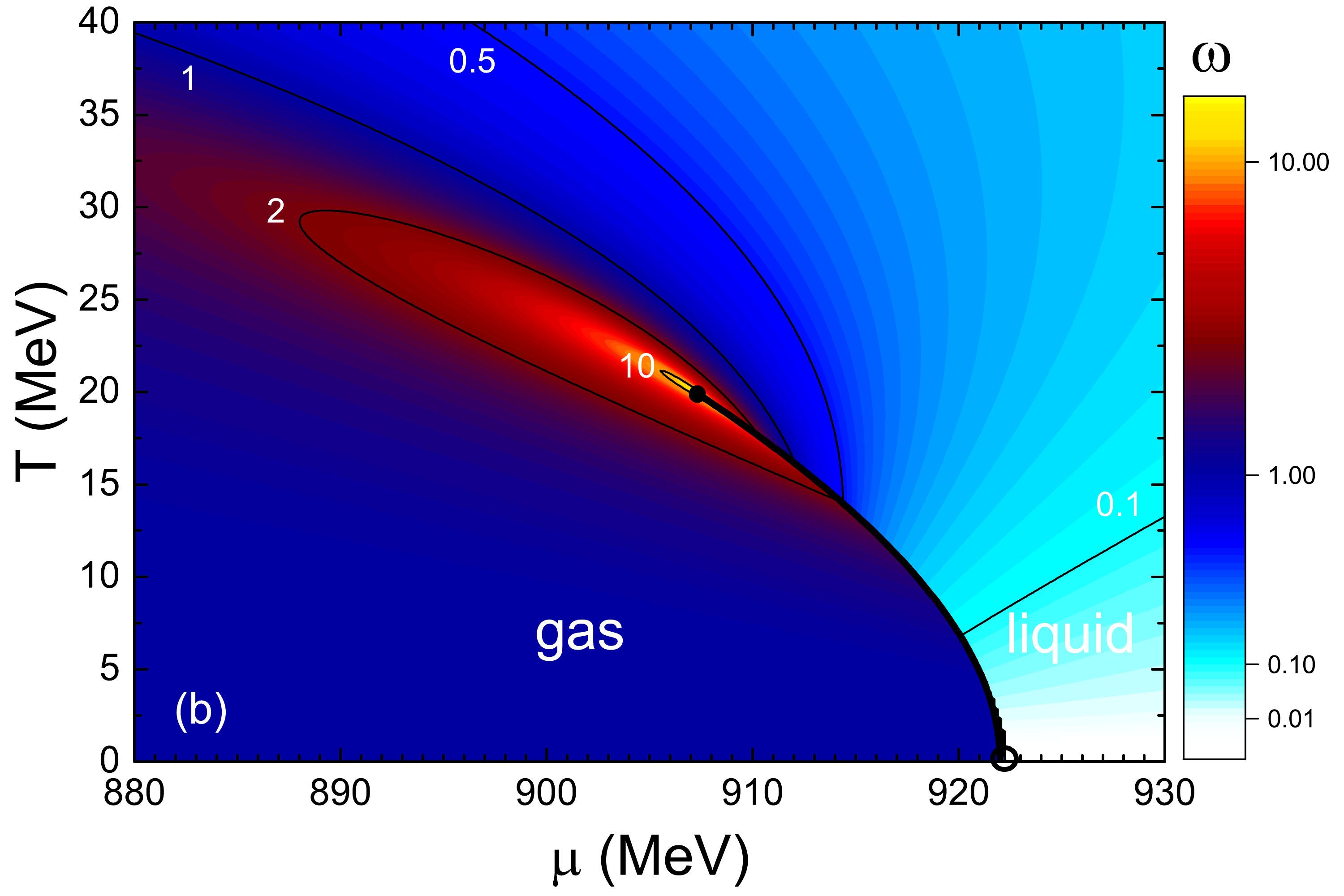}
\end{minipage}
\caption{
{\it Left:}
Scaled variance of the multiplicity distribution calculated within the
grand canonical formulation of the van der Waals model 
(massless particles obeying Boltzmann statistics) as a function
of temperature T and particle density $\rho$ scaled by the corresponding values
at the critical point (from Ref.~\cite{Vovchenko:2015xja}).
{\it Right:}
Scaled variance of the multiplicity distribution calculated within the
grand canonical formulation of the van der Waals model 
(nucleons obeying Fermi statistics) as a function
of temperature T and baryon chemical potential $\mu$
(from Ref.~\cite{Vovchenko:2015pya}).
\label{fig:vdw}
}
\end{center}
\end{figure}

\subsection{Central moments}\label{sec:sm}

The infinite correlation length at a second order phase transition is
expected to lead to divergence of  
the second central moment of the multiplicity distribution.
This was recently illustrated  
by Vovchenko, Anchinshkin, Gorenstein 
and Poberezhnyuk~\cite{Vovchenko:2015xja,Vovchenko:2015pya}
via analytical calculations performed in the GCE for a gas 
obeying the van der Waals equation of state.
As seen in Fig.~\ref{fig:vdw}
the scaled variance 
$\omega = (\langle N^2 \rangle - \langle N \rangle^2)
/\langle N \rangle$ increases when the critical point is approached and
it diverges at the
critical point.

Moments of other quantities are also expected to increase in the vicinity
of the critical point. In particular, these are 
event transverse momentum (vector magnitudes are summed) and  net charges of produced 
particles~\cite{Stephanov:1999zu}. Of course the latter have to be studied in a
phase space region which covers only a small fraction of all particles in order
to avoid suppression of fluctuations due to conservation laws
(see below for details).

It was pointed out by Stephanov~\cite{Stephanov:2008qz} that second moments of the
multiplicity distribution increase in proportion to the square of the correlation
length while moments of higher order are proportional to even higher powers of
the correlation length. Moreover, ratios of certain higher order cumulants are
expected to be independent of the correlation length 
and therefore allow a consistency test~\cite{Athanasiou:2010kw}. Based on QCD
model calculations the same paper also presents quantitative estimates of the size 
of the enhancement of multiplicity fluctuations caused by a critical point.

\subsection{Strongly intensive quantities}\label{sec:volume}

Since event-to-event volume fluctuations 
cannot be eliminated
in experimental studies of nucleus-nucleus collisions, it is important
to minimise their effect by defining suitable fluctuation measures.
It appears that using second and first moments of the distibution of two extensive
quantities 
(their first moments are proportional to  volume)
one can construct fluctuation measures which are, within
a statistical model of the ideal Boltzmann gas  in 
the grand canonical ensemble formulation (~SM(IB-GCE)~)~\cite{Gorenstein:2011vq},
independent of its fluctuations.
 
As the simplest example let us consider 
multiplicities
of two different types of hadrons, $A$ and $B$.
Their mean multiplicities 
are proportional to the system volume:
\begin{equation}
 \langle A \rangle \sim V ~, ~~~~~~~~
 \langle B \rangle \sim V ~. 
\label{eq:stat:mean}
\end{equation}
Obviously the ratio of mean multiplicities is independent of
the volume  $V$.
Moreover, the ratio $\langle A \rangle \ / \langle B \rangle $ is
independent of $P(V)$, where $P(V)$ is
the probability (density) distribution of $V$
for the considered set of collisions. 
Quantities which have the latter property are called strongly
intensive quantities~\cite{Gorenstein:2011vq}. Such quantities are recommended 
to be used in experimental studies 
of the system size dependence of fluctuations in A+A collisions as they eliminate the influence of
usually poorly known distributions of the system volume.

More generally, $A$ and $B$ can be any extensive
event quantities such as the sum of transverse momenta, the net charge or
the multiplicity of a particular type of particle.
The scaled variances of $A$ and $B$ and the mixed second
moment $\langle AB \rangle$ calculated within 
the SM(IB-GCE)~\cite{Gorenstein:2011vq} read:
\begin{equation}
 \omega[A] = \omega^*[A] + \langle A \rangle / \langle V \rangle \cdot\omega[V] ~, \\ 
\label{eq:stat:varA}
\end{equation}
\begin{equation}
 \omega[B] = \omega^*[B] + \langle B \rangle / \langle V \rangle \cdot\omega[V] ~, \\
\label{eq:stat:varB}
\end{equation}
\begin{equation}
 \langle AB \rangle = \langle AB \rangle^ * \langle V \rangle +
 \langle A \rangle  \langle B \rangle \langle V \rangle^2 \cdot
 (\langle V^2 \rangle - \langle V \rangle )~,
\label{eq:stat:AB}
\end{equation}
where quantities denoted by $^*$
are quantities calculated for a fixed value of the system volume.

From Eqs.~\ref{eq:stat:varA}-\ref{eq:stat:AB} 
follows~\cite{Gorenstein:2011vq,Gazdzicki:2013ana} that suitably
constructed functions of the second moments, namely
\begin{equation}\label{eq:delta}
 \Delta[A,B]
 ~=~ \frac{1}{C_{\Delta}} \Big[ \langle B\rangle\,
      \omega[A] ~-~\langle A\rangle\, \omega[B] \Big]
\end{equation}
and
\begin{equation}\label{eq:sigma}
  \Sigma[A,B]
 ~=~ \frac{1}{C_{\Sigma}}\Big[
      \langle B\rangle\,\omega[A] ~+~\langle A\rangle\, \omega[B] ~-~2\left(
      \langle AB \rangle -\langle A\rangle\langle
      B\rangle\right)\Big]
\end{equation}
are independent of $P(V)$ in the SM(IB-GCE).
Here the normalisation factors
$C_{\Delta}$ and $C_{\Sigma}$ are required to be proportional to first moments
of any extensive quantities.
In Ref.~\cite{Gazdzicki:2013ana} a specific choice of the $C_{\Delta}$ and
$C_{\Sigma}$ normalisation factors was proposed which makes the quantities
$\Delta[A,B]$ and $\Sigma[A,B]$ dimensionless and  leads to
$\Delta[A,B] = \Sigma[A,B] = 1$ in the independent particle
model (IPM). This normalisation is called here  the IPM normalisation and
unless otherwise stated the IPM normalisation is used.

Thus $\Delta[A,B]$ and $\Sigma[A,B]$ are strongly intensive quantities
which measure fluctuations of $A$ and $B$, 
i.e. they are sensitive to second moments of the distributions
of the quantities $A$ and $B$. 
Results on $\Delta[A,B]$ and $\Sigma[A,B]$ are referred to as
results on $A - B$ fluctuations, e.g. transverse momentum - multiplicity fluctuations.

For the case of 
multiplicity $A$ - multiplicity $B$ fluctuations the 
expressions for $\Delta[A,B]$ and $\Sigma[A,B]$ with the IPM normalisation
read:
\begin{equation}
 \Delta[A,B] = ( \langle B \rangle \omega[A] -
                 \langle A \rangle \omega[B] )~/~
               ( \langle B \rangle - \langle A \rangle )  
\label{eq:delta_n}
\end{equation}
and
\begin{equation}
 \Sigma[A,B] = ( \langle B \rangle \omega[A] +
                 \langle A \rangle \omega[B] - 
                 2 (\langle AB \rangle - \langle A \rangle \langle B \rangle) )~/~
               ( \langle B \rangle + \langle A \rangle )~.  
\label{eq:sigma_n}
\end{equation}

The $\Sigma$ quantity is a reincarnation of the popular
$\Phi$ measure of fluctuations~\cite{Gazdzicki:1992ri}.
The original definition of $\Phi$ is:
\begin{equation}\label{Phi}
\Phi_{x} =
\sqrt{\langle Z^2 \rangle \over \langle N \rangle } -
\sqrt{\overline{(x-\overline{x})^2}} \; ,
\end{equation}
where $x$ is a particle property and
\begin{equation}\label{z}
Z = \sum_{i=1}^{N}(x_i - \overline{x}) 
\end{equation}
with the sum running over the $N$ particles in the event.
$\Phi$ was shown to be related to $\Sigma$ as follows~\cite{Gorenstein:2011vq}:
\begin{equation}
  \Phi_{x} = \sqrt{\overline{x} \omega[x]} \left[\sqrt{\Sigma[X,N]}-1\right]
\label{eq:phi}
\end{equation}
with $X = \sum_{i=1}^{N}x_i$ and $C_\Sigma = \langle N \rangle \omega[x]$.

By construction strongly intensive measures of fluctuations
are always a functional of two extensive quantities. This, in general, hampers
a straight-forward interpretation of experimental results. However, under
certain conditions the $\Delta$ quantity can be used to
obtain the scaled variance of the extensive quantity $A$ separately.  

Let $A$ be the extensive quantity, e.g. selected for its sensitivity
to the critical fluctuations. Then choose a quantity
$B$ such which is proportional to the system volume 
$ B  \sim V$.
Then it is easy to show that the strongly intensive measures 
$\Delta_{B}[A,B]$ and $\Sigma_{B}[A,B]$
(equal to $\Delta[A,B]$ and $\Sigma_{B}[A,B]$ with the normalisation 
$C_{\Delta} = \langle B \rangle \sim V$)
obeys the relation:
\begin{equation}
    \Delta_{B}[A,B] = \Sigma_{B}[A,B] =  \omega^*[A] ~.
 \label{eq:delta_c}
\end{equation}
In the derivation of Eq.~\ref{eq:delta_c} one assumes 
the validity of Eq.~\ref{eq:stat:varA}
which needs to be investigated case-by-case. 
Thus $\Delta_{B}[A,B]$ is approximately equal to the scaled variance $\omega^*[A]$ of $A$
for a fixed system volume (see Eq.~\ref{eq:stat:varA}):
\begin{equation}
 \Delta_{B}[A,B_V] \approx \omega^*[A] ~,
\label{eq:delta_c_apprx}
\end{equation}

Suggestions of practical choices of $A$ and $B$ are: 
\begin{enumerate}[(i)]
\item
$A$   - multiplicity of hadrons which are sensitive to the critical
        behaviour, e.g. protons, sigma-mesons, pions 
        in the central rapidity window and
\item
$B$ - net electric charge in full phase space or 
        large acceptance excluding fragmentation
        regions of projectile and target nuclei; it is equal to the number
        of participant protons and thus approximately proportional to the
        volume of matter involved in the collision or \\
      - the number of projectile participants calculated as the difference
        between the number of nucleons in the beam nuclei and the
        number of projectile spectators measured by a "zero degree"
        calorimeter, e.g., the Projectile Spectator Detector of NA61/SHINE.
\end{enumerate}

Recently, strongly intensive measures which involve higher than second moments
were proposed~\cite{Sangaline:2015bma}. The next important step would be
to reformulate the critical properties of scaled factorial moments 
Eqs.~\ref{eq:p1} and~\ref{eq:p2} in terms of strongly intensive quantities.

\subsection{Critical hill}\label{sec:hill}

As discussed  in Sec.~\ref{sec:sm} moments of multiplicity distribution 
diverge when temperature and density
approach their critical values (see Fig.~\ref{fig:vdw}).
At this same point in the phase diagram the scaled factorial moments 
should obey the critical properties described by Eqs.~\ref{eq:p1} and~\ref{eq:p2}.

As previously argued the freeze-out temperature increases
with increasing collision energy and decreasing size of the colliding
nuclei. The maximum temperature, probably the closest to the phase
transition, is observed in p+p interactions.
However, the small volume and short life-time of the created matter
together with the conservation laws (see the next section),  
do not allow the divergence of the correlation length~\cite{Berdnikov:1999ph,Athanasiou:2010kw}.
Thus neither a divergence of fluctuation measures, e.g. of the scaled variance,
nor the appearance of critical behaviour of scaled factorial moments
are expected.

On the other hand, critical fluctuations developing in the system produced by
collisions of large nuclei may be erased by re-scattering processes between hadronisation
and kinetic freeze-out.
Therefore one can argue that the maximal signals of 
the critical point 
may be observed for collisions of medium mass nuclei.
In this case the volume and life-time of the created matter are 
large enough to allow for the critical behaviour to appear, and
the temperature is close enough to the critical temperature
to make the critical behaviour visible.
Thus observation of the ''critical hill'', as sketched in Fig.~\ref{fig:hill},
would provide convincing evidence for the existence and location of the critical point.

\subsection{Background fluctuations}\label{sec:background}

The critical hill, if observed, will rise above a  
background caused by  many different sources of fluctuations.
These are in particular:
\begin{enumerate}[(i)]
\item
volume fluctuations discussed in Sec.~\ref{sec:volume}, 
\item
conservation laws
discussed in Sec.~\ref{sec:conservation},
\item
formation and decay of resonances, 
\item
quantum statistical effects
(Bose-Einstein and Fermi-Dirac statistics).
\end{enumerate}
Their impact on fluctuations is discussed in the recent 
review~\cite{Gorenstein:2015ria} where also references to original papers are given.

In general the background fluctuations  are
not expected to lead to non-monotonic dependence of fluctuations
on collision energy and system size.

The effect of volume fluctuations  is addressed
in Sec.~\ref{sec:volume}.
As an additional example, the effect of conservation laws is discussed below
in more detail.

\subsection{Conservation laws}\label{sec:conservation}

Predictions of statistical models concerning the volume
dependence change qualitatively when material and/or motional
conservation laws are introduced, i.e. instead of the grand
canonical ensemble, the canonical (CE) or micro-canonical (MCE)
ensembles are used.
The effect of conservation laws has been extensively studied
for mean multiplicities since 
1980~(see e.g. Refs.~\cite{Rafelski:1980gk,Redlich:1979bf,Becattini:2003ft}) and
for second moments of multiplicity distributions since 
2004~(see e.g. Refs.~\cite{Begun:2004gs,Begun:2004pk}). 
An example is discussed below for illustration.

\begin{figure}[t]
\centering
\includegraphics[width=0.48\textwidth]{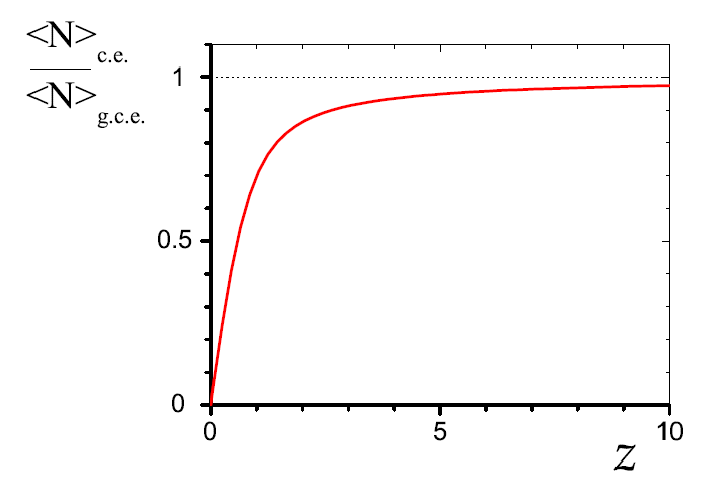}
\includegraphics[width=0.48\textwidth]{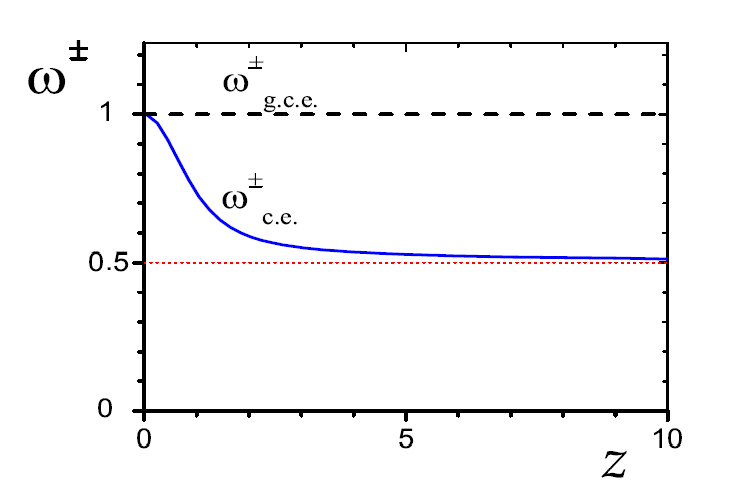}
\caption{
The ratio of mean multiplicities $\langle N \rangle = \langle N_+ \rangle$ or $\langle N_- \rangle$   
calculated within the SM(IB-CE) and the SM(IB-GCE) are plotted in 
the left panel
as a function of  mean multiplicity $z$
for the SM(IB-GCE), with $z$ being proportional to the system volume.
The corresponding scaled variances $\omega^+$, $\omega^-$ are shown 
in the right panel.
The ideal gas model of classical positively and negatively
charged particles was used for calculations.
The system net charge is assumed to be zero and thus
$\langle N_+ \rangle = \langle N_- \rangle$ and
$\langle \omega^+ \rangle = \langle \omega^- \rangle$.
The plots are taken from Ref.~\cite{Begun:2004gs}.
\label{fig:ideal}
}
\end{figure}

Figure~\ref{fig:ideal} 
taken from Ref.~\cite{Begun:2004gs}
presents the results of calculations
performed within the simplest model which allows to study
the effect of material conservation laws on mean multiplicity and
scaled variance of the multiplicity distribution.
In this model an ideal gas of classical positively and negatively 
charged particles is assumed.
The ratio of the mean multiplicities calculated within the SM(IB-CE) 
and the SM(IB-GCE) is
plotted in Fig.~\ref{fig:ideal} ($left$)
as a function of the mean multiplicity $z$ from the SM(IB-GCE), 
the latter being proportional
to the system volume.
The ratio approaches one with increasing volume. 
Thus for sufficiently large systems mean multiplicities obtained 
within the SM(IB-GCE) can be used instead of mean multiplicities
from the SM(IB-CE) and the SM(IB-MCE)~\cite{Begun:2004pk}.
This is however not the case for the scaled variance as illustrated
in Fig.~\ref{fig:ideal} ({\it right}).
The results for the SM(IB-CE) and the SM(IB-GCE) approach each other when the volume
decreases to zero. Of course, the scaled variance in the SM(IB-GCE) is  one
independent of volume. 
Different behaviour is observed for the scaled variance in the SM(IB-CE), where it 
decreases with increasing volume
and for a sufficiently large volume  approaches 0.5.

The above example clearly illustrates the importance of material
and motional conservation laws for hadron production
in nucleus-nucleus collisions at high energies.
In particular, the conservation laws may strongly affect fluctuations even
for large systems, leading to significant deviations from the simplest
reference model, i.e. independent particle production.

\section{Search results from experiments at the CERN SPS}\label{sec:results}

This section reviews the status of the experimental search for 
evidence of a second order phase transition and/or
the critical point at the CERN SPS
based on published results and preliminary data presented at conferences. \\

\subsection{Pioneering analyses}

\begin{figure}[htbp]
\includegraphics[width=0.6\columnwidth]{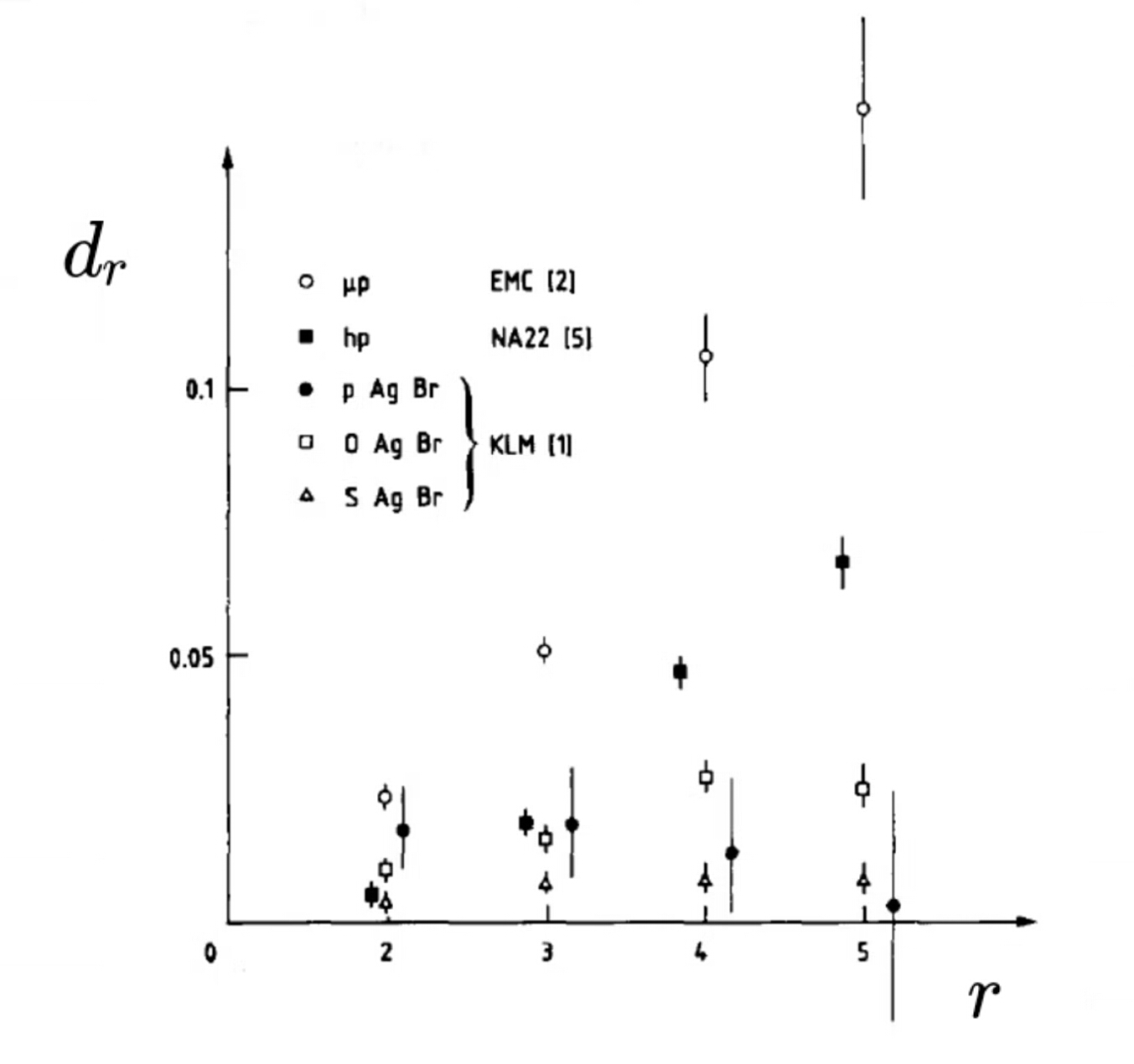}
\caption{
Anomalous dimensions $d_r = \phi_r / (r-1)$ of pseudo-rapidity spectra 
of hadrons produced in $\mu$+p, h+p, p+AgBr, O+AgBr and
S+AgBr collisions at
$\sqrt{s_{NN}} \approx 20$~GeV~\cite{Bialas:1990xd}.
\label{fig:bialas}
}
\end{figure}

The search for the critical behaviour of strongly interacting matter
at the CERN SPS started in 1990 from the pioneering paper of 
Bialas and Hwa~\cite{Bialas:1990xd}. In this work the authors 
compiled results from intermittency analysis performed by  
the EMC~\cite{Derado:1990fy}, 
NA22~\cite{Azhinenko:1990nn}
and KLM~\cite{Holynski:1988mr} 
experiments at the SPS. In these experiments intermittency indices
$\phi_r$  were determined from power-law fits to the dependence on
bin-size in (pseudo-)rapidity
of the scaled factorial moments of successive rank $r$. 
Figure~\ref{fig:bialas} shows that the anomalous dimensions, defined
as $d_r = \phi_r/(r-1)$, increase strongly with rank $r$ for the 
smaller reaction systems. Interestingly the anomalous dimension
stays constant for the heaviest system, for S+(Ag Br) collisions at 200$A$~GeV/c.
Bialas and Hwa interpreted this behaviour as an indication of a second order
phase transition. A quantitative prediction of $d_r$ was not provided.

\begin{figure}[htbp]
\begin{minipage}[b]{0.44\linewidth}
\centering
\includegraphics[width=0.95\columnwidth]{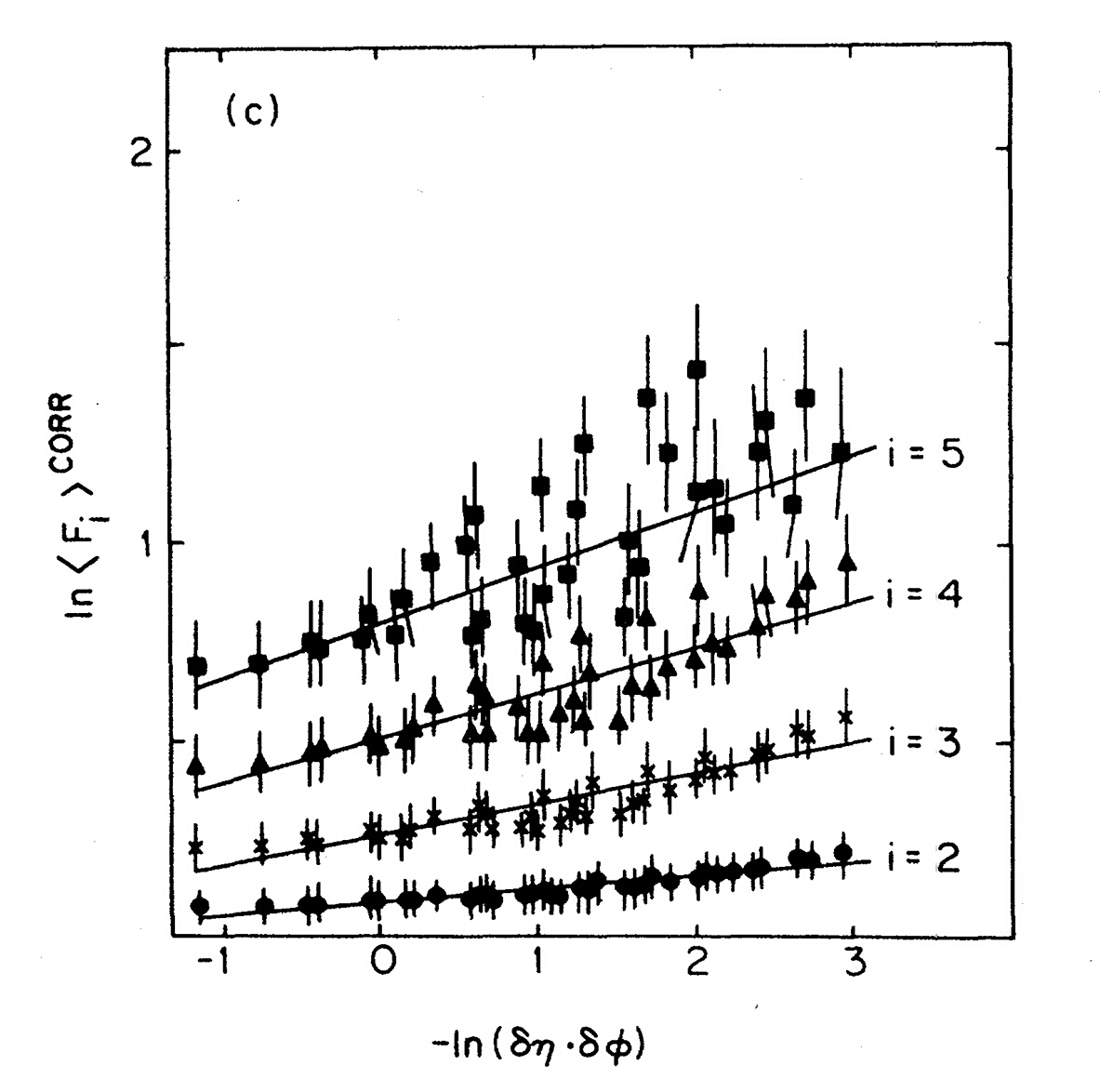}
\end{minipage}
\begin{minipage}[b]{0.53\linewidth}
\centering
\includegraphics[width=0.95\columnwidth,]{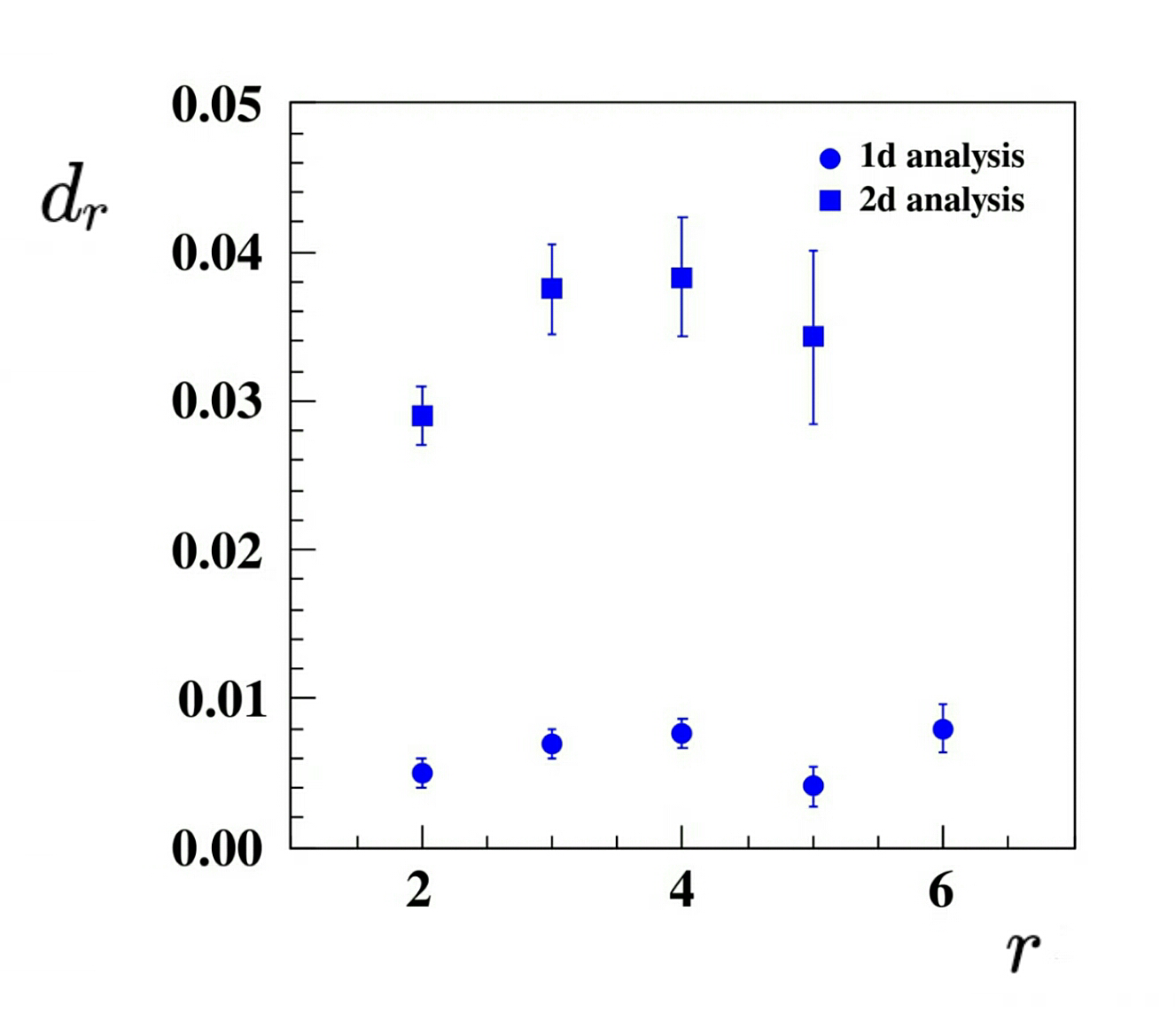}
\end{minipage}
\caption{
{\it Left:} Scaled factorial moments of rank i  from two-dimensional analysis
in pseudo-rapidity and azimuthal angle as a function of subdivision size
for the 19\% most central S+(Ag Br) collisions 
at 200$A$~GeV/c~\cite{Holynski:1989rn}.
{\it Right:} The corresponding 
anomalous dimensions $d_r$ versus order $r$ are shown by squares,
whereas the circles show results for the one-dimensional analysis
in pseudo-rapidity.
\label{fig:klm}
}
\end{figure}

An extended analysis of the emulsion data of the KLM collaboration
was published in Ref.~\cite{Holynski:1989rn}. Corrections were applied
for the non-uniform rapidity distribution and the intermittency
indices were determined both for 1-dimensional (pseudo-rapidity) and
2-dimensional (pseudo-rapidity and azimuthal angle) subdivisions of
phase space. The resulting anomalous dimensions for 
S+(Ag Br) collisions at 200$A$~GeV/c are plotted in Fig.~\ref{fig:klm}.
One observes that the values of $d_r$ are consistent with being independent
of $r$ for both analyses and confirm the earlier results. However, the
values of $d_r$ are roughly 5 times larger in the 2-dimensional analysis.
A strong reduction of the measured power $\phi_r$ with decreasing 
dimensionality of the analysis was explained by Bialas and Seixas~\cite{Bialas:1990gu}
as due to averaging of fluctuations via the projection procedure.
Thus the factorial moment analysis in three dimensions seems to be
mandatory in future searches for the critical behaviour.

Motivated by these results the WA80~\cite{Albrecht:1994rk} and
NA35~\cite{Bachler:1993nc} experiments at the SPS
revisited intermittency analysis in nucleus-nucleus collisions
at 200$A$~GeV/c.
WA80 did not have momentum measurement and inferior 2-track and angular
resolution compared to the emulsion experiment. They concluded that
they observed no significant intermittency effect in S+S and S+Au collisions
when taking into account statistical and systematic uncertainties.

The NA35 streamer chamber experiment performed momentum measurements in
central p+Au, O+Au, S+S and S+Au collisions which were subjected to a
fully differential 3-dimensional (rapidity, transverse momentum,
azimuthal angle) intermittency analysis. Although the factorial moments 
were found to increase with the number of subdivisions of phase space
this rise was not well described by a power law. Instead, a conventional
model supplemented by Bose-Einstein correlations provided a satisfactory
description.


In summary, the intermittency analyses of charged particle production 
in oxygen and sulphur induced reactions did
not lead to conclusive results on the existence of a second-order phase transition
in these reactions. More recent theoretical investigations suggest that when the
hadronization of a QGP occurs near the critical point the hadronization of the
chiral condensate will lead to intermittency in the production of protons
and low-mass $\pi^+\pi^-$ pairs with known intermittency index. A search for such
effects is in progress in the NA49 experiment and will be discussed below.  

\begin{figure}[htbp]
\includegraphics[width=0.5\columnwidth]{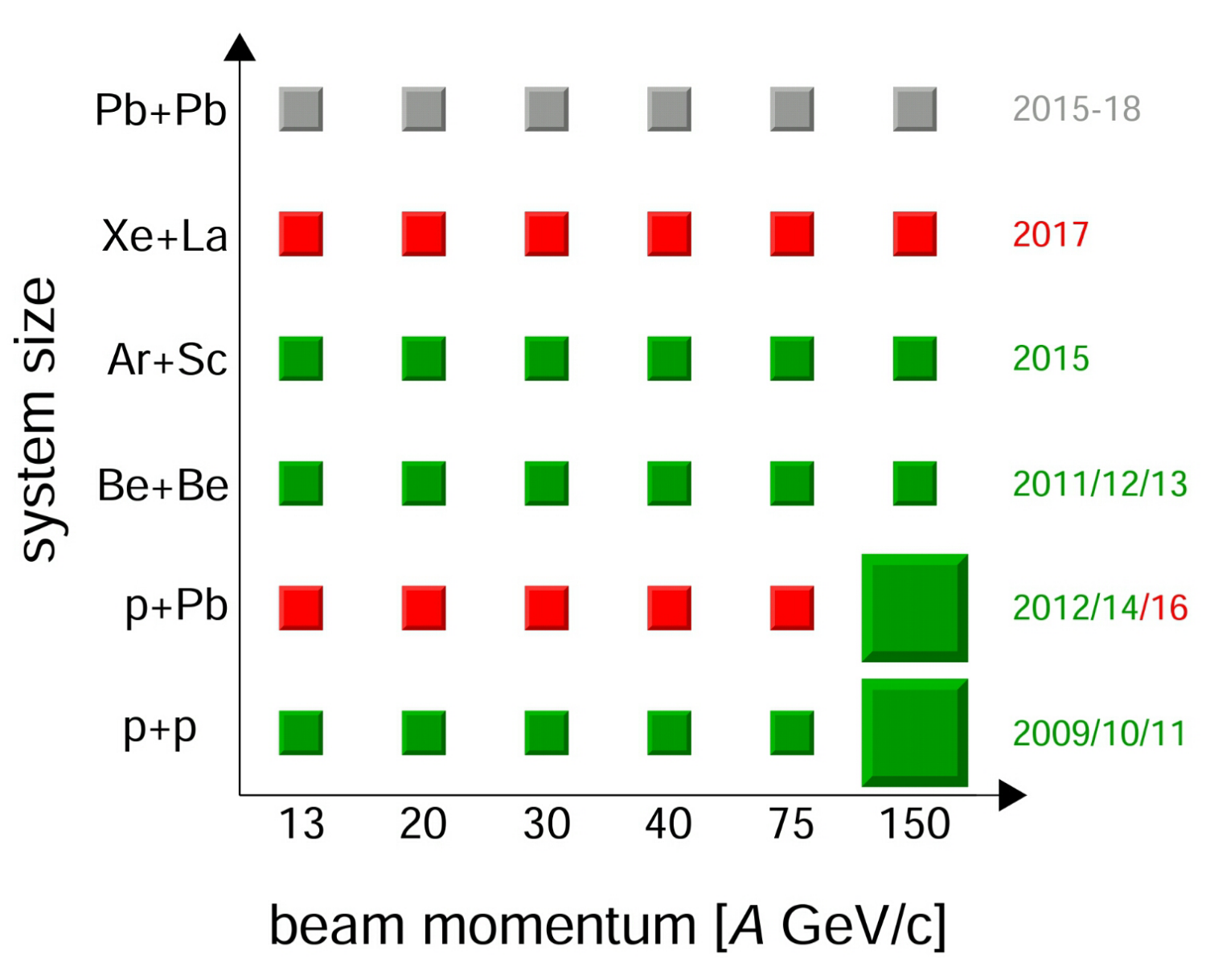}
\includegraphics[width=0.3\columnwidth]{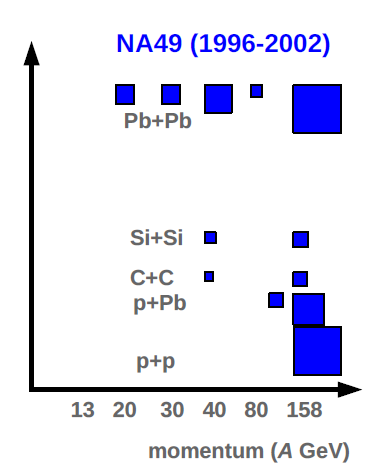}
\caption{Reactions and energies of the ongoing scan of the 
phase diagram by NA61 ({\it left}) 
(filled green squares denote completion of data taking)
and systems previously studied by NA49 ({\it right}). 
\label{na61_na49_scan}
}
\end{figure}

\subsection{Systematic fluctuation studies of NA49 and NA61/SHINE}

Search for evidence of critical behaviour of
strongly interacting matter was restarted at the beginning of 2000 
by the NA49 Collaboration~\cite{na49}. Fluctuations were analyzed in
a large number of
previously recorded data sets (see Fig.~\ref{na61_na49_scan} {\it right}).
A tantalizing increase of multiplicity and transverse momentum
fluctuations was found in collisions of medium size nuclei at 158$A$~GeV/c.
This motivated the ongoing measurements of 
NA61/SHINE~\cite{na61}, the successor of the NA49 experiment.
For the first time a systematic two-dimensional scan in system size and
collision energy is being performed (see Fig.~\ref{na61_na49_scan} $left$).
A search for the critical point is also in progress at RHIC by the
STAR collaboration~\cite{Adams:2005dq}. Relevant results will be mentioned
at the appropriate places. 
The results from the SPS obtained by NA49 and NA61/SHINE are reviewed below.


\begin{figure}[htbp]
\includegraphics[width=0.3\columnwidth]{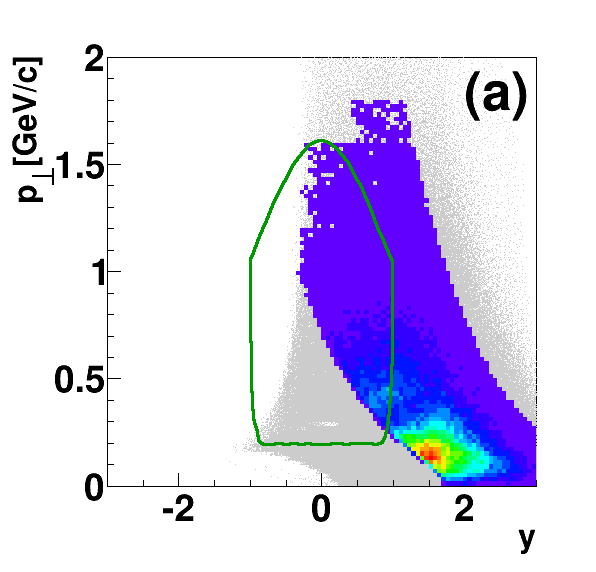}
\includegraphics[width=0.3\columnwidth]{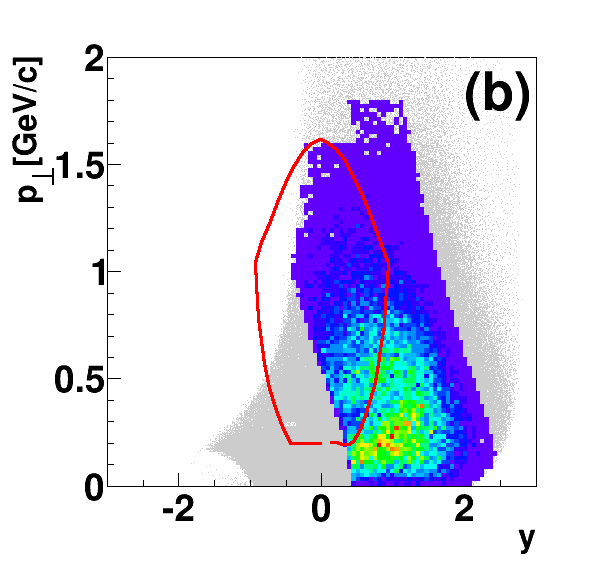}
\includegraphics[width=0.3\columnwidth]{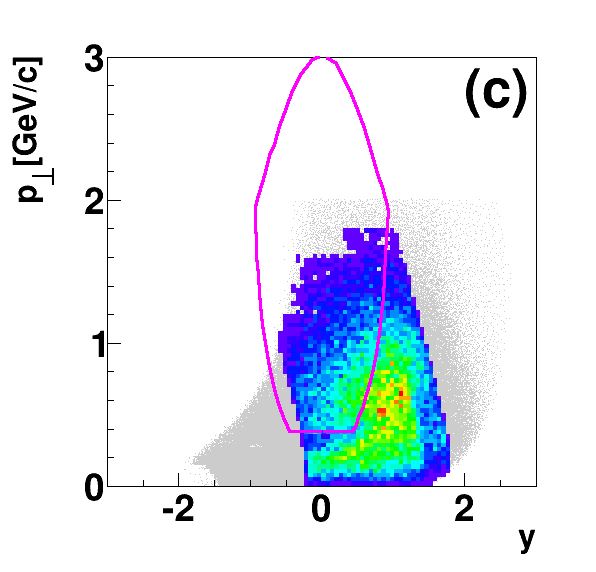}
\caption{Acceptance of the NA49 and NA61/SHINE experiments for pions (a),
kaons (b) and protons (c) at the SPS at 30$A$~GeV
(geometrical acceptance in grey, acceptance for identification in color), 
as well as that of the STAR detector at RHIC (lines) at the equivalent energy 
$\sqrt{s_{NN}}$~=~7.7~GeV.
\label{acceptance}
}
\end{figure}

When comparing results from the SPS to those from the RHIC-BES program
one should keep in mind that the event selection and acceptance of the experiments 
are significantly different. NA49~\cite{na49:det} and NA61/SHINE~\cite{na61:det} are fixed-target
spectrometers and cover mainly the forward region in the center-of-mass system, 
particularly when particle identification is required~(see Fig.~\ref{acceptance}).
An advantage of the fixed-target geometry
is that it allows to characterise the centrality of the collisions
by measuring the energy of the spectators from the beam nucleus independently from
the measurements performed on the produced particles. On the other hand, 
STAR at RHIC is a collider experiment with practically energy independent 
rapidity acceptance
$|y| \lesssim 0.7$, but without the low transverse momentum
region (see curves in Fig.~\ref{acceptance}). The track density in the detector
increases only moderately with collision energy. However, the projectile spectator regions 
are not accessible to measurement and the collision centrality selection has to be based 
on the multiplicity of produced particles.

\begin{figure}[htbp]
\centering
\includegraphics[width=0.3\columnwidth]{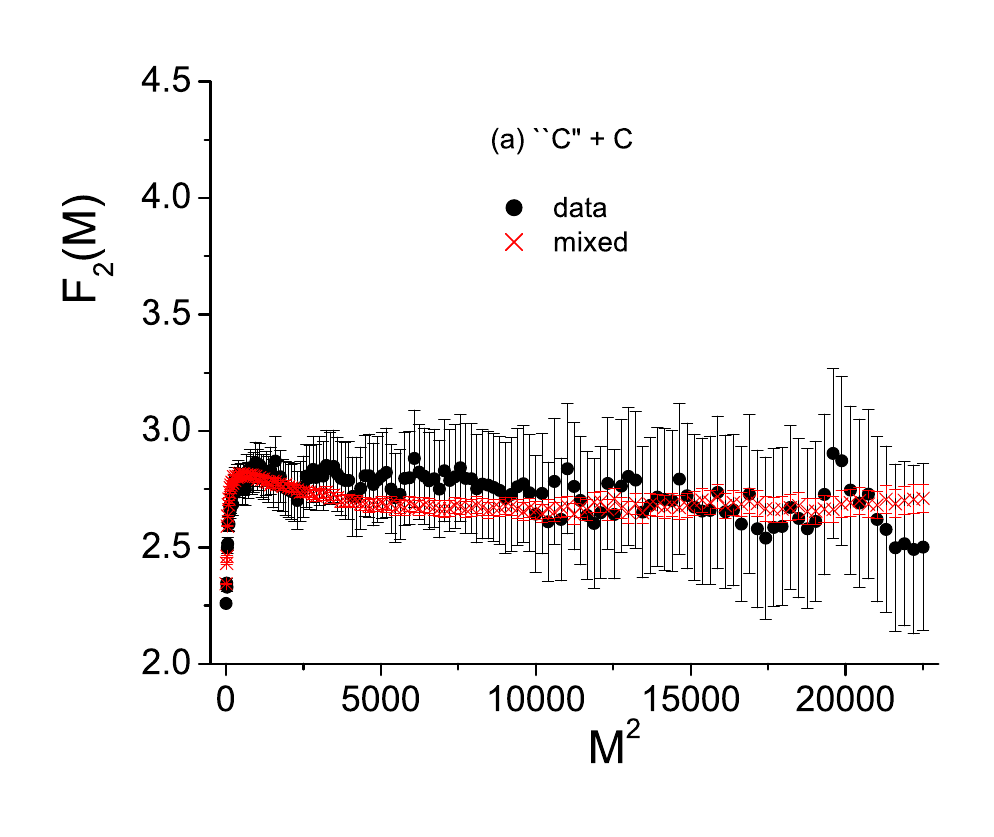}
\includegraphics[width=0.3\columnwidth]{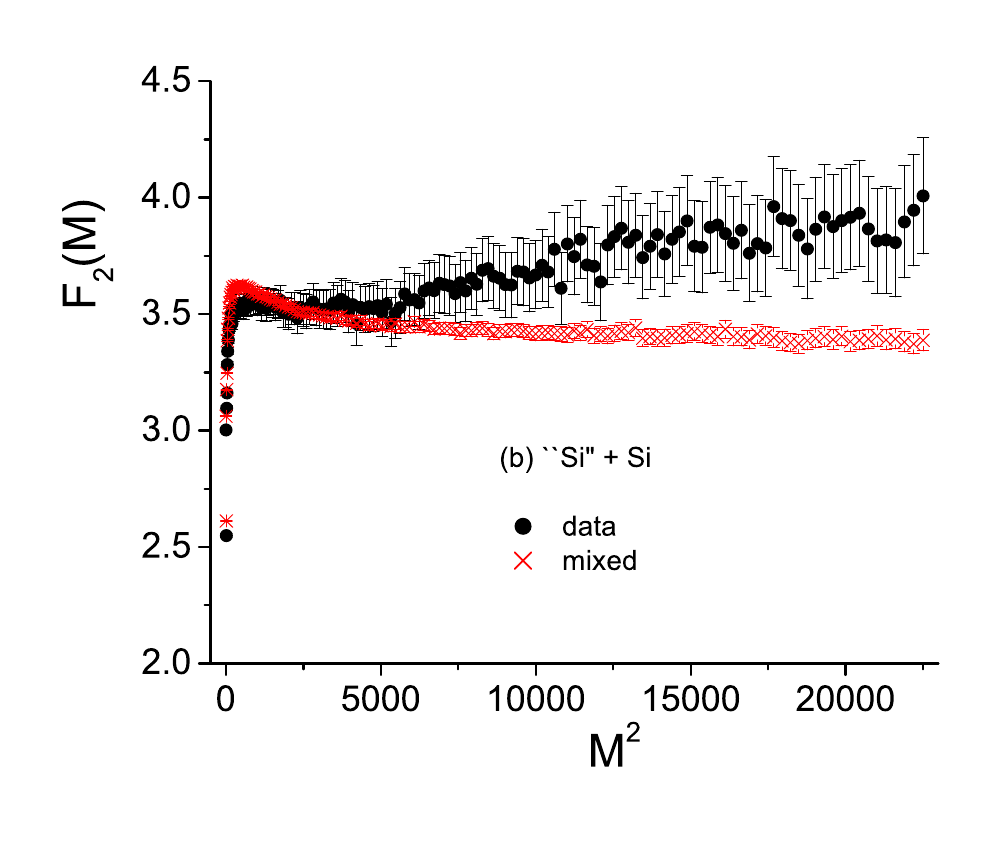}
\includegraphics[width=0.3\columnwidth]{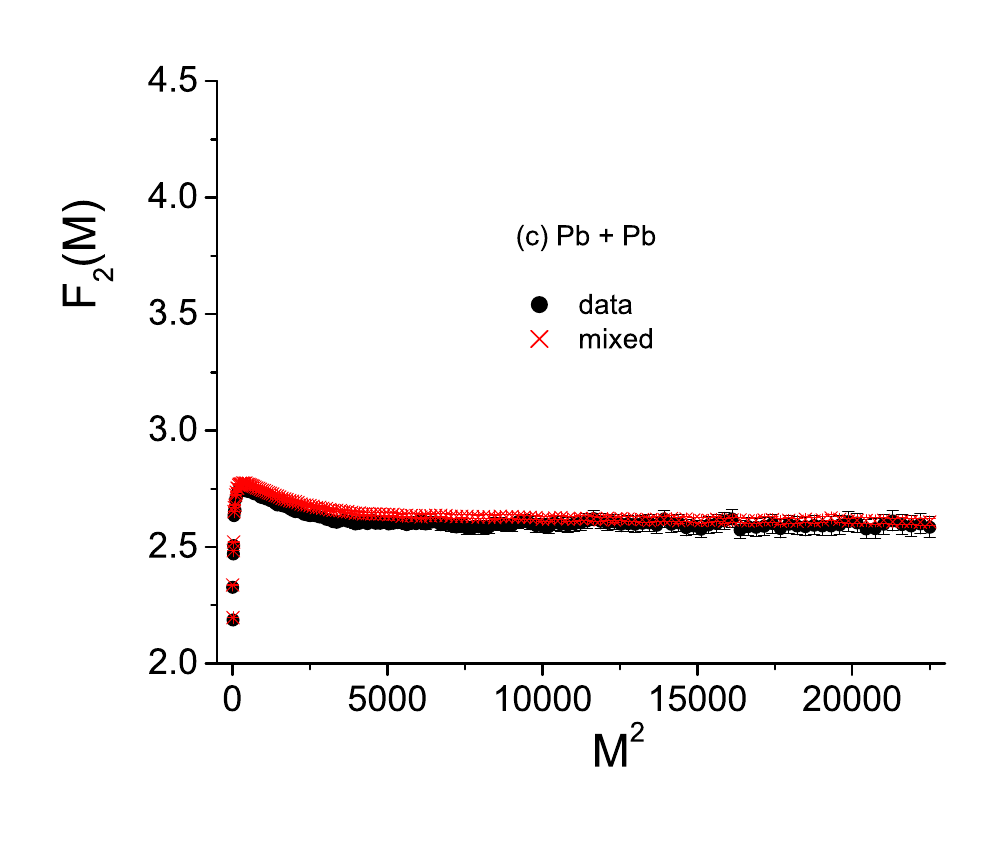}
\caption{Second scaled factorial moments $F_2(M)$ of the proton number in 
transverse momentum space at mid-rapidity ($-0.75 < y <0.75$) for the most central collisions  
of (a) ``C"+C (12\%), (b) ``Si"+Si (12\%), and (c) Pb+Pb (10\%) 
at $\sqrt{s_{NN}}=17.3$ GeV. The circles (crosses)  represent $F_2(M)$ 
of the data (mixed events) respectively. 
\label{fig:f2prot}
}
\end{figure}

\subsubsection{Scaled factorial moments}\label{proton_intermittency}

The NA49 experiment at the CERN SPS searched for an intermittency signal in the production
of proton~\cite{Anticic:2012xb} and low-mass $\pi^+\pi^-$ pairs~\cite{Anticic:2009pe} in the most
central collisions (12\%, 12\%, 10\%) of ``C", ``Si" and Pb nuclei on C (2.4\% interaction length),
Si (4.4\%) and Pb (1\%) targets, respectively, at beam energy of 158$A$~GeV ($\sqrt{s_{NN}} = 17.3$ GeV).
The analysis looked in transverse momentum space for a power law behaviour of the 
second scaled factorial moments (SSFMs, $F_2(M)$) defined in Eq.~\ref{eq:facmom} with $r = 2$.

{\it 4.1 Intermittency in proton production at mid-rapidity}

Protons were identified with a purity of above 80\% based on the ionization energy
loss of the tracks in the TPC detectors. Poorly measured and fake tracks were carefully removed by
suitable selection criteria since they can produce a spurious intermittency signal.
Since critical fluctuations originating from the CP are predicted to be strongest
in a region around mid-rapidity in the cms system, protons were selected in the rapidity
range $-0.75 < y <0.75$.

The strong background from mis-identified and non-critical protons was estimated and subtracted using mixed
events, which by construction do not contain critical fluctuations. The resulting SSFMs are plotted in
Fig.~\ref{fig:f2prot}. Evidently the mixed event background is consistent with the data for
``C"+C and Pb+Pb collisions. On the other hand, in ``Si"+Si reactions the values of $F_2(M)$ increase with $M^2$
while those of the mixed events remain nearly constant.

\begin{figure}[htbp]
\centering
\includegraphics[width=0.3\columnwidth]{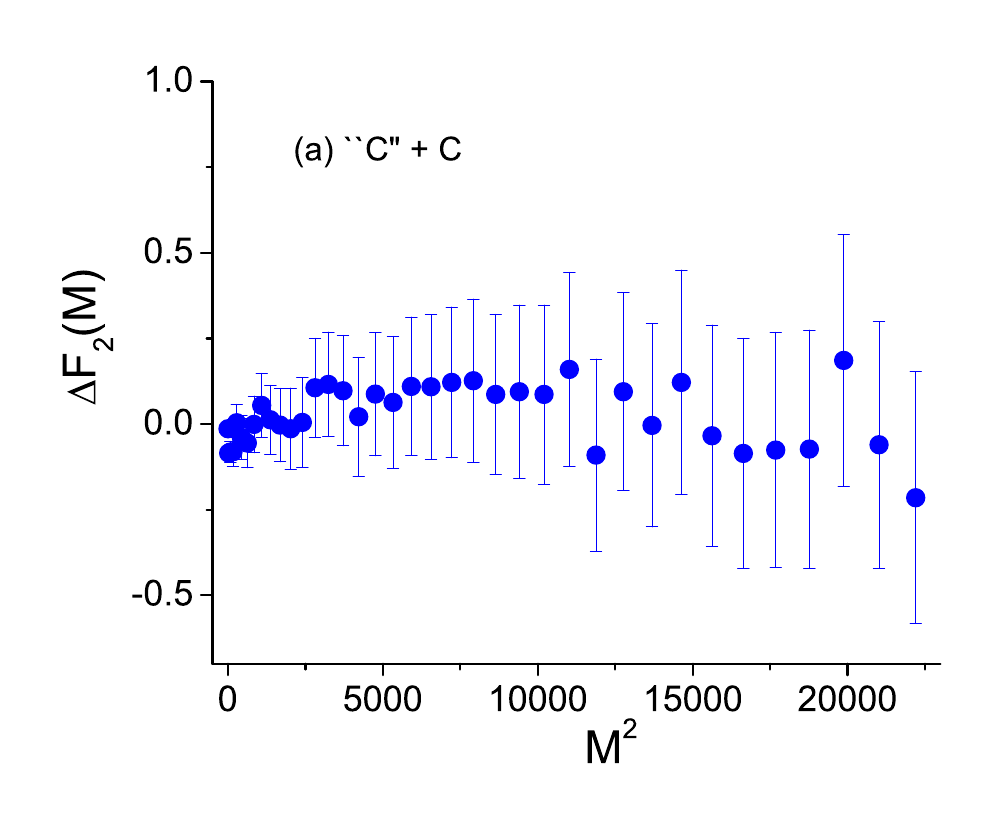}
\includegraphics[width=0.3\columnwidth]{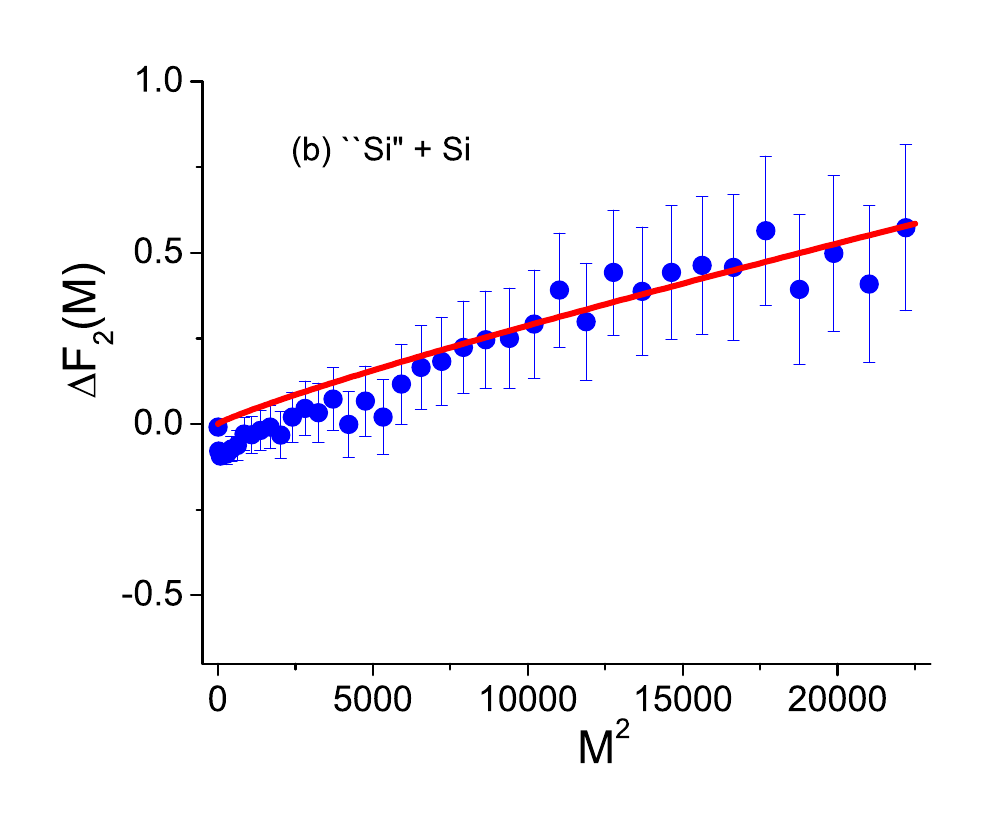}
\includegraphics[width=0.3\columnwidth]{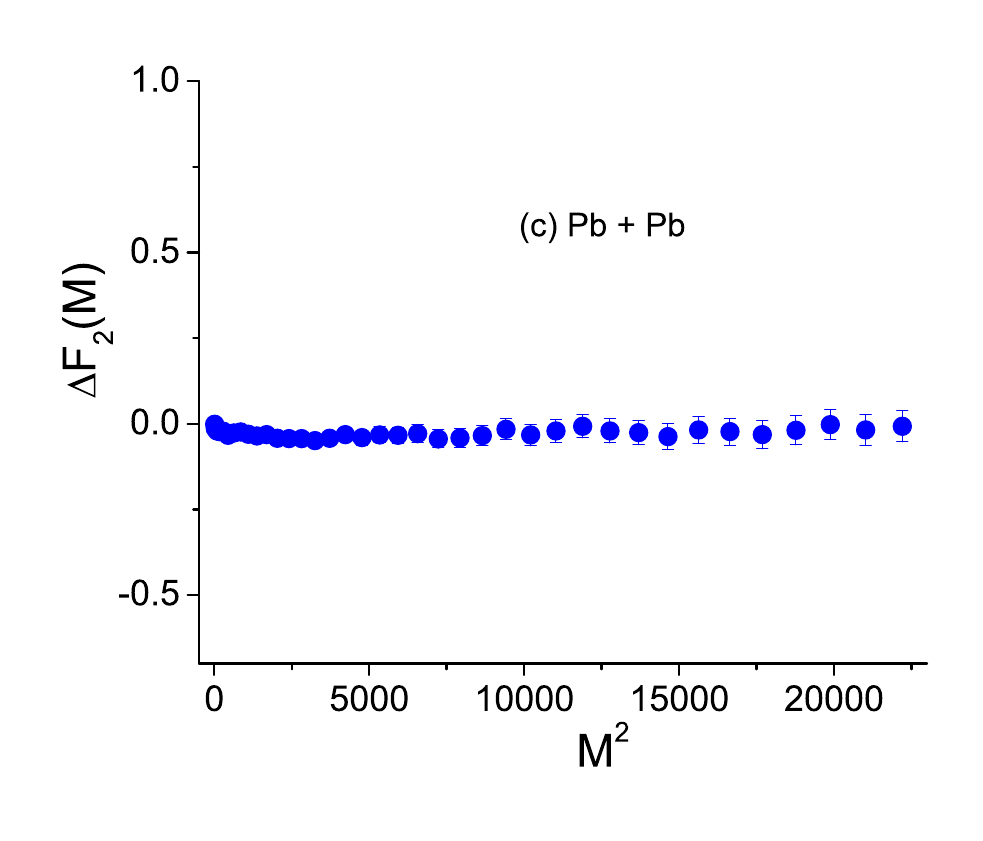}
\caption{The combinatorial background subtracted moments  
$\Delta F_2^{(e)}(M)$ corresponding to the moments of
Fig.~\ref{fig:f2prot} in bins of transverse momentum 
for the  most central collisions of (a) ``C"+C (centrality 12~\%),
(b) ``Si"+Si (centrality 12~\%)  and (c) Pb+Pb 
(centrality 10~\%) at $\sqrt{s_{NN}}=17.3$ GeV.
The line in the middle plot shows the result of 
a power-law fit for $M^2 > 6000$ with exponent 0.96. 
\label{fig:df2prot}
}
\end{figure}

Figure~\ref{fig:df2prot} shows the background subtracted SSFMs $F_2^{(e)}(M)$
\begin{equation}
\Delta F_2^{(e)}(M) = F_2^{(d)}(M)  - F_2^{(m)}(M)~,
\label{eq:estimator}
\end{equation}
where $F_2^{(d)}(M)$ and $F_2^{(m)}(M)$ correspond to the data and mixed-event background,
respectively. While the results for ``C"+C and Pb+Pb collisions scatter around zero, the
values for ``Si"+Si reactions rise with $M^2$. A power-law fit in the region $M^2 > 6000$ gave the result
$\phi_{2} = 0.96^{+0.38}_{-0.25}(stat)\pm0.16(syst)$ with $\chi^2/$dof $ \approx $ 0.09 - 0.51.
The re-sampling method~\cite{resampling}
was used to calculate the errors in order to take account of the strong correlation between successive (in $M$)
values of the SSFMs. Thus no intermittency signal is present for proton production
in ``C"+C and Pb+Pb collisions whereas power-law fluctuations are observed in ``Si"+Si reactions
with an exponent consistent with the CP prediction.

{\it 4.2 Intermittency in low-mass $\pi^+ \pi^-$ pair production}

Results of a search for critical fluctuations in the chiral condensate
via a similar intermittency study of low-mass $\pi^+ \pi^-$ pairs was published by the NA49 collaboration
in Ref.~\cite{Anticic:2009pe}. The chiral condensate is believed to
decay into $\pi^+ \pi^-$ pairs near the mass threshold when deconfined matter
hadronises. As in the case for protons
these fluctuations may be detectable by studying SSFMs of the $\pi^+ \pi^-$ pair
number, provided the combinatorial background can be sufficiently reduced.

Pions were required to have laboratory momenta exceeding 3 GeV/c and identification
was based on the ionization energy loss of the tracks in the TPC detectors.
For further analysis low-mass $\pi^+ \pi^-$ pairs satisfying
\begin{equation}
2 m_{\pi} +\epsilon_1 \leq m_{\pi^+\pi^-} \leq 2 m_{\pi} + \epsilon_2
\label{eq:epswin}
\end{equation}
were considered, where $m_{\pi^+\pi^-}$ is the pair invariant mass,
$\epsilon_1 = 5 $~MeV was chosen to remove the
enhancement of pairs from Coulomb attraction and $\epsilon_2 = 34, 24, 1 $~MeV
for ``C"+C, ``Si"+Si, and Pb+Pb respectively was optimised to reduce the combinatorial
background. Finally the remaining background was estimated by pairs from mixed events
which were made to satisfy the same criteria.

\begin{figure}[htbp]
\centering
\includegraphics[width=0.3\columnwidth]{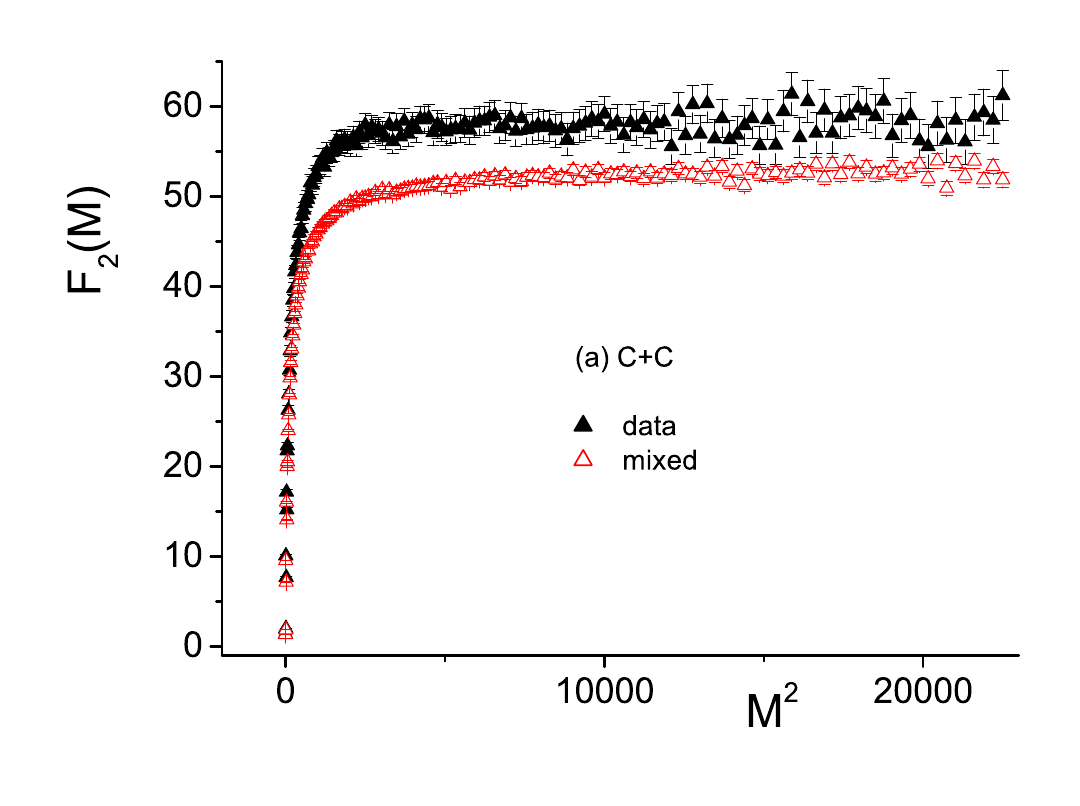}
\includegraphics[width=0.3\columnwidth]{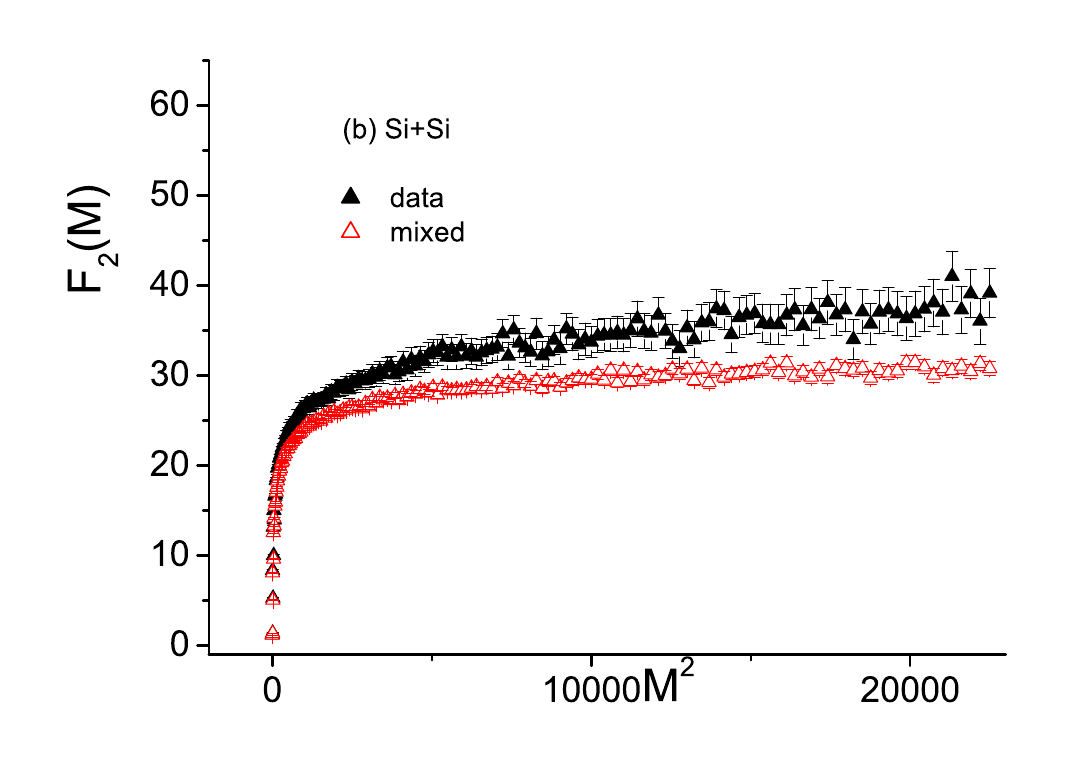}
\includegraphics[width=0.3\columnwidth]{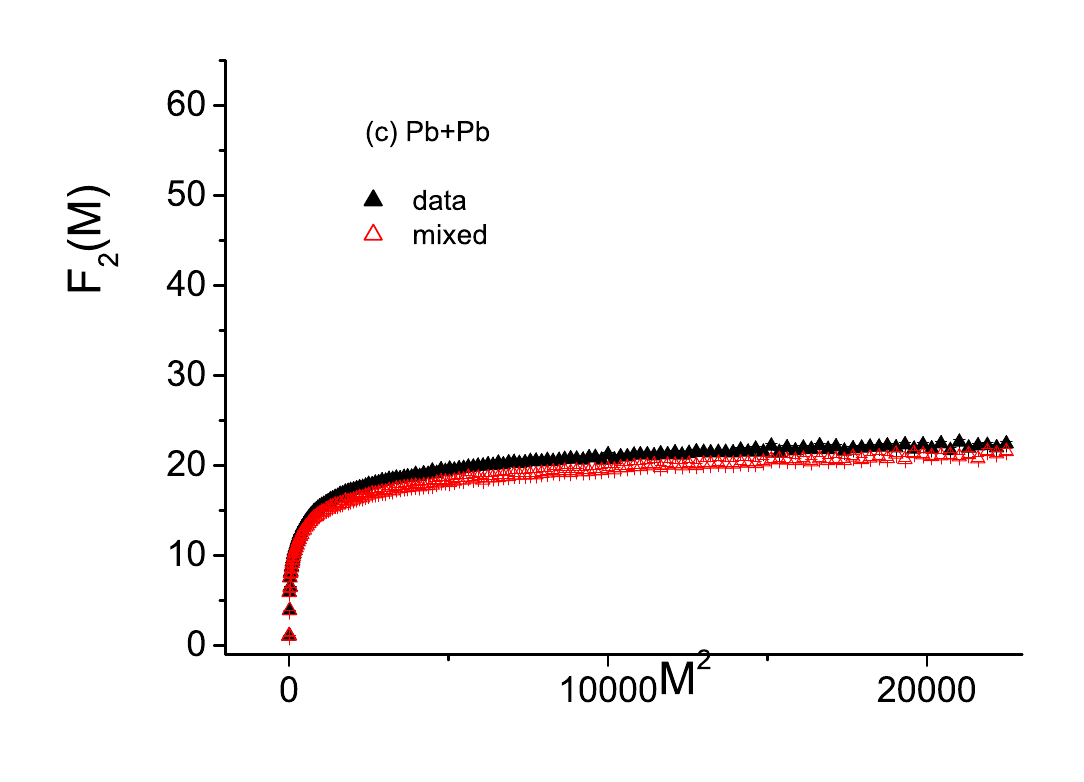}
\caption{The second factorial moment $F_2(M)$ in transverse momentum space for: 
(a) C+C (window of analysis $[285,314]$~MeV), 
(b) Si+Si (window of analysis $[300.9,304]$~MeV) and 
(c) Pb+Pb (window of analysis $[285,286]$~MeV) systems. 
The full triangles represent the moments of NA49 data while 
the open triangles the moments for the corresponding mixed events.
\label{fig:fig8abcd}
}
\end{figure}

\begin{figure}[htbp]
\centering
\includegraphics[width=0.3\columnwidth]{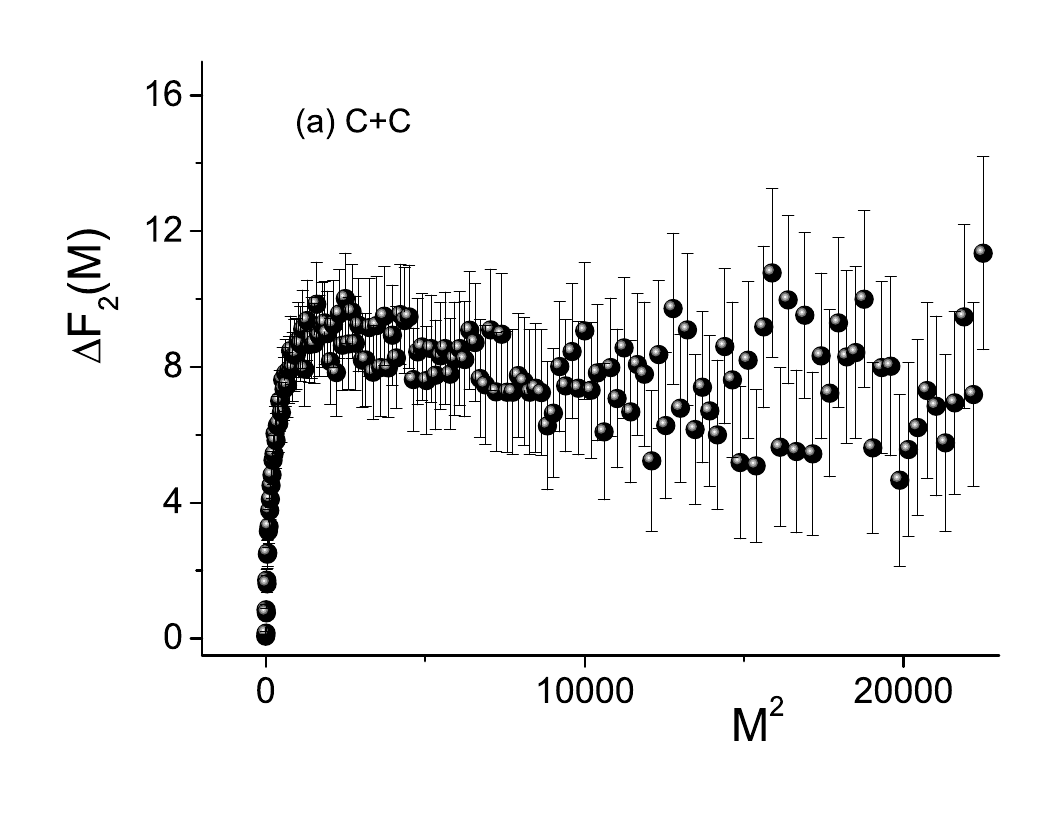}
\includegraphics[width=0.3\columnwidth]{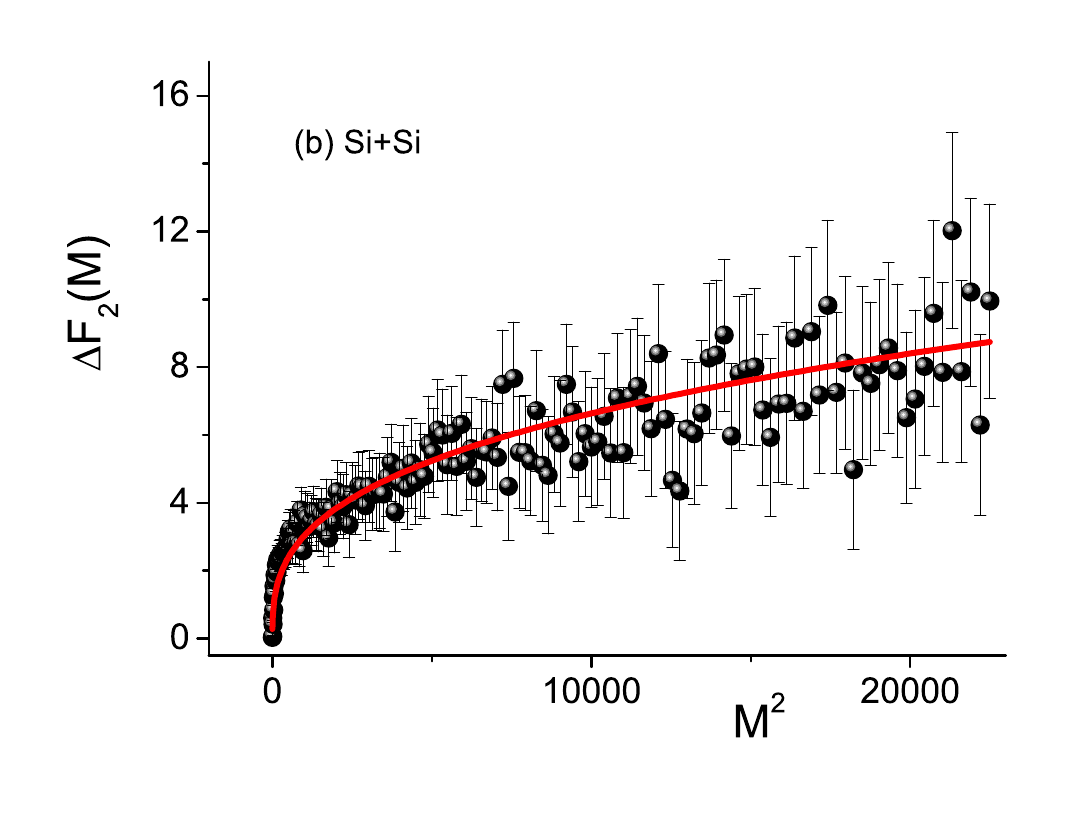}
\includegraphics[width=0.3\columnwidth]{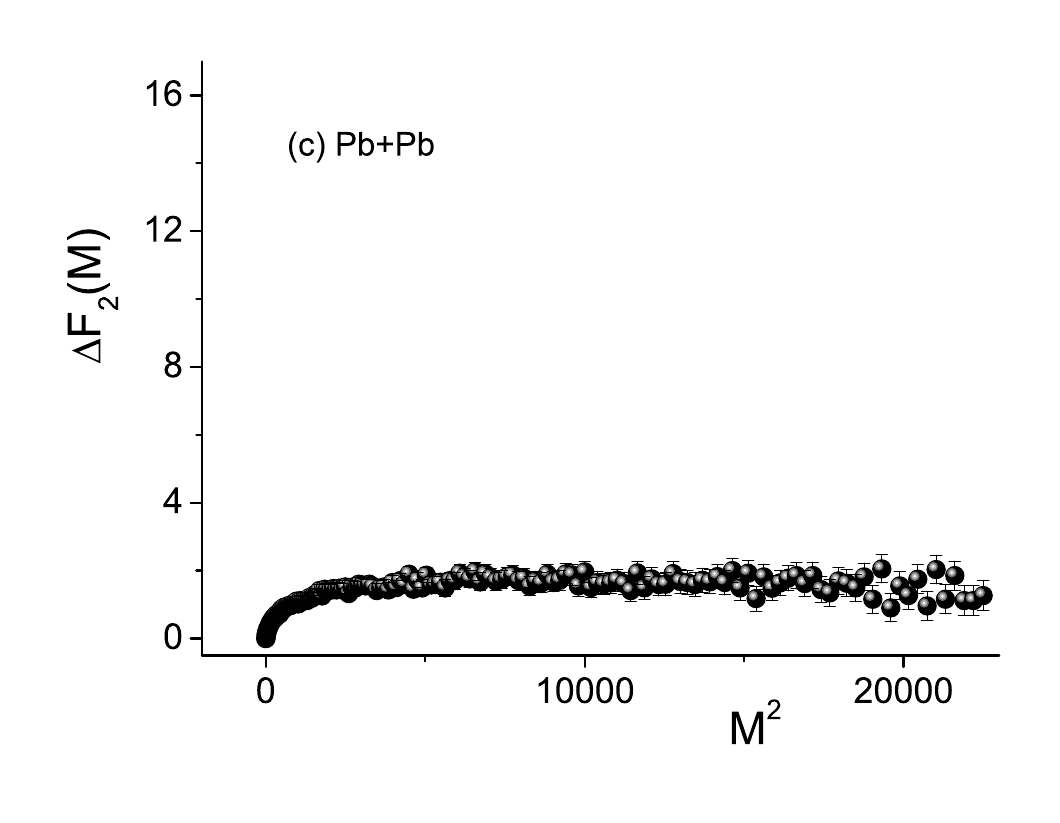}
\caption{The combinatorial background subtracted moments $\Delta F_2$ in transverse momentum space 
for: (a) C+C, (b) Si+Si and (c) Pb+Pb systems. The line in (b) shows the result of a power-law fit 
for $M^2 > 2000$ with exponent 0.33. 
\label{fig:fig9abcd}
}
\end{figure}

The SSFMs  of the $\pi^+ \pi^-$ pair multiplicity distribution 
 versus the number of subdivisions 
$M^2$ of the transverse momentum phase space are shown in Fig.~\ref{fig:fig8abcd} for pairs from data and
from mixed events. The rapidity region covered by the selected pairs essentially
extends forward of $y \gtrsim 0.5$. One observes that only for ``Si"+Si the SSFMs rise
faster for the data than for the mixed events. The combinatorial background subtracted
moments $\Delta F_2(M)$ are plotted versus $M^2$ in Fig.~\ref{fig:fig9abcd}.
At larger values of $M^2$ the results are consistent with being constant for ``C''+C and 
Pb+Pb whereas one finds an increase for the ``Si''+Si system. 
Here a power law function provides a good fit ($\chi^2/$dof $ \approx $ 0.3)
with an exponent $\Phi_2 = 0.33$  $\pm $ 0.04 where the error was estimated by exploiting the
subsample method. The extracted exponent indicates a significant intermittency effect, but is smaller than 
the expectation for the CP of $\Phi_2 = 0.67$. This might well be a consequence of the difficulty of isolating
the $\pi^+ \pi^-$ pairs from the $\sigma$ decays.

Recently another analysis method was developed to estimate critical exponents of
fluctuations arising from the existence of a CP. This technique 
studies finite size scaling of the particle source size parameters
as obtained from Bose-Einstein interferometry analysis~\cite{Lacey:2014wqa}.
The results based on RHIC and LHC measurements were interpreted as a possible
indication of the CP, however located at a value of $\mu_B$ beyond the range
accessible in the SPS energy range.

\subsubsection{Fluctuations of charged particle multiplicity}\label{mfluct}

\begin{figure}
\centering
\includegraphics[width=0.80\columnwidth]{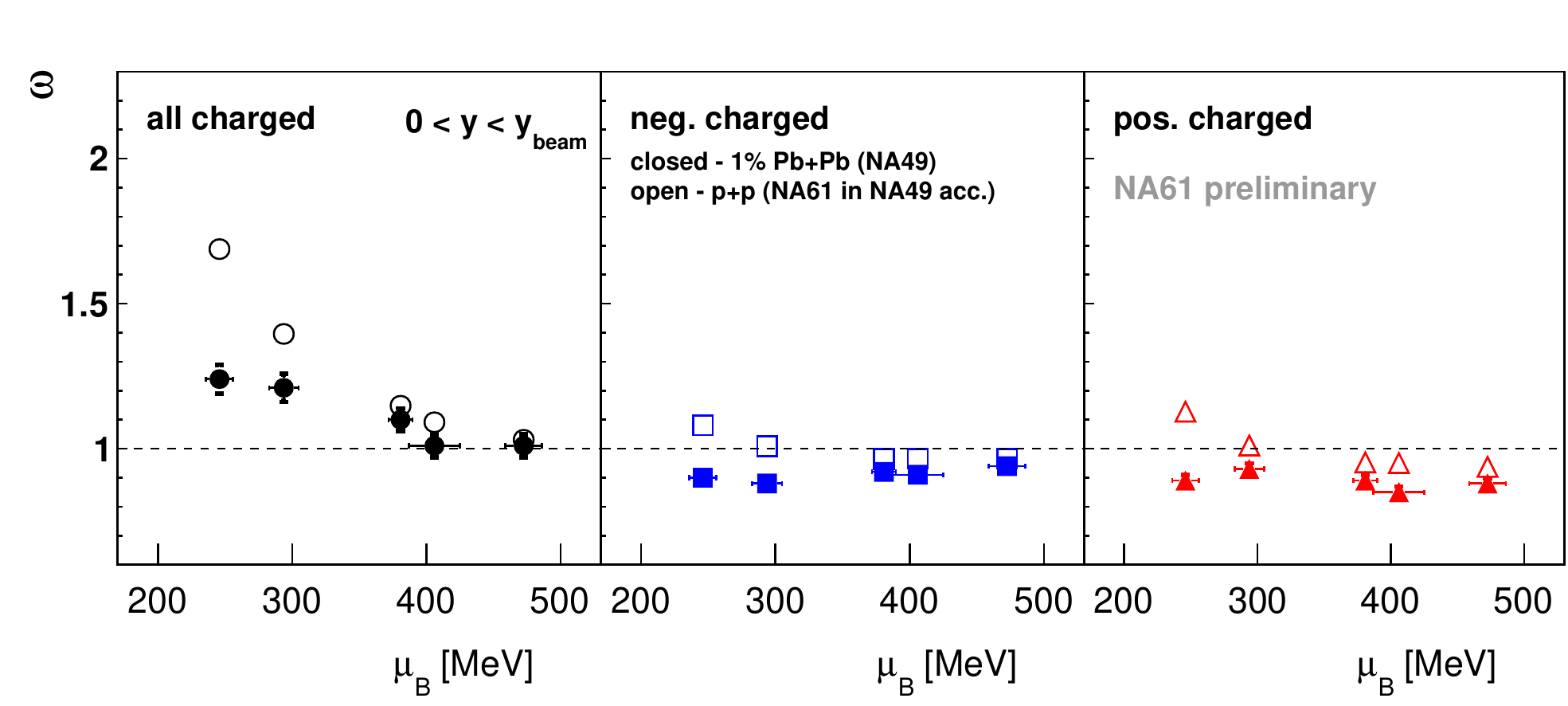}
\caption{Scaled variance $\omega$ of the multiplicity distribution of 
charged particles versus $\mu_B$ for the 1~\% most central Pb+Pb collisions 
and inelastic p+p reactions
for $0 < y < y_{beam}$ (assuming the pion mass). Full symbols show results of 
NA49~\cite{Alt:2007jq}, 
open symbols preliminary measurements of NA61/SHINE~\cite{Czopowicz:2015mfa,Aduszkiewicz:2015jna}.
\label{omega:mub_full}
}
\end{figure}

The signature of the CP is expected to be primarily an increase of multiplicity 
fluctuations~\cite{Stephanov:1999zu} which are usually quantified by the 
scaled variance $\omega = (\langle N^2 \rangle - \langle N \rangle^2)/\langle N \rangle$
of the distribution of particle multiplicities $N$ produced in the collisions. The measure $\omega$
is intensive, i.e. it is independent of the system volume
in statistical models within the GCE formulation.
However, $\omega$ is sensitive to the unavoidable volume fluctuations~\cite{Gorenstein:2011vq}.
Therefore the measurements were restricted to the 1~\%  most central collisions.
This selection is based on the energy deposited by beam spectator nucleons
in the forward calorimeter. Although this tightly constrains the number of projectile
participants, small fluctuations of the number of the target participants remain
(see model calculations of Ref.~\cite{Konchakovski:2005hq}). 
   
\begin{figure}
\centering
\includegraphics[width=0.80\columnwidth]{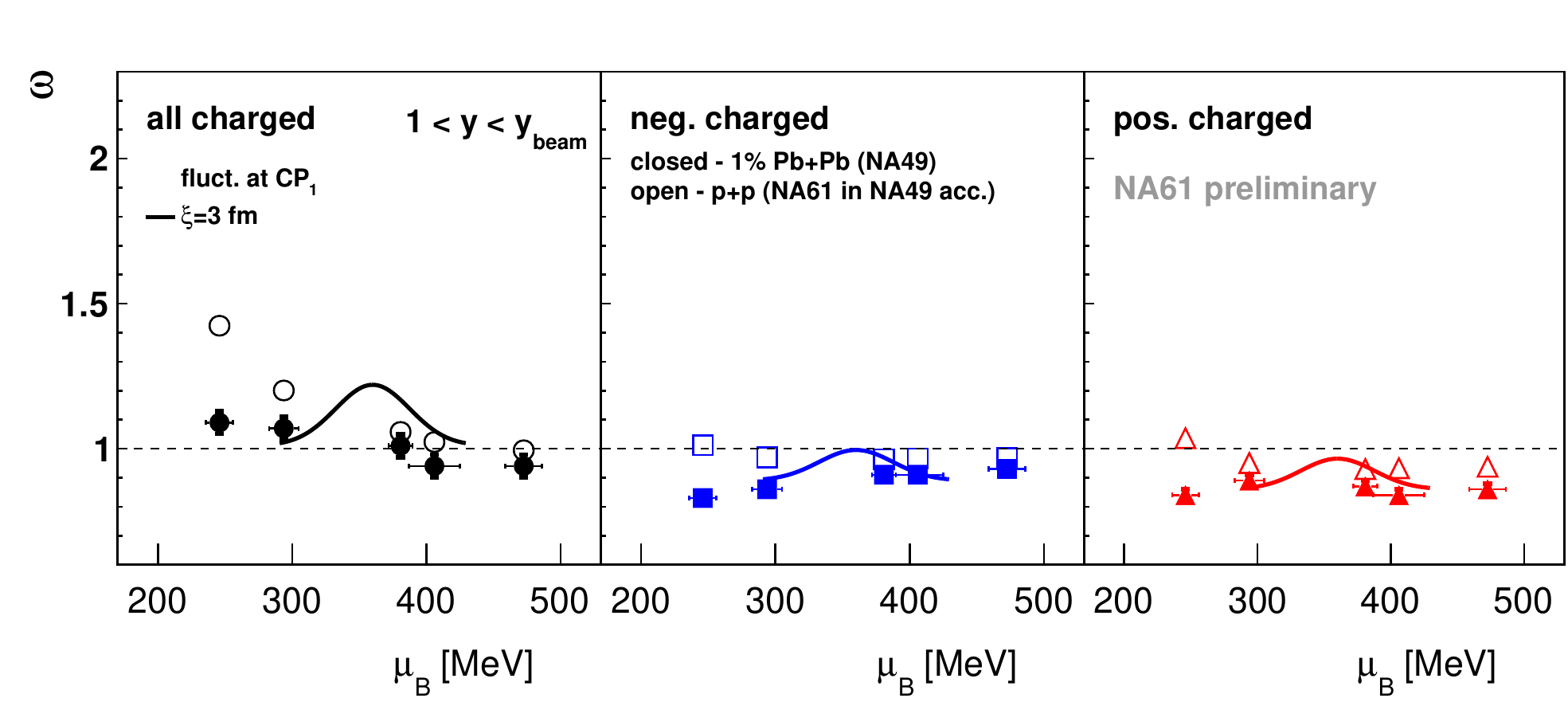}
\caption{Scaled variance $\omega$ of the multiplicity distribution of charged particles
versus $\mu_B$ for the 1~\% most central Pb+Pb collisions and inelastic p+p reactions
for $1.0 < y < y_{beam}$ (assuming the pion mass). Full symbols show results of 
NA49~\cite{Alt:2007jq}, open symbols 
preliminary measurements of NA61/SHINE~\cite{Czopowicz:2015mfa,Aduszkiewicz:2015jna}.
Curves illustrate the effect of a critical point~\cite{Grebieszkow:2009jr}.
\label{omega:mub_forward}
}
\end{figure}

Results for $\omega$ of charged particles in Pb+Pb collisions (NA49~\cite{Alt:2007jq}) are shown 
in Fig.~\ref{omega:mub_full} versus $\mu_B$\footnote{$\mu_B$ was obtained from statistical model fits to yields
of different particle types at the various collision energies~\cite{Becattini:2005xt}. 
It is a monotonically decreasing function of collision energy} 
and compared to preliminary NA61 results from 
p+p interactions~\cite{Czopowicz:2015mfa,Aduszkiewicz:2015jna}. Evidently,
the energy dependence of $\omega$ for the forward hemisphere 
(which represents almost the full
acceptance of NA49 and NA61) is smooth without significant maxima.
Note that at high SPS energies 
$\omega$ measured in inelastic p+p interactions is significantly
larger than the one for central Pb+Pb collisions.
The interpretation of this effect is under 
discussion~\cite{Konchakovski:2007ss,Aduszkiewicz:2015jna}.

Figure~\ref{omega:mub_forward} presents the scaled variance $\omega$ of charged particles
in the rapidity interval $1.0 < y < y_{beam}$ where the azimuthal acceptance of the detector
is high and uniform at all energies. Again no significant irregularities are observed.  

\begin{figure}
\centering
\includegraphics[width=0.40\columnwidth]{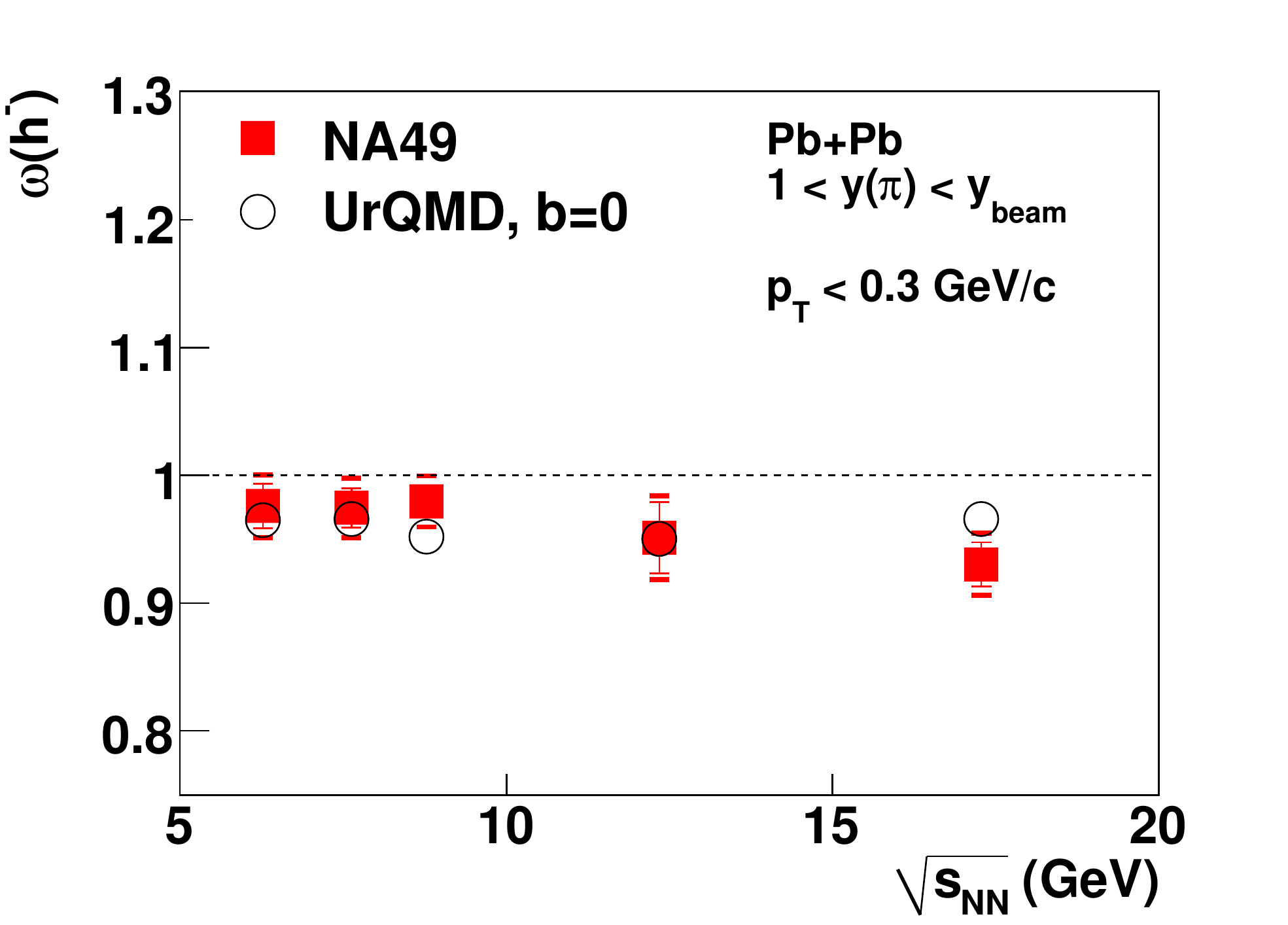}
\includegraphics[width=0.40\columnwidth]{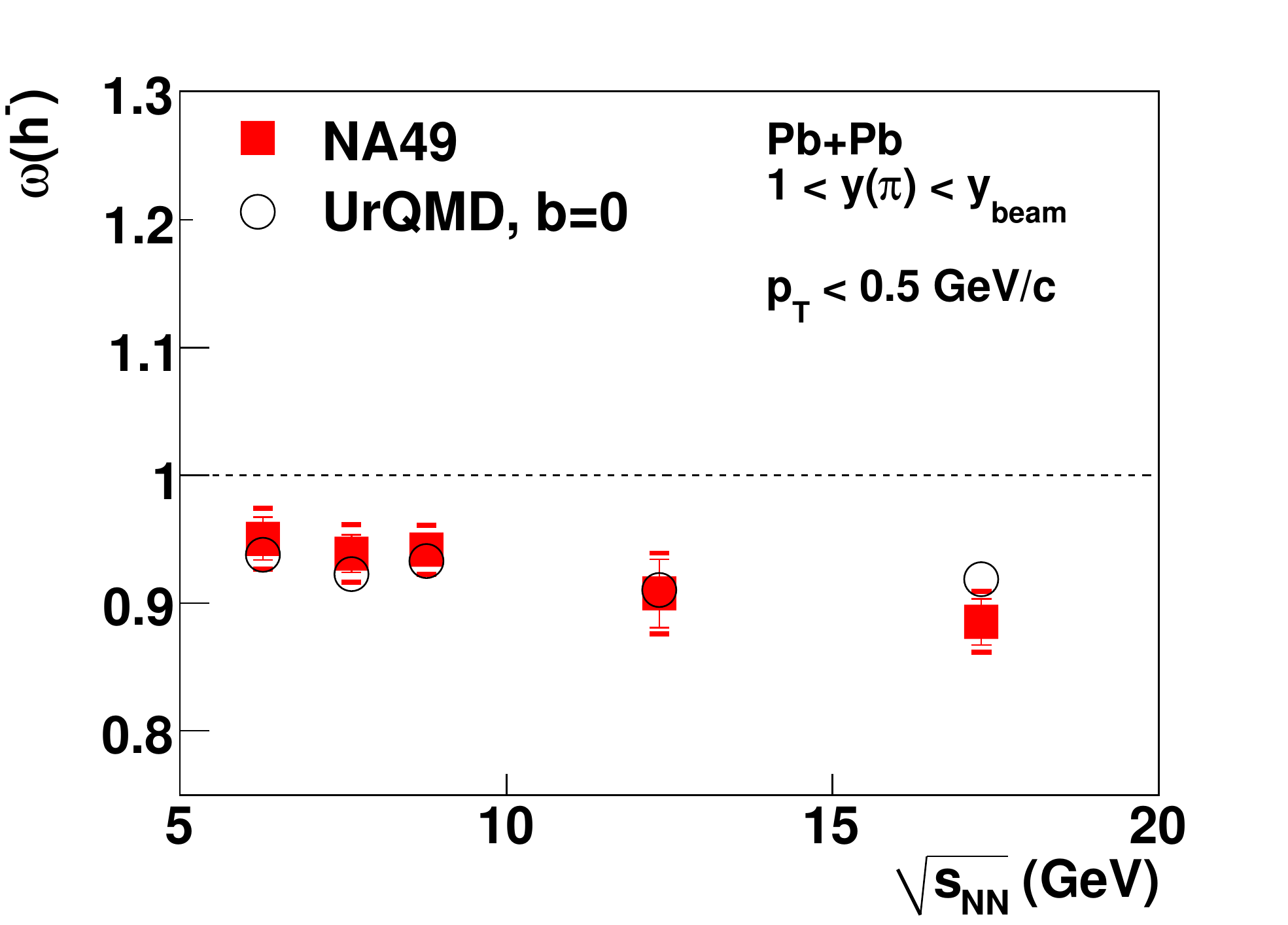}
\caption{Scaled variance $\omega$ of the multiplicity distribution of negatively 
charged particles versus $\mu_B$ in the 1~\% most central Pb+Pb collisions
for $1.0 < y < y_{beam}$ (assuming the pion mass)~\cite{Alt:2007jq}. 
{\it left:} $p_T < 0.3$~GeV/c. 
{\it right:} $p_T < 0.5$~GeV/c. 
Full symbols show data, open symbols are results from the UrQMD model.
\label{omega:ptcut}
}
\end{figure}

Fluctuations induced by the CP are expected to be stronger at low transverse momenta $p_T$~\cite{Stephanov:priv}. 
Therefore $\omega$ was also calculated for $p_T < 0.3$~GeV/$c$ 
and $< 0.5$~GeV/$c$. The results are plotted in
Fig.~\ref{omega:ptcut}~\cite{Alt:2007jq} and turned out to be very similar.

The effects otf a hypothetical critical point at $\mu_B = 360$~MeV are illustrated
by the curves in Fig.~\ref{omega:mub_forward}~\cite{Grebieszkow:2009jr}. 
Assuming $\xi$ = 3~fm the value of $\omega$ was estimated to
increase by 0.5 (respectively 0.25) for all charged (negatively or positively) 
charged particles~\cite{Stephanov:1999zu,Stephanov:priv}
with respect to the value 
expected for the background fluctuations.
The limited acceptance of the detector is expected to reduce the increase
by a factor of the order of 0.6~\cite{Alt:2007jq}. Guided by the
considerations of Ref.~\cite{Hatta:2002sj} on the region over which the effects of the CP
increase the fluctuations, a parameterization 
by a Gaussian function in $\mu_B$ was chosen with $\sigma(\mu_B) \approx$ 30~MeV. 
Evidently, the data do not support a maximum as might be expected for a 
CP~(see solid curves in Fig.~\ref{omega:mub_forward}).

\begin{figure}
\centering
\includegraphics[width=0.80\columnwidth]{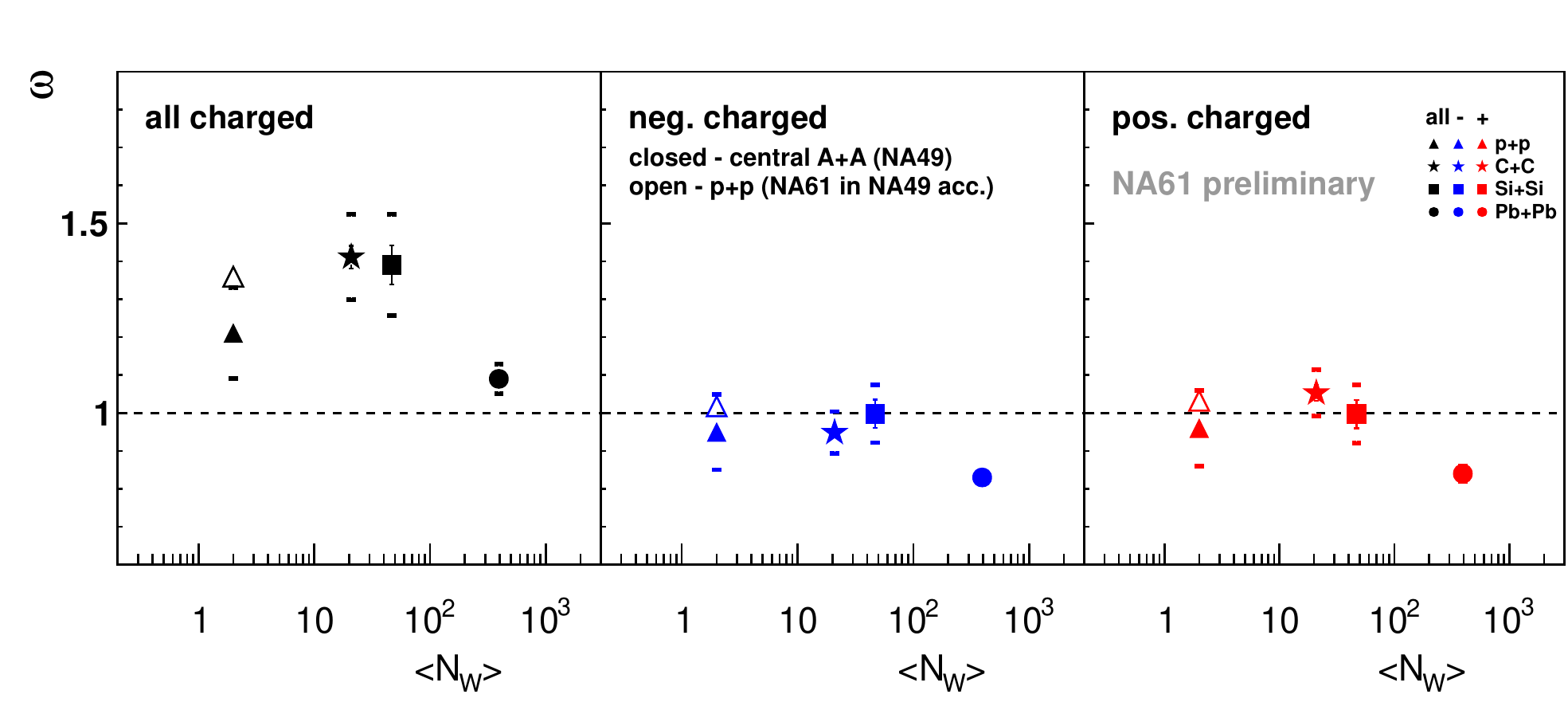}
\caption{Scaled variance $\omega$ of the multiplicity distribution of charged particles
versus the number of wounded nucleons $N_W$ in inelastic p+p ($1.1 < y < 2.6$) 
and the 1~\% most central ``C"+C, ``Si"+Si and Pb+Pb collisions 
at 158$A$ GeV ($1.0 < y < y_{beam}$). 
Full symbols show results of NA49~\cite{Alt:2007jq}, open symbols 
NA61/SHINE~\cite{Czopowicz:2015mfa,Aduszkiewicz:2015jna}.
\label{omega:nw}
}
\end{figure}

NA49 also obtained $\omega$ for smaller size nuclei at the top SPS energy of 158$A$ GeV~\cite{Alt:2006jr}.
The results together with those for inelastic p+p collisions from NA49 and NA61 are plotted in Fig.~\ref{omega:nw}.
Interestingly, there may be an indication of a maximum for medium size nuclei. 

\begin{figure}
\centering
\includegraphics[width=0.80\columnwidth]{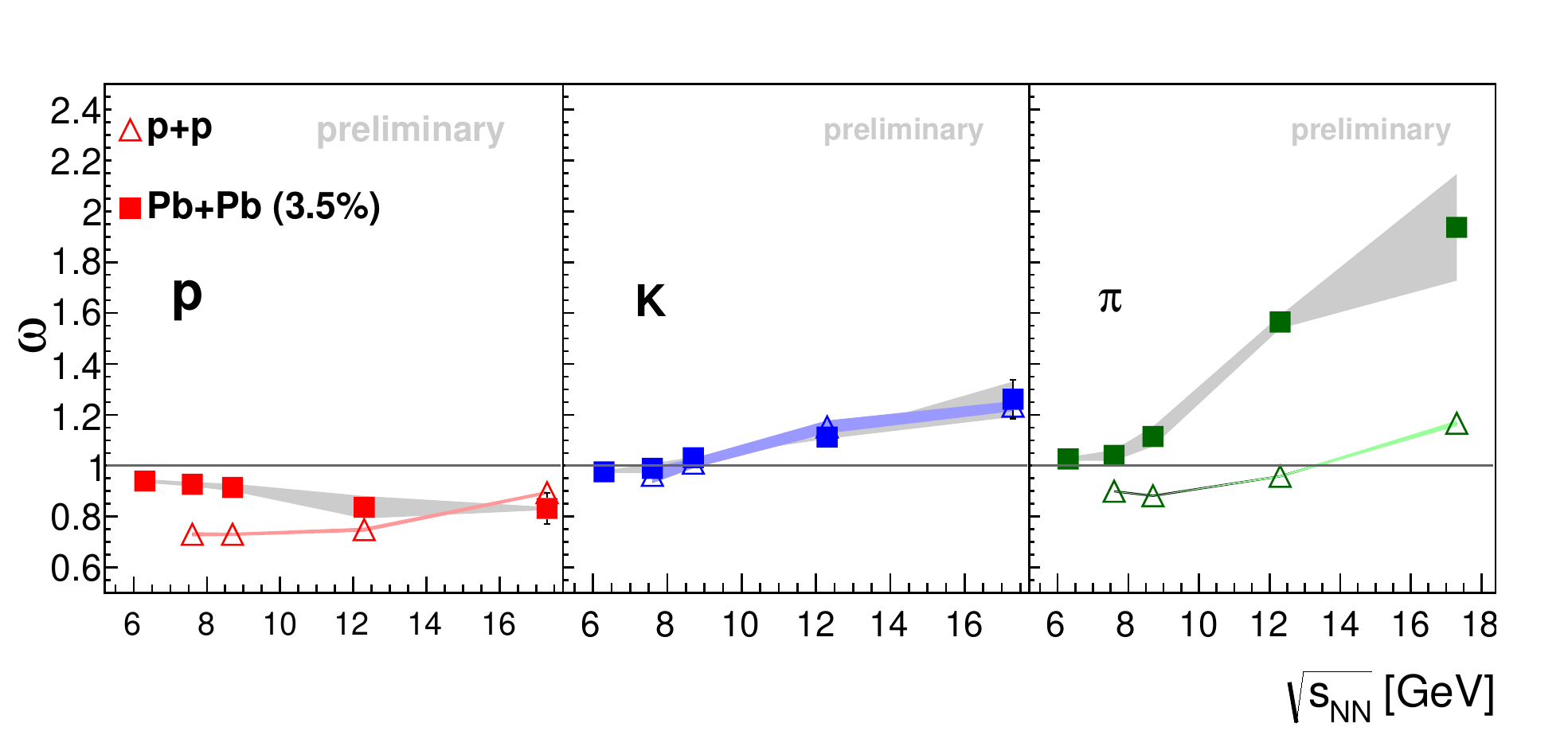}
\caption{Scaled variance $\omega$ of the multiplicity distribution of protons, kaons, and pions
versus nucleon-nucleon cms energy $\sqrt{s_{NN}}$ for the 3.5~\% most central Pb+Pb collisions 
(NA49, full symbols) and inelastic p+p reactions (NA61, open symbols). 
Large fluctuations for pions in central Pb+Pb collisions at high energies
are likely due to
volume fluctuations.
Results are preliminary~\cite{Seyboth:ismd2013}.
\label{omega:ident}
}
\end{figure}

A new identification procedure (the identity method~\cite{Gazdzicki:2011xz,Gorenstein:2011hr}) was developed
which allows to determine the second and third moments of 
the multiplicity distribution when the particle identification
is not unique but can only be done on a statistical basis. 
Applying this method NA49 and NA61/SHINE  determined
the scaled variance of the
multiplicity distribution of identified protons, kaons and pions in inelastic p+p and 3.5\% most central
Pb+Pb collisions. The results are shown in Fig.~\ref{omega:ident}. As in the case of $\omega$
for unidentified charged particles
no indication of the CP is found. It was pointed out that higher moments of the multiplicity distributions
are more sensitive to effects of the CP~\cite{Stephanov:2008qz}. The STAR collaboration at RHIC performed
such a study for the net-proton multiplicity in central Au+Au collisions at energies in the range
$\sqrt{s_{NN}}$~=~7.7~-~200~GeV~\cite{Adamczyk:2013dal}, but also found no evidence for the CP.

\subsubsection{Fluctuations of the net charge}\label{charge_fluct}

Fluctuations of the net charge were originally studied in collisions of
heavy nuclei in an attempt to find evidence for a deconfined phase. The NA49 collaboration concluded
that the measurements at CERN SPS energies~\cite{Alt:2004ir} were not sensitive to the initial
fluctuations in the QGP since they get masked by the effects of resonance
decays. The search for the CP has rekindled interest in this observable
because the CP might enhance net-charge fluctuations when the system freezes out in 
its vicinity. 

\begin{figure}
\centering
\includegraphics[width=0.60\columnwidth]{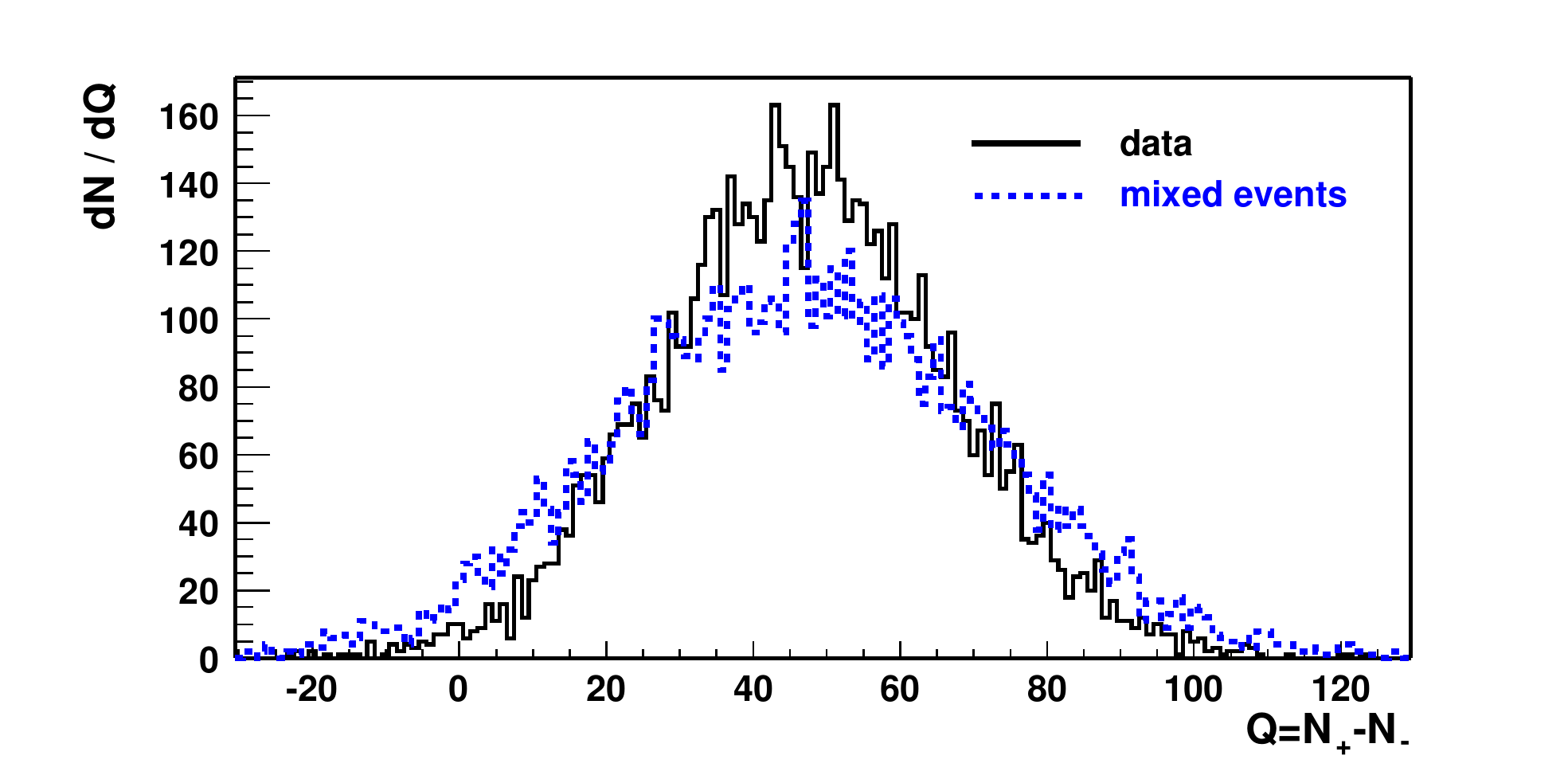}
\caption{
The distribution of the net-charge for central Pb+Pb collisions
at 158$A$~GeV (solid line) and the corresponding distribution obtained
for mixed events (dotted line) in the maximum rapidity interval
$\Delta y = 3$.
\label{fig:NetQ}
}
\end{figure}

Fluctuations of the net charge were quantified by the measure $\Phi$ proposed in
Ref.~\cite{Gazdzicki:1992ri} and defined in Eq.~\ref{Phi}.
For the case of charge fluctuations $x$ is taken to be the electric charge $q$ 
and the measure is called
$\Phi_q$.

The distribution of net-charge $Q = \sum_{i=1}^{N}q_i$ in real
and mixed events is shown in Fig.~\ref{fig:NetQ} where the sum runs over
the $N$ particles of the individual events. One observes a clear narrowing
effect due to charge conservation, which needs to be corrected.
In a scenario in which particles are correlated only by
global charge conservation (GCC)
the value of $\Phi_q$ is given by
\begin{equation}\label{Phiq1}
\Phi_{q,\rm{GCC}}=\sqrt{1-P}-1 ,
\end{equation}
where
\begin{equation}\label{P}
P=\frac {\langle N_{ch}\rangle }{\langle N_{ch} \rangle _{tot}}
\end{equation}
with $\langle N_{ch}\rangle$ and $\langle N_{ch} \rangle _{tot}$
being the mean charged multiplicity in the detector
acceptance and in full phase space (excluding spectator nucleons),
respectively.
In order to remove the sensitivity to  GCC the measure $\Delta\Phi_{q}$ is
defined as the difference:
\begin{equation}\label{deltaPhi}
\Delta \Phi_q =\Phi_q-\Phi_{q,\rm{GCC}}\;.
\end{equation}
By construction, the value of $\Delta \Phi_q$ is zero
if the particles are correlated by global charge conservation only.
It is negative in case of an additional correlation between positively
and negatively charged particles, and it is positive if the positive
and negative particles are anti-correlated.

\begin{figure}
\centering
\includegraphics[width=0.40\columnwidth]{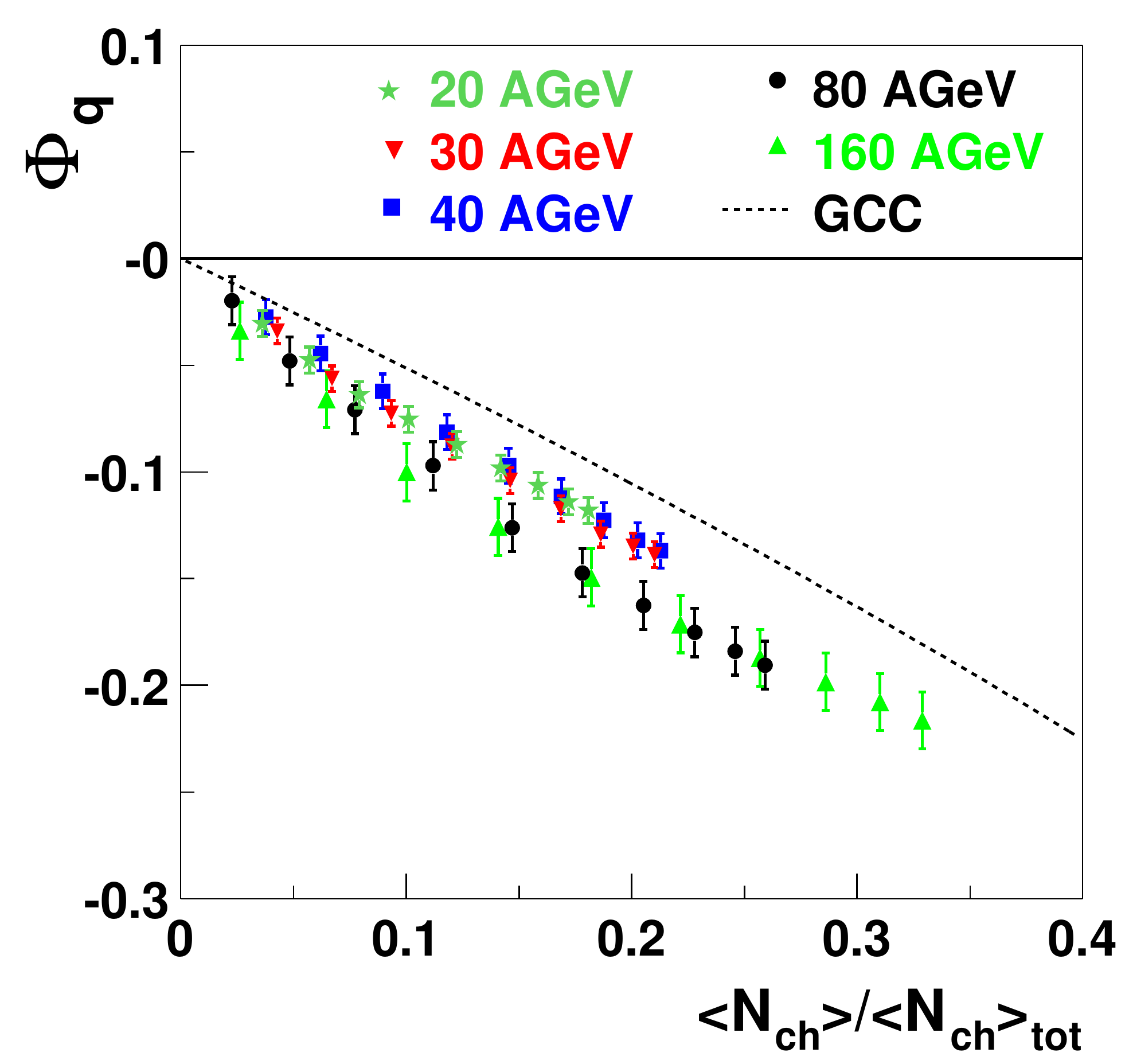}
\includegraphics[width=0.40\columnwidth]{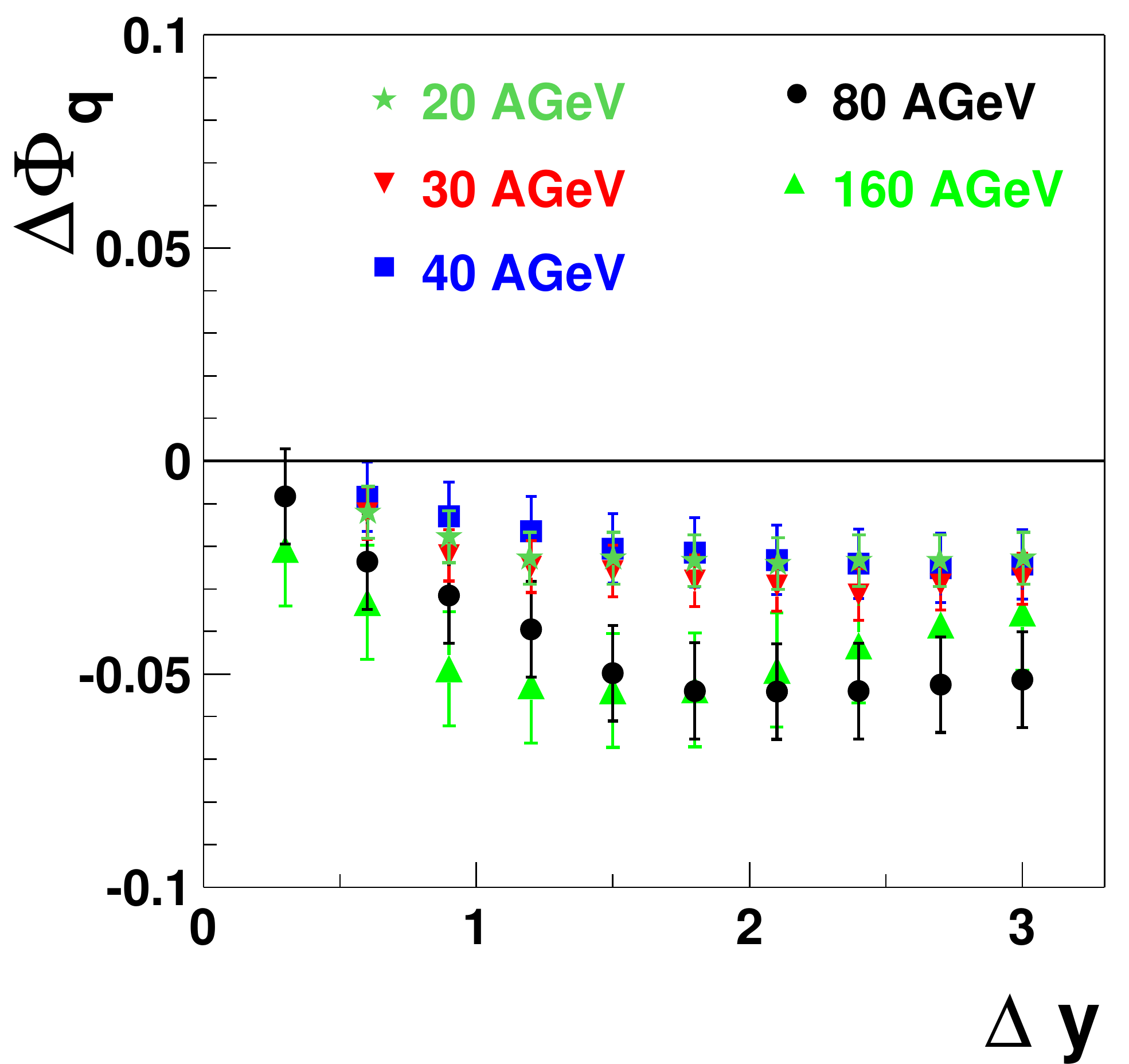}
\caption{
{\it Left:} The dependence of the measure $\Phi_q$ on the fraction
of accepted particles for central Pb+Pb collisions at 20$A$-158$A$~GeV.
The dashed line shows the dependence expected for the case when the only source
of particle correlations is global charge conservation.
{\it Right:} The dependence of $\Delta\Phi_{q}$
on the width of the rapidity interval $\Delta y$ for central Pb+Pb collisions
at 20$A$-158$A$~GeV.
Note that experimental points for a given energy are correlated as
the data used for a given rapidity interval
 are included in the broader intervals.
\label{fig:PhiQ}
}
\end{figure}

\begin{figure}
\centering
\includegraphics[width=0.80\columnwidth]{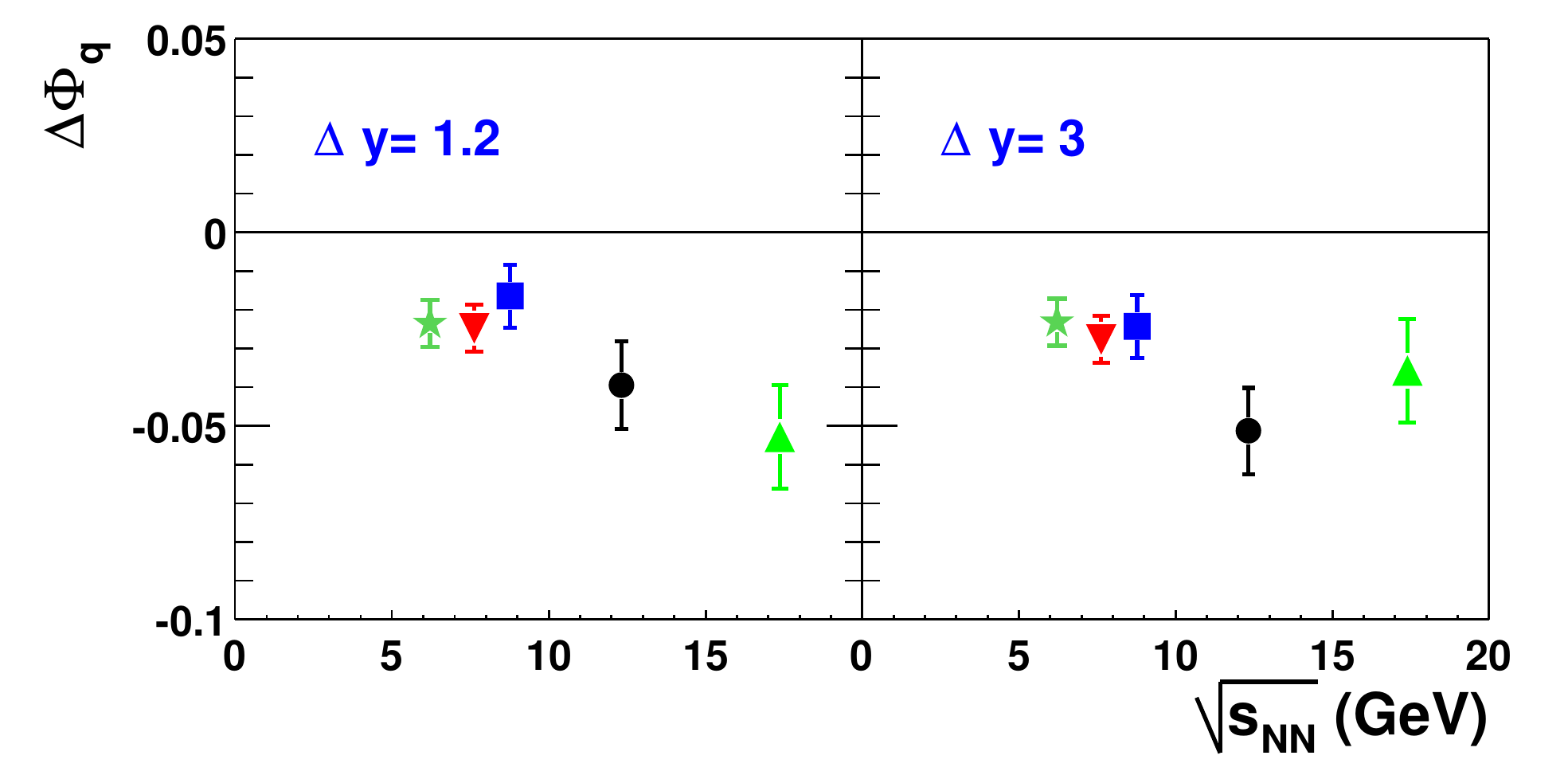}
\caption{
The energy dependence of $\Delta\Phi_{q}$ measured in central Pb+Pb collisions
at 20$A$, 30$A$, 40$A$, 80$A$ and 158$A$~GeV for a narrow rapidity interval 
$\Delta y = 1.2$~({\it left}) and a broad rapidity interval $\Delta y = 3$~({\it right}).
\label{fig:Delta_PhiQ_vs_sqrts}
}
\end{figure}

Figure~\ref{fig:PhiQ} demonstrates that the observed net-charge fluctuations are close
to the expectation for global charge conservation for all the selected bands of rapidity
(the charged particle multiplicity in the band was used as abscissa). The values $\Delta\Phi_{q}$
stay small at all SPS energies. As shown
by the energy dependence of $\Delta\Phi_{q}$ for two rapidity intervals in Fig.~\ref{fig:Delta_PhiQ_vs_sqrts}
one does not observe a peak structure. The STAR collaboration at RHIC performed
a study of the higher moments of the net-charge distribution in central Au+Au collisions at energies in the range
$\sqrt{s_{NN}}$~=~7.7~-~200~GeV~\cite{Adamczyk:2014fia} and so far also found no evidence for the CP.

\subsubsection{Transverse momentum - multiplicity fluctuations}\label{ptfluct}

Enhancement is also expected for transverse momentum - multiplicity fluctuations  
when the freeze-out occurs close to the CP~\cite{Stephanov:1999zu}. 
For the case of transverse momentum - multiplicity fluctuations
the fluctuation measures $\Delta$ and $\Sigma$ with the IPM normalisation 
(see Sec.~\ref{sec:volume}) read:
\begin{equation}
        \Delta[P_{T},N] =
        \frac {1}{\langle N \rangle \omega[p_{T}]}[\langle N \rangle \omega[P_{T}] - 
  \langle P_{T} \rangle \omega[N]]
        \label{eq:delta1}
\end{equation}
and
\begin{equation}
  \Sigma[P_{T},N] =
  \frac{1}{\langle N \rangle \omega[p_{T}]}[\langle N \rangle \omega[P_{T}] +
  \langle P_{T} \rangle \omega[N] - 2(\langle P_{T}N \rangle - 
  \langle P_{T} \rangle \langle N \rangle )]\ ,
\label{eq:sigma1}
\end{equation}
where
\begin{equation}
  \omega[P_{T}] = \frac{\langle P_{T}^2 \rangle  - 
  \langle  P_{T} \rangle ^2}{\langle P_{T} \rangle }
\end{equation}
and
\begin{equation}
  \omega[N] = \frac{\langle N^2 \rangle  - \langle N \rangle ^2}
  {\langle N \rangle }
\end{equation}
are the scaled variances of the two fluctuating extensive event quantities $P_{T}$, the sum of the absolute
values of transverse momenta $p_{T}$, and $N$, the number of particles, respectively. 
The quantity $\omega[p_{T}]$ is the scaled variance of
the inclusive $p_{T}$ distribution (summation runs over all particles and all events)
\begin{equation}
  \omega[p_{T}] =
  \frac{\overline{p_{T}^2} - \overline{p_{T}}^2}{\overline{p_{T}}}.
\end{equation}

There is an important difference between $\Delta[P_{T},N]$ and $\Sigma[P_{T},N]$.
Only the first two moments: $\langle P_{T} \rangle $, $\langle N \rangle$,
and $\langle P_{T}^2 \rangle $, $\langle N^2 \rangle $ are
required to calculate $\Delta[P_{T},N]$,
whereas $\Sigma[P_{T},N]$ includes the correlation term
$\langle P_{T} \cdot N \rangle$.
Thus the measures $\Delta[P_{T},N]$ and $\Sigma[P_{T},N]$
can be sensitive to specific fluctuations in different ways. Both measures are
dimensionless and have a common scale required for a quantitative comparison of fluctuations
of different, in general dimensional,
extensive quantities. The values of $\Delta$ and $\Sigma$
are equal to zero in the absence of event-by-event fluctuations and equal
to one for fluctuations given by the model of independent particle production.

\begin{figure}
\centering
\includegraphics[width=0.80\columnwidth]{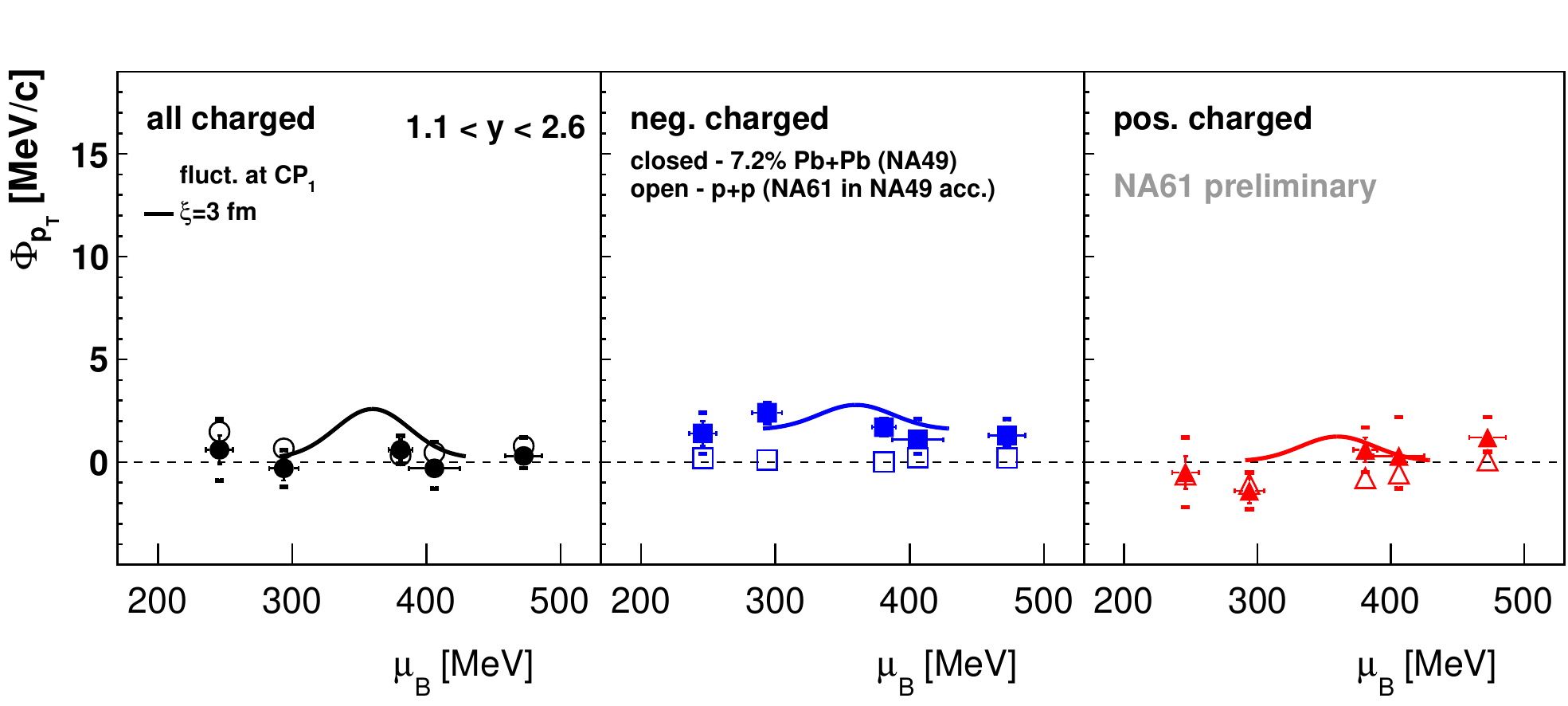}
\includegraphics[width=0.80\columnwidth]{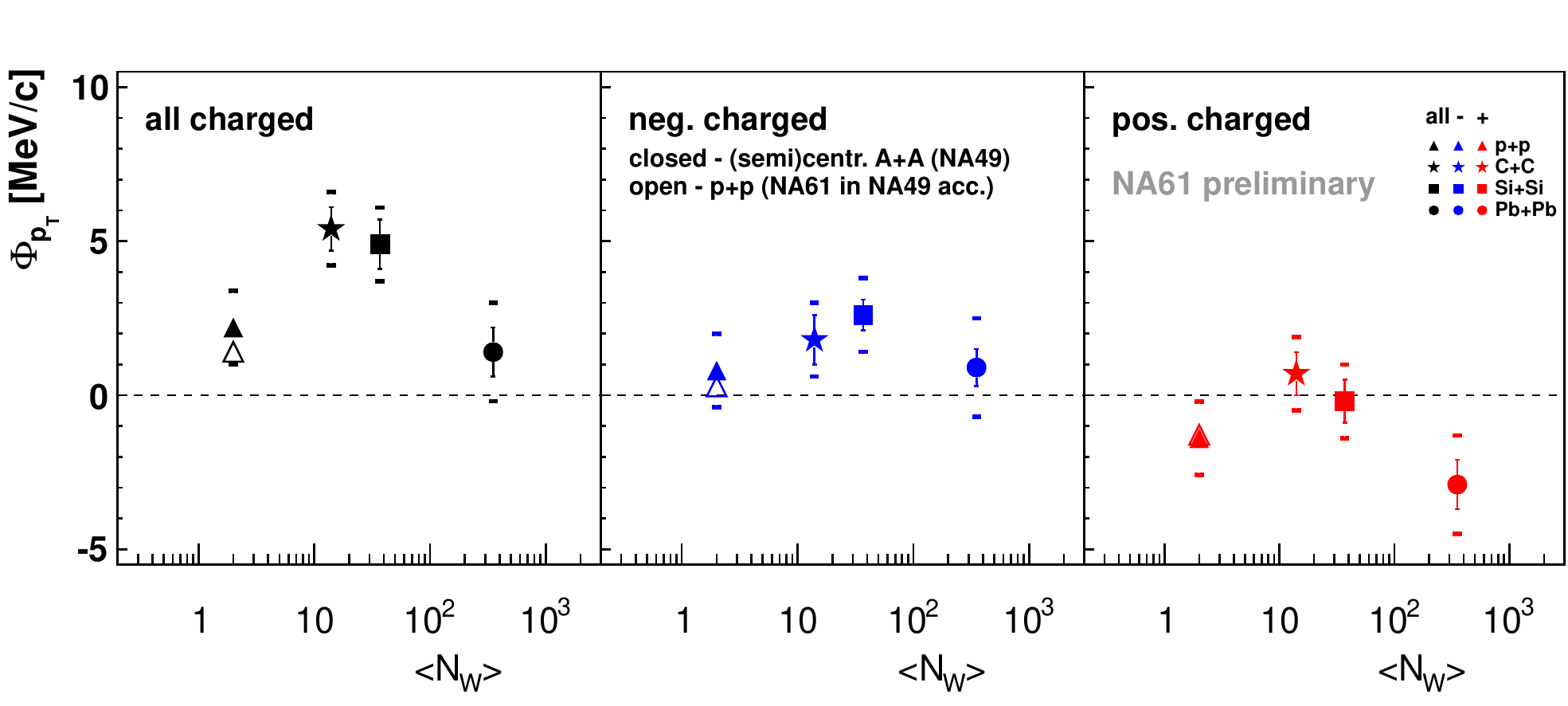}
\caption{Measure $\Phi_{p_T}$ of transverse momentum - multiplicity fluctuations
 of charged particles.
{\it Top:} versus $\mu_B$ for the 7.2~\% most central Pb+Pb collisions (full symbols, NA49~\cite{Anticic:2008aa})
and inelastic p+p reactions (open symbols, NA61/SHINE~\cite{Czopowicz:2015mfa,Aduszkiewicz:2015jna}). 
{\it Bottom:} versus the number of wounded nucleons $N_W$
in central C+C, Si+Si and Pb+Pb collisions at 158$A$ GeV (NA49~\cite{Anticic:2003fd}) and inelastic p+p interactions
(NA61/SHINE preliminary~\cite{Czopowicz:2015mfa,Aduszkiewicz:2015jna}). Results are for cms rapidity $1.1 < y < 2.6$ assuming the pion mass.
Curves illustrate the effect of the critical point~\cite{Grebieszkow:2009jr}.
\label{phipt}
}
\end{figure}

The measure $\Phi_{p_T}$ is related to the quantity $\Sigma$ (see Eq.~\ref{eq:phi}): 
\begin{equation}
  \Phi_{p_T} = \sqrt{\overline{p_T} \omega[p_T]} \left[\sqrt{\Sigma[P_{T},N]}-1\right].
\end{equation}
Results on the dependence of $\Phi_{p_T}$ on $\mu_B$ ($\sqrt{s_{NN}}$) 
in central Pb+Pb (NA49~\cite{Anticic:2008aa}) and
inelastic p+p collisions (NA61/SHINE preliminary~\cite{Czopowicz:2015mfa,Aduszkiewicz:2015jna}) 
are plotted in Fig.~\ref{phipt}~(top). The measurements
are compared to expectations for the CP~(solid curves in 
Fig.~\ref{phipt}~$top$~\cite{Grebieszkow:2009jr})
which were obtained in a similar manner like the predictions for $\omega$ under the assumption that the increase of
$\Phi_{p_T}$ at the CP amounts to 10~MeV/$c$ for a correlation length of $\xi$ = 3~fm.
However, 
more recent theoretical estimates~\cite{Athanasiou:2010kw} found much less sensitivity 
of $p_T$ fluctuations to the CP. 

\begin{figure}
\centering
\includegraphics[width=0.40\columnwidth]{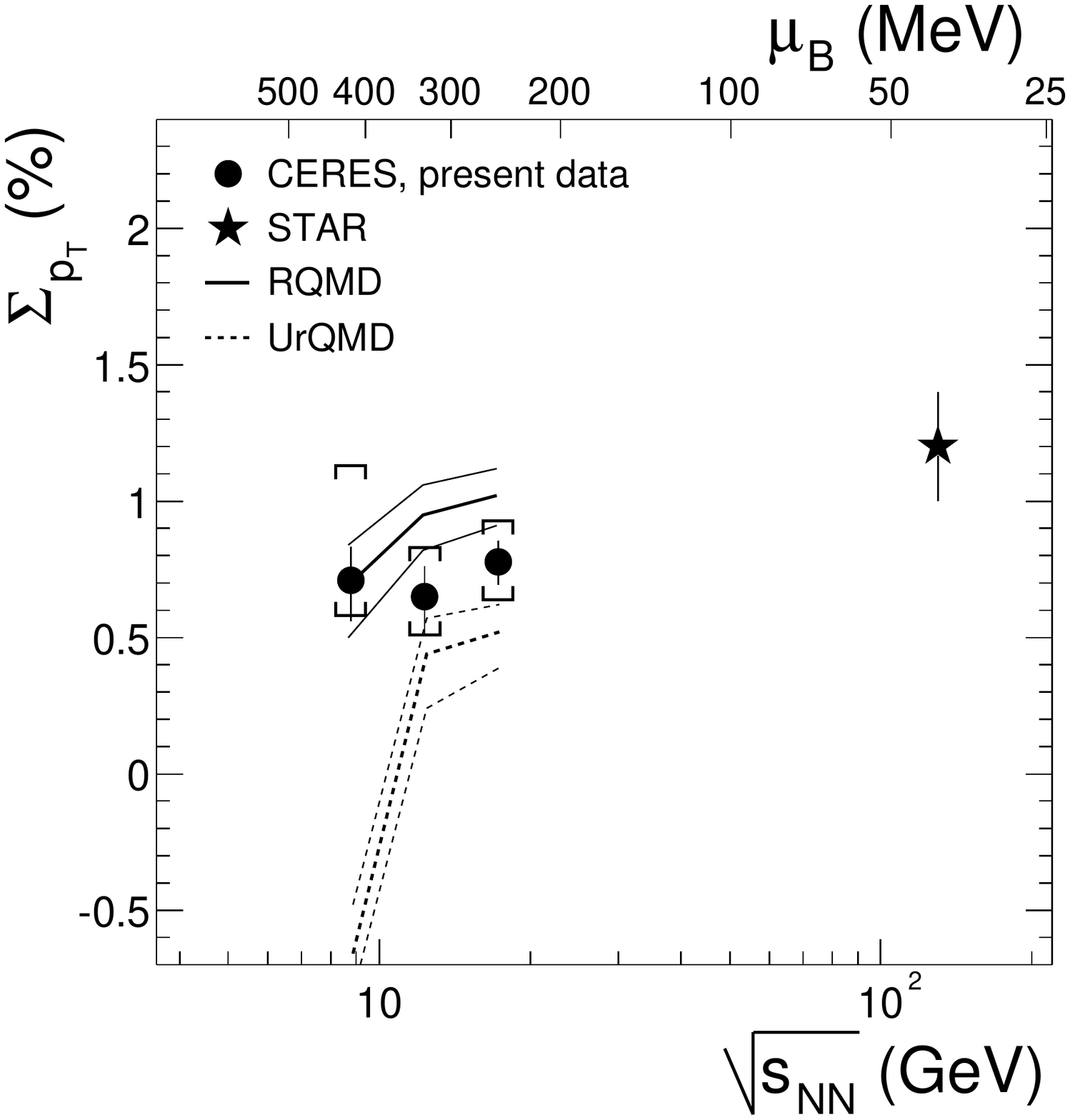}
\caption{The fluctuation measure $\Sigma_{p_{T}}$
as function of $\sqrt{s_{NN}}$ and of $\mu_B$~\cite{Adamova:2003pz}.
The full circles show CERES results (after short range correlation (SRC) removal) in central Au+Pb collisions
at 40, 80, and 158$A$~GeV/$c$ in the pseudo-rapidity range $2.2 < \eta < 2.7$. 
The brackets indicate the systematic errors.
Also shown is the STAR result~\cite{Voloshin:2001ei} at $\sqrt{s_{NN}}$ = 130~GeV
which is not corrected for SRC. Results and statistical errors from {\sc rqmd} and
{\sc urqmd} calculations (with re-scattering) are indicated as solid and dashed
lines, respectively.
\label{ceres:sigmapt}
}
\end{figure}

Some results on $\Phi_{p_T}$ for charged particles from central Au+Pb collisions were published by the
CERES experiment~\cite{Adamova:2003pz} at the SPS for beam energies of 40$A$, 80$A$ and 158$A$~GeV.
The results in the pseudo-rapidity acceptance of the experiment ($2.2 < \eta < 2.7$) are $1.1 \pm 0.4$,
$2.3 \pm 0.8$ and $3.3 \pm 0.7$ MeV, respectively for the 5~\% most central collisions with systematic
uncertainties of the order of 1.5 MeV. 
In order to account for a possible change of mean $p_{T}$ at
different beam energies, CERES defined a
dimensionless measure, the "normalised dynamical fluctuation"~$\Sigma_{p_T}$:
\begin{equation}
\Sigma_{p_T} \equiv
{\rm sgn}({\sigma^{2}_{p_{T},dyn}})\cdot \frac{\sqrt{|\sigma^{2}_{p_{T},dyn}|}}{\overline{p_{T}}}.
\end{equation}
Figure~\ref{ceres:sigmapt} shows that there is no significant energy dependence of this measure.

Measurements by NA49 for different
size nuclei at the top SPS energy of 158$A$~GeV are shown 
in Fig.~\ref{phipt}~({\it bottom}). 
As found for multiplicity fluctuations there may also be 
a maximum of transverse momentum fluctuations 
in medium-size nuclei.

\begin{figure}
\centering
\includegraphics[width=0.80\columnwidth]{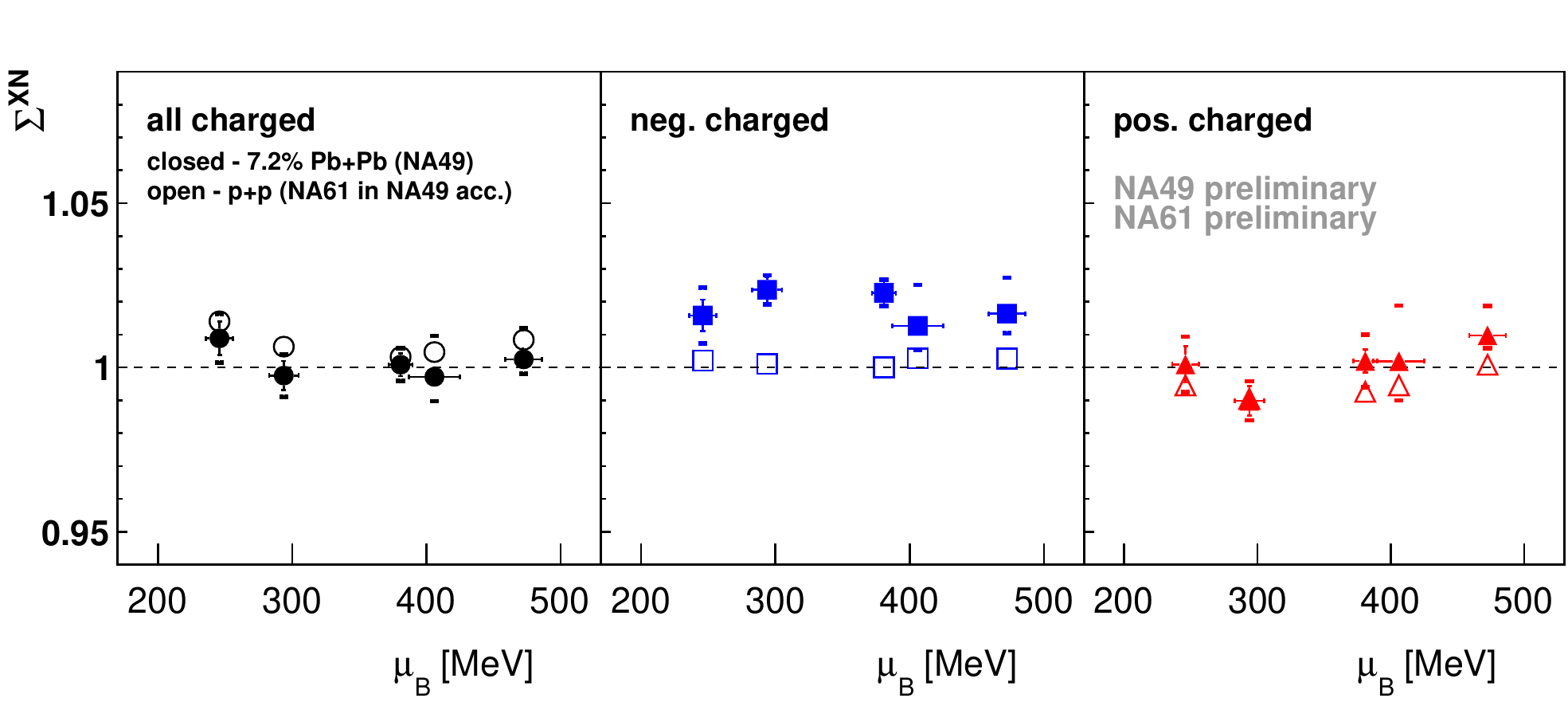}
\includegraphics[width=0.80\columnwidth]{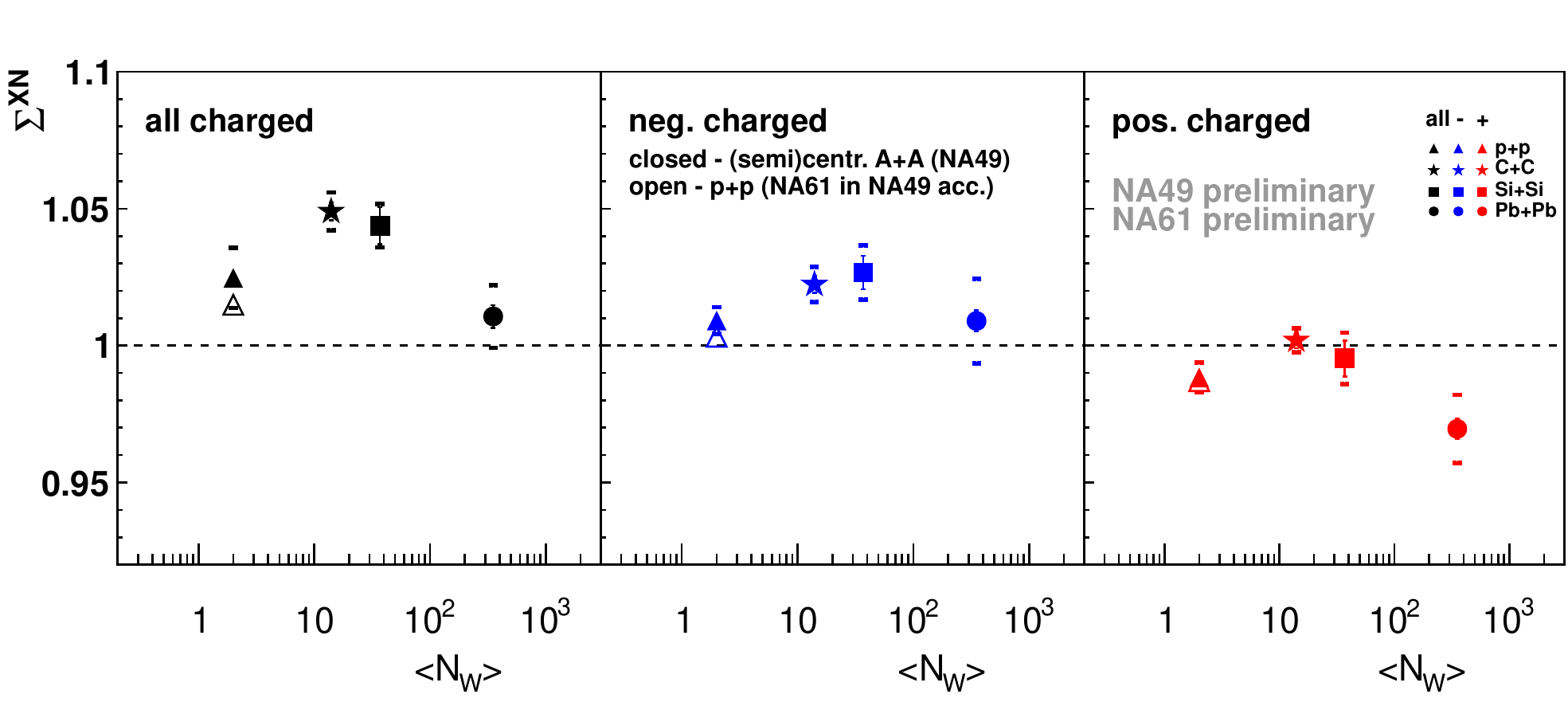}
\caption{Measure $\Sigma[P_T,N]$ (= $\Sigma^{P_T,N}$) of 
transverse momentum - multiplicity fluctuations
of charged particles.
{\it Top:} versus $\mu_B$ for the 7.2~\% most central Pb+Pb collisions (full symbols)
and inelastic p+p reactions (open symbols). {\it Bottom:} versus the number of wounded nucleons $N_W$
in inelastic p+p and central "C"+C, "Si"+Si and Pb+Pb collisions at 158$A$~GeV.
Results are for cms rapidity $1.1 < y < 2.6$ assuming the pion mass. 
(NA49 and NA61/SHINE~\cite{Czopowicz:2015mfa,Aduszkiewicz:2015jna}).
\label{Sigma}
}
\end{figure}

\begin{figure}
\centering
\includegraphics[width=0.80\columnwidth]{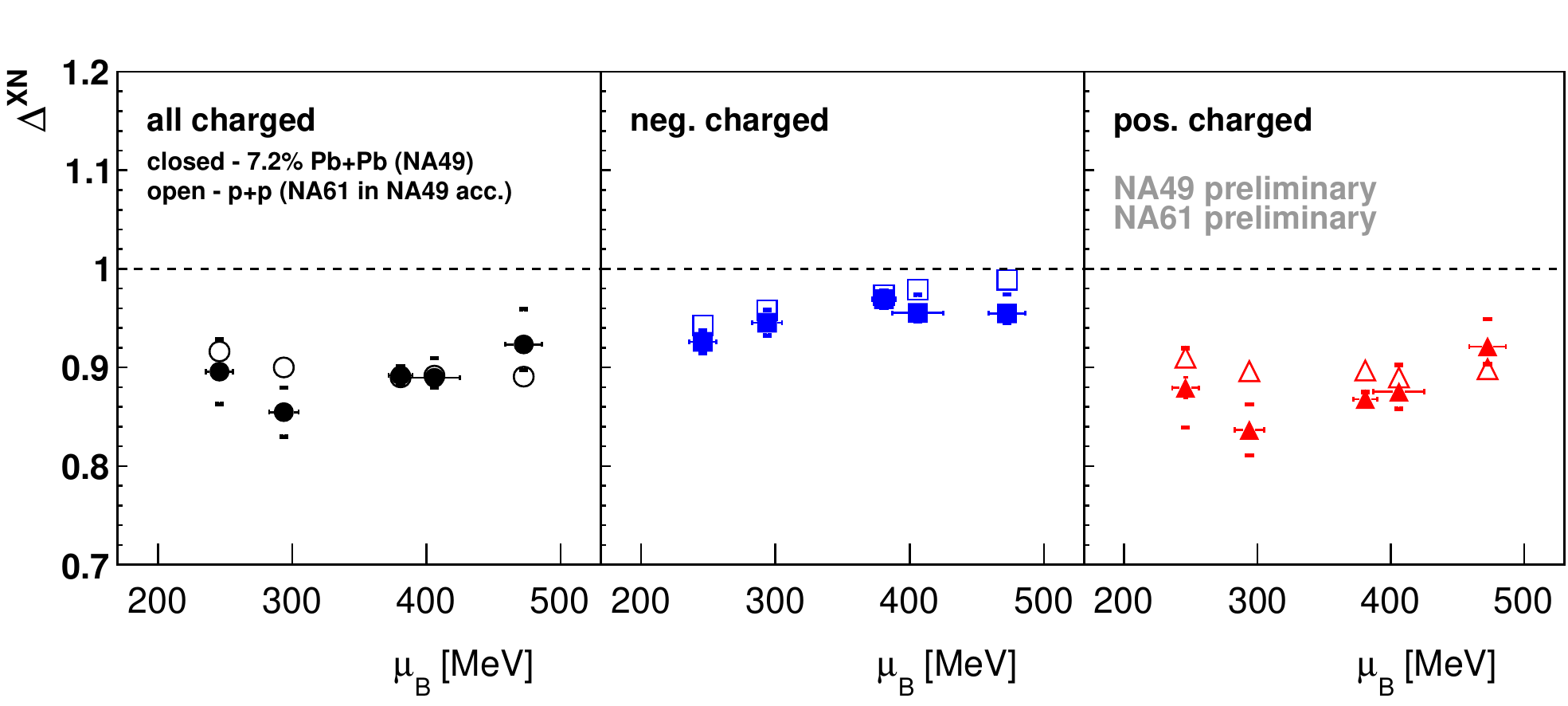}
\includegraphics[width=0.80\columnwidth]{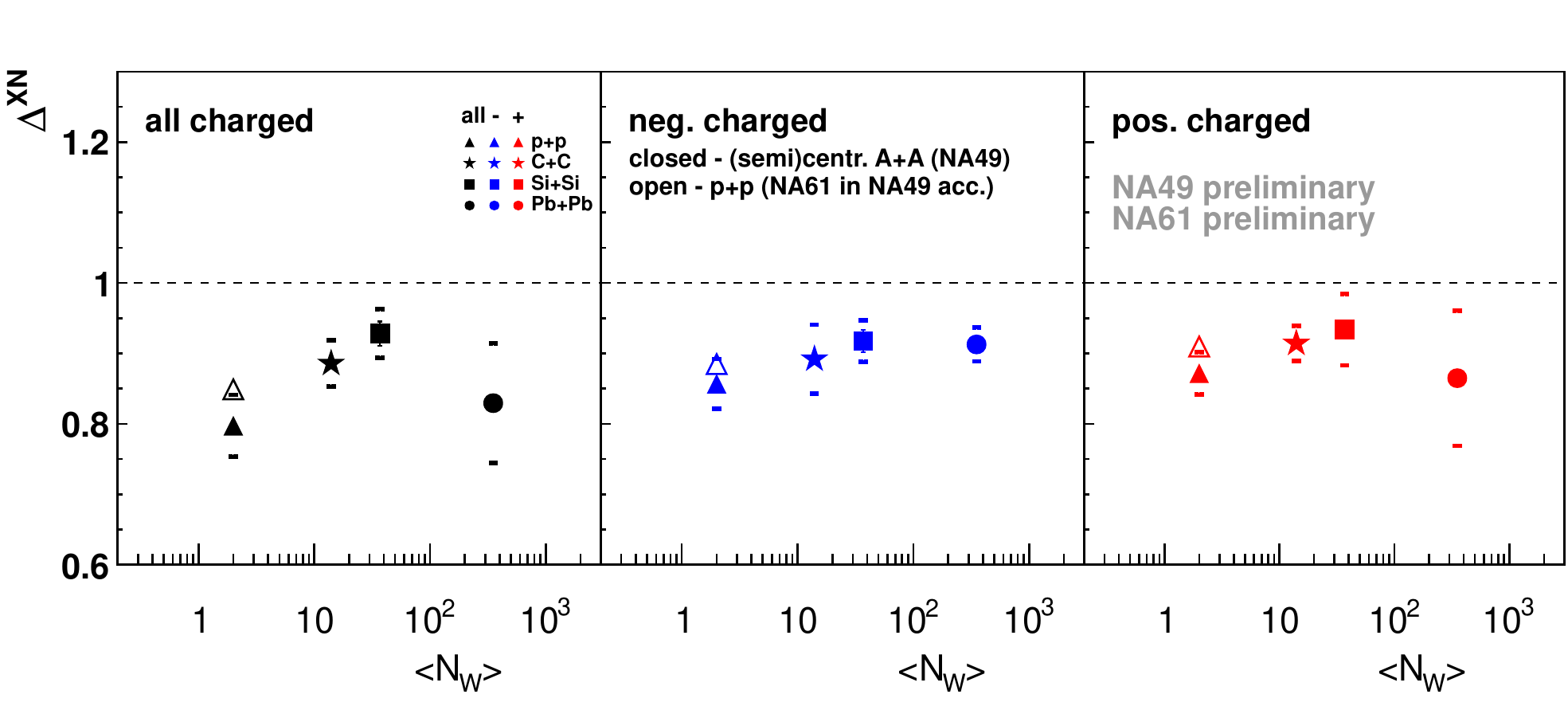}
\caption{Measure $\Delta[P_T,N]$ ( = $\Delta^{P_T,N}$) 
of transverse momentum - multiplicity 
fluctuations of charged particles.
{\it Top:} versus $\mu_B$ for the 7.2~\% most central Pb+Pb collisions (full symbols)
and inelastic p+p reactions (open symbols). 
{\it Bottom:} versus the number of wounded nucleons $N_W$
in inelastic p+p and central C+C, Si+Si and Pb+Pb collisions at 158$A$~GeV.
Results are for cms rapidity $1.1 < y < 2.6$ assuming the pion mass. 
(NA49 and NA61/SHINE~\cite{Czopowicz:2015mfa,Aduszkiewicz:2015jna}).
\label{Delta}
}
\end{figure}

The corresponding
results on $\Sigma[P_T,N]$~\cite{Seyboth:ismd2013} are presented
in Fig.~\ref{Sigma}. As expected from the close relation with $\Phi_{p_T}$
they indeed show behaviour consistent with that of $\Phi_{p_T}$. Results on
$\Delta[P_T,N]$~\cite{Seyboth:ismd2013} are shown in Fig.~\ref{Delta}. At present 
there are no predictions for the effect of the CP in these observables.

\begin{figure}[ht]
  \centering
  \includegraphics[width=0.48\textwidth]{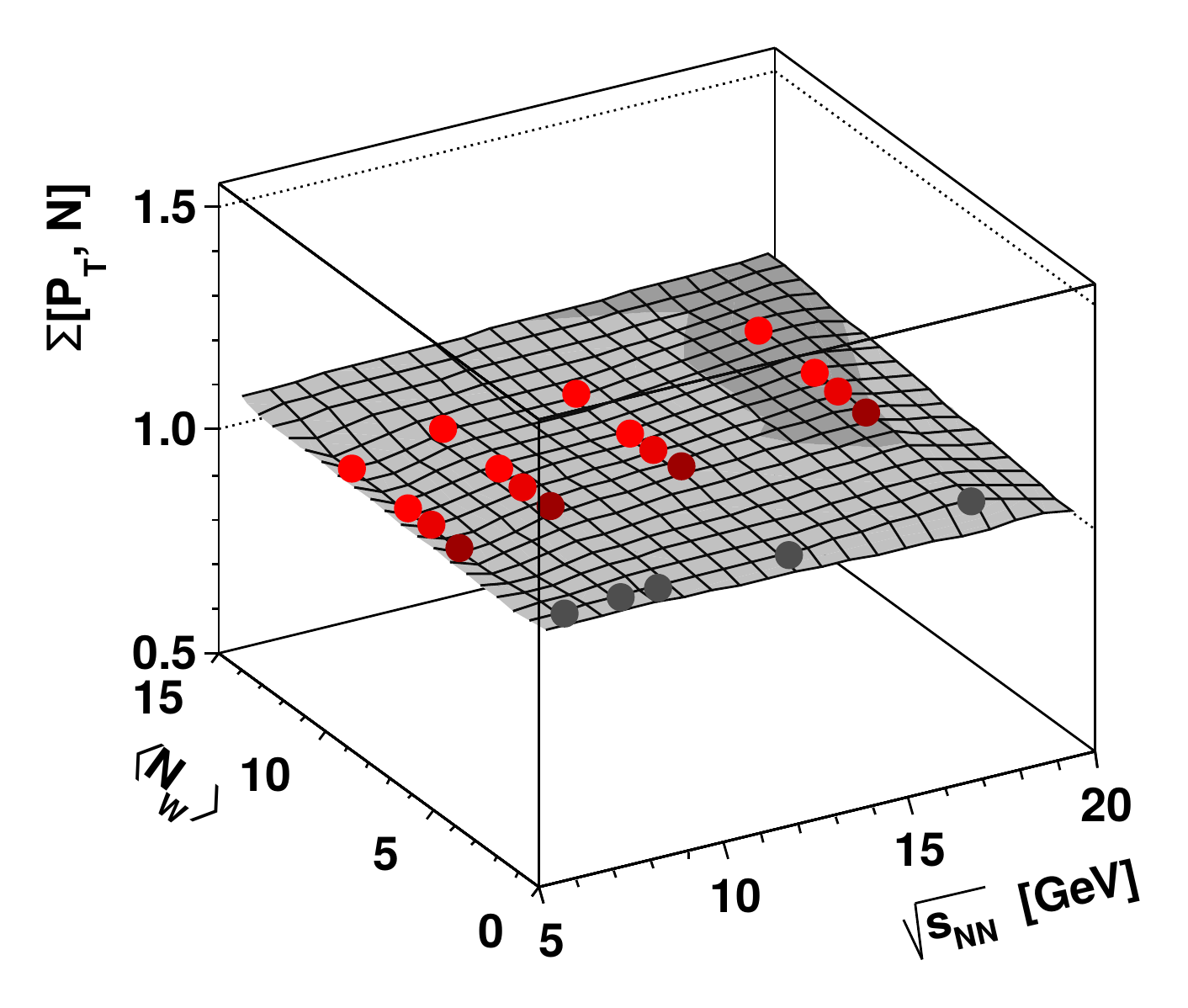}
  \includegraphics[width=0.48\textwidth]{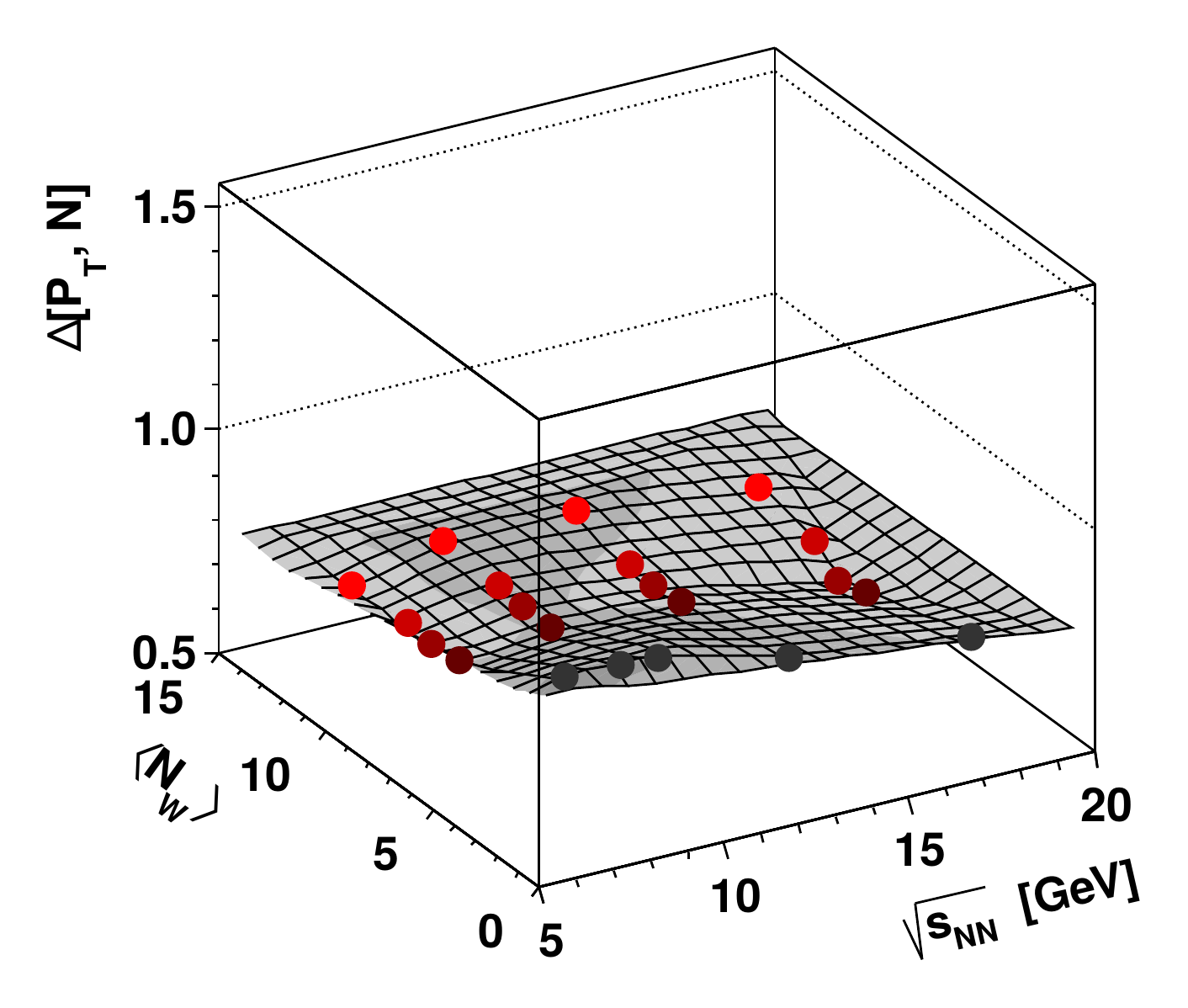}
  \caption{
Preliminary NA61/SHINE results on $\Sigma[P_T,N]$ and 
$\Delta[P_T,N]$ 
for Be+Be collisions of several collision centralities and energies
compared with the corresponding results 
from the energy scan with inelastic p+p interactions~\cite{Czopowicz:2015mfa,Aduszkiewicz:2015jna}.
\label{fig:2D}
  }
\end{figure}

Finally NA61/SHINE results~\cite{Czopowicz:2015mfa,Aduszkiewicz:2015jna} 
from the two-dimensional scan
in system size and collision energy are presented in Fig.~\ref{fig:2D}.
The data come from inelastic p+p interactions and centrality selected
Be+Be collisions. No indication of of the ''critical hill'' is observed.

Data on Ar+Sc collisions at 13$A$, 19$A$, 30$A$, 40$A$, 75$A$ and 150$A$~GeV/c
are already recorded by NA61/SHINE. They may lead to the discovery of the
critical point of strongly interacting matter as possibly suggested by the
first indications seen by the NA49 experiment and discussed in this review.

\clearpage

\section{Conclusions and Outlook}\label{sec:outlook}

The continuing search in nucleus-nucleus reactions for the maximum of 
fluctuations predicted
for the critical point of strongly interacting matter has not yet turned 
up firm evidence in collisions of heavy nuclei either at the CERN SPS 
(central Pb+Pb collisions) or in the RHIC BES program (Au+Au collisions). 

It is intriguing that both the fluctuations of 
quantities integrated over the full experimental acceptance 
(event multiplicity and transverse momentum) as well as
the bin size dependence of the second factorial moment of pion and proton
multiplicities    
in medium-sized Si+Si collisions
at 158$A$~GeV/c possibly suggest critical behaviour of the created matter.

These results provide strong motivation for the ongoing systematic
scan of the phase diagram by the NA61/SHINE experiment at the SPS
(see Fig.~\ref{na61_na49_scan})
and the continuing search at the Brookhaven Relativistic Hadron Collider.


\vspace{1cm}
{\bf Acknowledgements}

\vspace{0.2cm}
We are grateful to Andrzej Bialas, Tobiasz Czopowicz,
Katarzyna Grebieszkow and Mark Gorenstein 
for help and critical comments.
We would like to thank Helmut Satz for the motivation
to write this review.

This work was supported by
the National Science Centre of Poland (grant
UMO-2012/04/M/ST2/00816),
the German Research Foundation (GA 1480\slash 2-2).

\end{document}